\documentclass{article}

\PassOptionsToPackage{table}{xcolor}
\usepackage[T1]{fontenc}
\usepackage{iclr2027_conference,times}

\usepackage{microtype}
\usepackage{graphicx}
\usepackage{subcaption}
\usepackage{booktabs} %
\usepackage{float}    %
\usepackage{placeins} %
\usepackage{listings}
\usepackage{xcolor}

\usepackage{hyperref}
\usepackage{url}

\ifdefined\HiLiftAnonymous\else
\iclrfinalcopy
\fi

\usepackage{amsmath}
\usepackage{amssymb}
\usepackage{mathtools}
\usepackage{amsthm}
\usepackage[version=4]{mhchem}
\usepackage{siunitx}
\usepackage{longtable,tabularx}
\usepackage[normalem]{ulem}

\definecolor{mlScoreBest}{RGB}{126,194,126}
\definecolor{mlScoreGood}{RGB}{198,226,198}
\definecolor{mlScoreMid}{RGB}{244,244,244}
\definecolor{mlScoreBad}{RGB}{248,197,176}
\definecolor{mlScoreWorst}{RGB}{238,133,100}

\newcommand{\natAoafourGeotSurfCpRelLtwo}{4.91}
\newcommand{\natAoafourGeotSurfTauRelLtwo}{9.31}

\newcommand{\natAoafourGeotVolPRelLtwo}{4.34}

\newcommand{\natAoafourGeotVolVRelLtwo}{5.99}

\newcommand{\natAoatwelveGeotSurfCpRelLtwo}{3.00}
\newcommand{\natAoatwelveGeotSurfTauRelLtwo}{8.17}

\newcommand{\natAoatwelveGeotVolPRelLtwo}{4.09}

\newcommand{\natAoatwelveGeotVolVRelLtwo}{7.49}

\newcommand{\natAoatwentytwoGeotSurfCpRelLtwo}{11.65}
\newcommand{\natAoatwentytwoGeotSurfTauRelLtwo}{14.08}

\newcommand{\natAoatwentytwoGeotVolPRelLtwo}{14.03}

\newcommand{\natAoatwentytwoGeotVolVRelLtwo}{17.83}

\newcommand{\natangleGeometryAoafourGeotSurfCpRelLtwo}{3.59}
\newcommand{\natangleGeometryAoafourGeotSurfTauRelLtwo}{8.04}
\newcommand{\natangleGeometryAoafourGeotVolPRelLtwo}{3.49}
\newcommand{\natangleGeometryAoafourGeotVolVRelLtwo}{5.13}

\newcommand{\natangleGeometryAoatwelveGeotSurfCpRelLtwo}{2.67}
\newcommand{\natangleGeometryAoatwelveGeotSurfTauRelLtwo}{7.68}
\newcommand{\natangleGeometryAoatwelveGeotVolPRelLtwo}{3.10}
\newcommand{\natangleGeometryAoatwelveGeotVolVRelLtwo}{5.43}

\newcommand{\natangleGeometryAoatwentytwoGeotSurfCpRelLtwo}{8.55}
\newcommand{\natangleGeometryAoatwentytwoGeotSurfTauRelLtwo}{10.52}
\newcommand{\natangleGeometryAoatwentytwoGeotVolPRelLtwo}{9.66}
\newcommand{\natangleGeometryAoatwentytwoGeotVolVRelLtwo}{11.88}

\usepackage[capitalize,noabbrev]{cleveref}

\theoremstyle{plain}

\theoremstyle{definition}

\theoremstyle{remark}

\usepackage[disable,textsize=tiny]{todonotes}

\title{HiLiftAeroML: A High-Fidelity\\Computational Fluid Dynamics Dataset\\for High-Lift Aircraft Aerodynamics}

\author{%
\textbf{Neil Ashton\textsuperscript{\(\dagger\)}\thanks{Correspondence to \texttt{neil@neilashton.co.uk}.}\quad
Adam Clark\textsuperscript{\(\ddagger\)}\quad
Konrad Goc\textsuperscript{\(\ddagger\)}\quad
Liam Heidt\textsuperscript{\S}\quad
Christopher Ivey\textsuperscript{\S}} \\
\textbf{Rahul Agrawal\textsuperscript{\S}\quad
Sanjeeb T Bose\textsuperscript{\S}\quad
Corey Adams\textsuperscript{\(\dagger\)}\quad
Peter Sharpe\textsuperscript{\(\dagger\)}\quad
Daniel Leibovici\textsuperscript{\(\dagger\)}} \\
\textbf{Semih Akkurt\textsuperscript{\(\dagger\)}\quad
Sheel Nidhan\textsuperscript{\(\dagger\)}\quad
Rishi Ranade\textsuperscript{\(\dagger\)}\quad
Jean Kossaifi\textsuperscript{\(\dagger\)}} \\
\textnormal{\textsuperscript{\(\dagger\)}NVIDIA, Santa Clara, US} \\
\textnormal{\textsuperscript{\(\ddagger\)}The Boeing Company, Seattle, US} \\
\textnormal{\textsuperscript{\S}Cadence Design Systems, Santa Clara, US}
}

\begin{document}

\maketitle

\begin{abstract}
HiLiftAeroML is, to our knowledge, the first open high-fidelity computational fluid dynamics dataset dedicated to high-lift aircraft aerodynamics. It contains 1,800 simulations spanning 180 variants of the NASA Common Research Model high-lift configuration and ten angles of attack from \(4^\circ\) to \(22^\circ\). Each case was generated with a GPU-accelerated explicit wall-modeled large-eddy simulation approach on solution-adapted grids of 300--500 million cells, covering attached, separated, and post-stall flow conditions. Comparisons with wind-tunnel measurements for reference landing configurations show good agreement in integrated loads and sectional pressures, with grid adaptation substantially improving drag and pitching-moment predictions. The CC-BY-4.0 release includes geometries, time-averaged surface and volume fields, integrated loads, validation material, and deterministic benchmark splits. Initial GeoTransolver and Transolver baselines, evaluated on the complete native surface and volume support of every held-out case rather than on a sampled subset, reconstruct the fields well for several interpolation and held-out-geometry tests, while high-angle separated flow, limited-data training, and out-of-distribution flow-regime shifts remain substantially harder. The dataset and baselines provide a common resource for developing and assessing data-driven models for realistic high-lift aerodynamics.
\end{abstract}

\section{Introduction}
High-lift aircraft configurations produce three-dimensional separation, transitional boundary layers, and vortex interactions during take-off and landing. These flows remain difficult for Reynolds-Averaged Navier--Stokes (RANS) methods, sustaining reliance on wind-tunnel testing even as CFD has become established for cruise design~\citep{Slotnick2014,clark2025HLPW5}. Scale-resolving approaches, including hybrid RANS--LES and wall-modeled large-eddy simulation (WMLES), improve predictions in this regime, although flap separation and slat transition remain challenges~\citep{ashton2023b,cetin2023b,hlcrmkonrad}. Their computational cost makes broad sweeps over geometry and operating conditions expensive~\citep{Appa2021,nielsen2024,hosseinverdi2025rapidus}. 

\subsection{Machine Learning and Related Work}

Surrogate models can accelerate repeated evaluations across flap/slat settings and angles of attack (AoA), helping screen candidates before further CFD or experiments~\citep{ashton2025fluidintelligenceforwardlook}. Their usefulness depends on both the fidelity and coverage of the training data and their generalisation to new conditions.

Graph neural networks~\citep{nabian2024x} and transformer-based neural operators~\citep{Bleeker2024,alkin2025abuptautomotiveaerospaceapplications,wen2025geometryawareoperatortransformer,adams2025geotransolverlearningphysicsirregular} provide geometric representations for aerodynamic surrogate models. Their assessment depends on datasets that expose physically meaningful changes in geometry and flow regime.

Table~\ref{table:aerodatasets} compares representative open aircraft datasets. These resources cover transonic wings, wing--body configurations, and blended-wing-body designs, using steady RANS simulations with varying geometric and operating-condition coverage. HiLiftAeroML complements them with WMLES of deployed high-lift systems across attached, separated, and post-stall conditions. This follows the release of scale-resolving automotive datasets such as DrivAerML, WindsorML, and AhmedML~\citep{ashton2024drivaer,ashton2024windsorCorrected,ashton2024ahmedCorrected}. Independent use is already demonstrated by AeroJEPA~\citep{giral2026aerojepa}, which evaluates aerodynamic field models on HiLiftAeroML.

\begin{table}[t!] 
\centering
\resizebox{\textwidth}{!}{%
\begin{tabular}{l l l l l l l} 
 \toprule
 \textbf{Dataset} & \textbf{Samples} & \textbf{Type} & \textbf{Turbulence Model} & \textbf{Mesh} & \textbf{Mach} & \textbf{AoA Range} \\
 
 \midrule
 \multicolumn{7}{c}{\textit{Wing-Only Datasets}} \\
 \midrule
 SuperWing \cite{yang2025superwingcomprehensivetransonicwing} & 28,856 & Surf+Vol & SA RANS & $\approx$ 3.6M & 0.75-0.9 & +$2^\circ$ - +$12^\circ$ \\
 Emmi-Wing \cite{paischer2025goingspeedsoundpushing} & 30,000 & Surf+Vol & SA RANS & n/a & 0.44-0.88 & -$10^\circ$ - +$10^\circ$ \\
 
 \midrule
 \multicolumn{7}{c}{\textit{Full Aircraft Datasets}} \\
 \midrule
 DLR \cite{bekemeyer2025} & 149 & Surface & SA-QCR RANS & $\approx 43$M & 0.5--0.88 & $-2.5^\circ$ to $+7.5^\circ$ \\
 ONERA \cite{PETER2025106838} & 468 & Surface & SA-QCR RANS & $\approx 39$M & 0.3--0.96 & $-15^\circ$ to $+15^\circ$ \\
 SHIFTWing \cite{shift_wing_2025} & 1698 & Surf+Vol & SA-RANS & $\approx 6$M & 0.5, 0.85 & $0^\circ$ to $+4^\circ$  \\ 
 BlendedNet++ \cite{sung2025blendednetlargescaleblendedwing} & 12,490 & Surface & SA-RANS & $\approx 11$M & 0.05--0.5 & $-8^\circ$ to $+16^\circ$ \\
 \textbf{HiLiftAeroML} & \textbf{1,800} & \textbf{Surf+Vol} & \textbf{WMLES} & $\mathbf{300-500M}$ & \textbf{0.2} & $\mathbf{+4^\circ}$ \textbf{to} $\mathbf{+22^\circ}$ \\
 \bottomrule
\end{tabular}%
}
\caption{Comparison of Aircraft CFD datasets, categorized by geometry type. HiLiftAeroML targets high-lift configurations with WMLES; the other listed datasets use RANS.}
\label{table:aerodatasets}
\end{table}

\subsection{Dataset and Benchmark Contributions}
The geometry parametrization, WMLES workflow, experimental validation, and initial aerodynamic characterization of HiLiftAeroML were reported by \citet{ashton2026hiliftgeneration}. That study introduced the 1,800-case simulation campaign and explicitly deferred surrogate-model benchmarking to subsequent work.

This paper establishes the machine-learning benchmark built on that campaign. Its contributions are 14 deterministic split families for interpolation, data efficiency, unseen geometries, and distribution shifts; a complete-native-support evaluation protocol with explicit field, integrated-load, and profile metrics; and GeoTransolver and Transolver baselines across 12 split families, comprising 23 model--split records. Beyond these benchmarks, the paper substantially expands the methodological and dataset documentation, including the numerical formulation, time-averaging and post-processing procedures, field and reference conventions, release corrections and validity masks, native-point surface quadrature, measured computational costs, and guidance on data access, versioning, and reuse. The experiments distinguish generalization across geometric settings from extrapolation across flow regimes: high-deflection tests remain close to in-distribution accuracy, whereas AoA and stall extrapolation produce severe errors. The benchmark, evaluation protocol, empirical findings, and expanded documentation extend the earlier data-generation study.

\section{Data Generation}
This section summarizes the simulation campaign introduced by \citet{ashton2026hiliftgeneration}, including its geometry parametrization, WMLES workflow, and experimental validation, to establish the provenance of the benchmark data.

\subsection{Reference Geometry} \label{sec:refGeom}

\begin{figure}[!htb]
\centering
\includegraphics[width=0.60\linewidth]{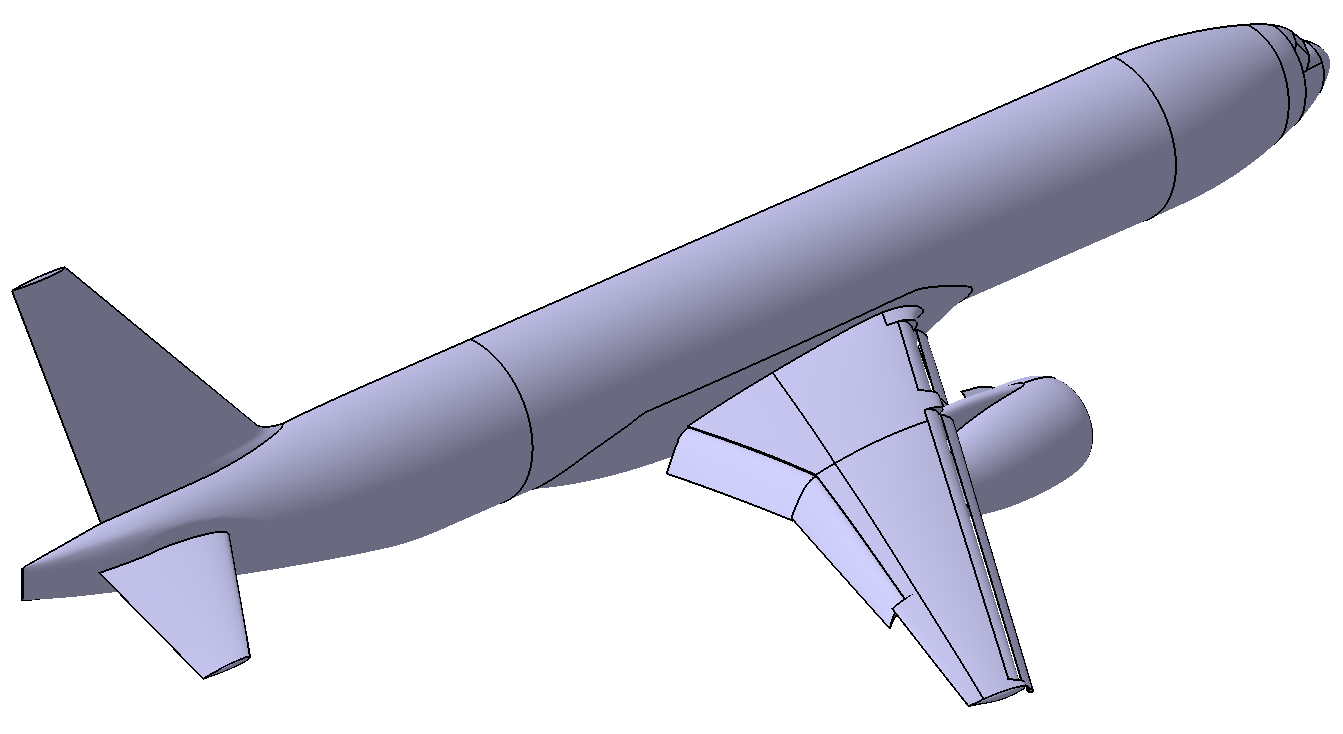}
\caption{CRM-HL Geometry shown in the Reference Landing Configuration\label{fig:CRMHL_geom_main}}
\end{figure}

The geometric foundation for this dataset is the high-lift variant of the NASA Common Research Model (CRM-HL). Established by \citet{lacy2016development} as a representative baseline for commercial transport aircraft, the CRM-HL iterates upon the transonic CRM \cite{vassberg2008crm} by integrating a new wing design with complex high-lift systems. The configuration includes inboard and outboard leading-edge slats, trailing-edge flaps, flap support fairings, and a nacelle-pylon assembly. Following an extensive testing campaign \cite{lacy2020geom}, the definition was expanded to include vertical and horizontal tail surfaces and standardized settings for takeoff and landing (see Fig. \ref{fig:CRMHL_geom_main}).

A key methodological distinction in this work is the use of the theoretical reference geometry rather than a specific wind-tunnel model. While previous AIAA High-Lift Prediction Workshops relied on physical model replications, such as the ONERA LRM or the NASA 5.2\% scale semi-span model \cite{clark2025HLPW5}, our focus on geometric parameterization necessitates a different approach. Specifically, mechanical features such as slat brackets and support fairings must dynamically track the movement of high-lift surfaces. To facilitate this, we replace model-specific hardware details with simplified, parametric equivalents, thereby decoupling the simulation geometry from the constraints of physical wind-tunnel hardware.

\subsection{Geometry parameterization and boundary condition variation}\label{sec:geometry_parameterization}
The inboard and outboard slats and flaps each have an independently varied deflection and gap multiplier, giving eight geometry parameters (Table~\ref{table:dataset_parameters}, left). Slat deflections range from \(10^\circ\) to \(35^\circ\), and flap deflections from \(10^\circ\) to \(45^\circ\). Gap multipliers span 0.5--1.5 relative to deflection-dependent schedules through the reference take-off and landing positions. Slat height and flap overlap follow those schedules without independent variation. The device-positioning diagram and detailed schedules are provided in Appendix~\ref{app:geometry_schedules} (Fig.~\ref{fig:CRMHL_pos_main}).

\begin{table}[!htb]
\centering
\caption{Geometry parameter ranges (left) and reference conditions (right). IB: inboard; OB: outboard; MAC: mean aerodynamic chord.}
\label{table:dataset_parameters}
\small
\setlength{\tabcolsep}{4pt}
\begin{tabular}{@{}lcc@{\hspace{1.5em}}lrll@{}}
\toprule
\multicolumn{3}{c}{Geometry ranges} & \multicolumn{4}{c}{Reference conditions} \\
\cmidrule(r){1-3}\cmidrule(l){4-7}
Parameter & Min. & Max. & Variable & Value & Units & Description \\
\midrule
IB Slat Deflection & $10^\circ$ & $35^\circ$ & $c_{ref}$ & 275.80 & inches & MAC \\
OB Slat Deflection & $10^\circ$ & $35^\circ$ & $b_{ref}$ & 1156.75 & inches & 1/2 wingspan \\
IB Flap Deflection & $10^\circ$ & $45^\circ$ & $s_{ref}$ & 297360.0 & inches$^2$ & planform half-area \\
OB Flap Deflection & $10^\circ$ & $45^\circ$ & M & 0.2 & & Mach Number \\
IB Slat Gap Multiplier & 0.5 & 1.5 & Re & $1.6 \times 10^{6}$ & & Reynolds Number \\
OB Slat Gap Multiplier & 0.5 & 1.5 & AoA & 4--22 & degrees & Angle of Attack \\
IB Flap Gap Multiplier & 0.5 & 1.5 & $T_{ref}$ & 518.67 & Rankine & Temperature \\
OB Flap Gap Multiplier & 0.5 & 1.5 & $P_{atm}$ & 14.696 & Psi & Atmospheric Pressure \\
\bottomrule
\end{tabular}
\end{table}

Each geometry is simulated at ten angles of attack (AoA), from $4^\circ$ to $22^\circ$ in $2^\circ$ increments, to capture pre-stall and post-stall aerodynamic characteristics across the flap and slat configurations.

\subsection{Reference case setup}
The baseline case setup follows the 5th High-Lift Workshop, with reference dimensions and flow conditions summarized in Table~\ref{table:dataset_parameters} (right). The simulations are performed at a chord based Reynolds number, $Re_{MAC} = 1.6 \times 10^6$ to aid the ongoing work relating to the impact of slat-transition, and also minimize its aerodynamic impacts. The semi-span airframe is mounted on the symmetry plane of a hemispherical domain with a radius of $\approx 75\times MAC$. Following \citet{hlcrmkonrad,agrawal2023arbntf}, at the inlet (the forward half of the hemisphere), a uniform freestream is prescribed at the case angle of attack. All solid boundaries on the aircraft model are treated viscously with the equilibrium wall model. In the outlet region, a characteristic non-reflecting boundary condition is specified with an outlet pressure \citep{poinsot1992boundary}. The symmetry plane is treated with a no-stress boundary condition. A schematic is provided in Appendix~\ref{app:boundary_conditions}. \\

\subsection{Flow Solver \& Discretization}\label{sec:flow_solver}

The simulations use the Fidelity Charles flow solver, which is an explicit, unstructured, finite-volume solver for the compressible Navier-Stokes equations. The solver is  2\textsuperscript{nd}-order accurate in space and 3\textsuperscript{rd}-order accurate in time. More details of the solver, as well as relevant validation cases on aircraft flows, can be found in \citet{bres2018large} and \citet{goc2021large,agrawal2025Flap,transonickonrad} as well as in Appendices~\ref{app:numMeth}--\ref{app:validation}. The solver uses operators that are formally skew-symmetric in order to discretely conserve kinetic energy. The numerical discretization also approximately preserves entropy. \citet{konradthesis,agrawalthesis} have highlighted that dynamic subgrid-scale models provide improved accuracy over constant coefficient models, especially in the context of flow separation. Informed by these investigations, the dynamic Smagorinsky subgrid-scale model \citep{moin1991dynamic} is employed in this work. Wall shear stress is supplied by an equilibrium wall model \citep{lehmkuhl2018large}. It is remarked that no explicit treatment for the flow-transition is utilized in the present simulations. The simulations are performed in a constant CFL mode. \\

\noindent
Fidelity Stitch constructs unstructured Voronoi meshes~\citep{bres2018}, with approximately isotropic cells, Cartesian packing in the freestream, and refinement near solid boundaries. Solution adaptation then determines the local resolution for each geometry and angle of attack. Appendix~\ref{app:numMeth} gives the grid illustration (Fig.~\ref{fig:baselinemesh_main}), numerical equations, wall treatment, and boundary conditions.

\subsection{Workflow}

First, CAD geometry is defined in a fully parameterized fashion in CATIA. For a given case, the native CATPart is converted into IGES using CADfix by ITI Global. From there, the IGES file describing the particular case is read into HeldenMesh by Helden Aerospace, cleaned up, and run to generate both a coarse and a refined surface only triangulation of the geometry. The coarse is subsequently converted into an STL for use in downstream inference. The refined surface triangulation is read into ``Fidelity Surfer'', where the outer domain is added and the resulting surface (triangulation) is saved for volume grid generation. Next, an initial surface and volume mesh is created resulting in a baseline grid, approximately $100 \times 10^6$ control volumes. An initial solution was computed on this mesh for 15 convective time units (CTU). Subsequently, a solution adaptation algorithm \citep{agrawal2023reynolds,agrawal2026aiaa} was used to adapt the surface and consequently the volume grid (uniquely for each geometry and angle of attack). The flow was further simulated on the adapted mesh for a nominal 15 CTU for AoA $\leq 12^\circ$ (where separation is minimal) and 30-200 CTU (dependent on each case) for simulations with AoA $> 12^\circ$. Time-averaging was performed after flushing out initial transients (10 CTU) in all cases. Simulations were typically each run on 8 NVIDIA GB200 GPU nodes. It is acknowledged that for post-stall conditions, more CTU may be required but a compromise was taken for computational resource efficiency. The surface and volume solution were then exported from Charles to the widely used .vtu format (see Appendix~\ref{app:numMeth} for details of the volume export process) \\

\subsection{Validation}
\begin{figure}[!htb]
    \centering
    \includegraphics[width=\linewidth]{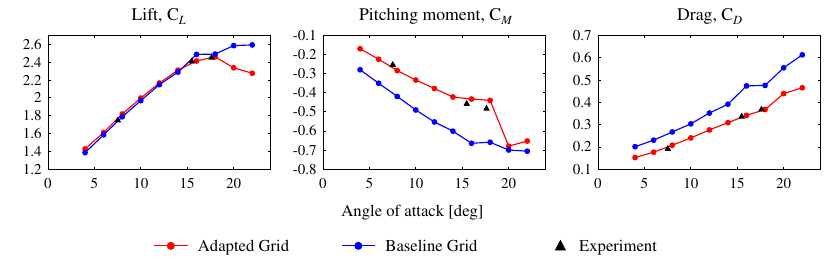}

    \caption{Comparison of predicted lift (left), pitching moment (centre), and drag (right) across the angle-of-attack sweep with experiments \citep{mouton2024testing} for the with-tail and with-stabilizer configuration at $Re_{MAC} = 1.6 \times 10^6$.}
    \label{fig:forces}
\end{figure}

To establish the baseline accuracy of the simulations, we compare the integrated forces from the present wall-modeled LES with the experiments \citep{mouton2024testing} in a landing configuration with vertical and horizontal stabilizers (experimentally identified as LDG-HV). This configuration was embedded into the training dataset as a blind validation test based on the semi-automated solution procedure that is detailed above.   For the set of conditions selected (at the chosen Reynolds number and angle of attack range for the LDG-HV geometry), there are three angles of attack measured from the 
experiment.  Fig. \ref{fig:forces} confirms that the integrated forces (lift, pitching moment and drag) are reasonably predicted relative to the experiments, in particular, on the adapted grid.  While the lift 
is not greatly changed due to the grid adaptation process, significant 
improvements in the drag coefficient and pitching moment are observed
with favorable comparisons against the experiments on the adapted grid \citep{agrawal2023reynolds}.
Note that some of the non-smoothness observed on the coarse grid at 
high angles of attack (particularly in the lift polar) can be partially
attributed to the relatively short time history (as this grid is only 
used to drive the adaptation process).  \\

\noindent
The near-stall comparison also captures the observed outboard wedge-shaped separation and separation near the Yehudi break. Appendix~\ref{app:validation} contains the LDG-HV oil-flow comparison (Fig.~\ref{fig:tau18}) and sectional-pressure validation for the related LDG configuration without the tail. These checks do not establish uniform accuracy across every released geometry.

\section{Dataset}
\subsection{Design and release}
Latin-hypercube sampling using SciPy~\citep{2020SciPy-NMeth} generates 180 variants across the eight geometry parameters in Table~\ref{table:dataset_parameters}. Simulating each variant at ten angles of attack gives 1,800 cases. Each case includes STEP/STL geometry, time-averaged surface and volume fields, integrated forces and moments, reference quantities, convergence plots, and native-point surface quadrature weights.

The CC-BY-4.0 release is available at \url{https://huggingface.co/datasets/nvidia/HiLiftAeroML}. It occupies approximately 66.9~TB as stored on the Hub; users can begin with metadata and individual cases using the revision-pinned download examples in Appendix~\ref{app:dataset-access}. Appendix~\ref{app:dataset-dataDesc} gives the file and field catalogue, and Appendix~\ref{app:detailedml} specifies benchmark inputs and evaluation.

\subsection{Aerodynamic coverage}
\begin{figure}[!htb]
\begin{center}
\includegraphics[width=\textwidth]{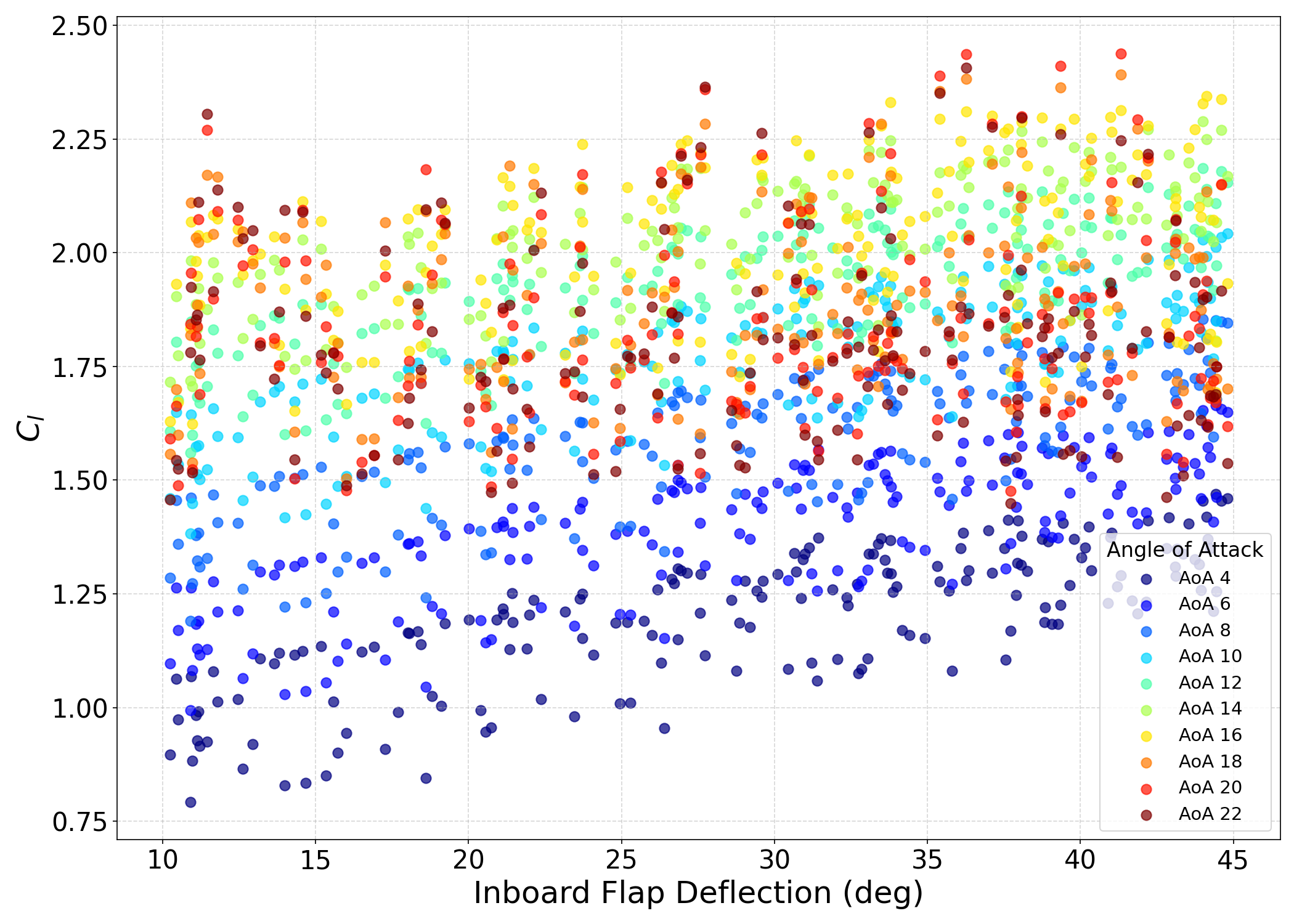}
\caption{Lift Coefficient vs IB (inboard) flap deflection angle for all angles of attack\label{fig:ibflap}}
\end{center}
\end{figure}

Figure~\ref{fig:ibflap} shows \(C_L\) across the sampled geometries, coloured by AoA. The overall range is approximately 0.79--2.44, and high-angle cases show a wider spread consistent with configuration-dependent separation and stall. Because all eight geometry parameters vary, this scatter does not isolate the causal effect of inboard flap deflection.

At fixed AoA, different geometries can also occupy different regimes. LHC029 and LHC013 are largely attached at \(8^\circ\), whereas at \(18^\circ\) LHC013 has extensive separation and LHC029 remains attached over much of the wing. These cases differ in several deflections and gap multipliers. Their parameter values, force sweeps, and skin-friction comparison are given in Appendix~\ref{app:dataset} (Fig.~\ref{fig:highlow2_main}).

\section{Machine-Learning Evaluation}\label{sec:ml_evaluation}

HiLiftAeroML provides 14 deterministic split families that test interpolation, data efficiency, unseen geometries, fixed angles of attack (AoA), and out-of-distribution (OOD) shifts. We use two similar-capacity headline models: GeoTransolver~\citep{adams2025geotransolverlearningphysicsirregular} with its three-scale local-feature branch enabled (17.06M active parameters), and Transolver~\citep{wu2024transolver} (18.54M).

Surface and volume networks are trained separately, and training presents stochastic samples of at most 200k entities per case. Evaluation is instead exhaustive: each retained native surface point and each valid native volume point of every held-out case is scored exactly once, using bounded, globally mixed contexts of at most 200k entities. This is a bounded-memory exhaustive traversal of the stored solution, not a single whole-mesh attention pass and not a mesh-convergence claim. Relative field metrics are computed after reversing the training normalization, in nondimensional physical variables (\(C_p\), wall-shear coefficient vector, and velocity vector \(\mathbf{U}/|U_\infty|\)), so that a common multiplicative scale cannot change a relative error. Table~\ref{tab:ml_native_benchmark} reports relative \(L_2\) as a percentage together with case-set coefficient \(R^2\); Appendix~\ref{app:detailedml} gives the exact supports, weights, equations, integrated-load conventions, checkpoints, and provenance. Absolute root-mean-square errors (RMSE) in engineering units are reported in Appendix~\ref{app:ml_absolute_errors} (Table~\ref{tab:ml_native_rmse}). Twelve of the 14 split families have completed surface/volume records, giving 23 model--split records and \(8{,}142\) case--model--split evaluations involving \(1{,}355\) distinct CFD cases.

\ifdefined\HiLiftAnonymous\else
The reported scores use a preliminary version of the FluidsBench evaluation metrics, frozen for this study; these metrics may differ from the final FluidsBench release. Appendix~\ref{app:ml_provenance} identifies the implementation and companion artifacts.
\fi

\begin{table}[!htb]
\centering
\caption{Complete-native-support evaluation of the two similar-capacity headline
models across all 12 represented split families (23 completed surface/volume
records), with surface and volume results from separately trained networks.
Relative $L_2$ values are percentages under the equal-case macro reduction of
Equation~\eqref{eq:macro}. Bold marks the better value within a completed
same-split model pair. Test populations differ by split, so models should be
compared within a split rather than across splits. Surface checkpoints are
epoch 1000 except Scarce GeoTransolver (869), Geometry Transolver (987), and
OOD-deflection Transolver (943); all volume checkpoints are epoch 1000.
Appendix~\ref{app:detailedml} gives the
supports, weights, metric definitions, and provenance.}
\label{tab:ml_native_benchmark}
\footnotesize
\renewcommand{\arraystretch}{1.05}
\setlength{\tabcolsep}{3.5pt}
\begin{tabular}{@{}llrrrrrrr@{}}
\toprule
 & & \multicolumn{5}{c}{Surface models} & \multicolumn{2}{c}{Volume models} \\
\cmidrule(lr){3-7}\cmidrule(l){8-9}
Split & Model & \(C_p\) & \(\tau_w\) & \(C_L\) & \(C_D\) & \(C_M\) & Pressure & Velocity \\
 & & rel. \(L_2\) & rel. \(L_2\) & \(R^2\) & \(R^2\) & \(R^2\) & rel. \(L_2\) & rel. \(L_2\) \\
\midrule
AoA $=4^\circ$ & GeoTransolver & \textbf{4.91} & \textbf{9.31} & \textbf{0.987} & \textbf{0.998} & 0.985 & \textbf{4.34} & \textbf{5.99} \\
AoA $=4^\circ$ & Transolver & 6.21 & 10.14 & 0.974 & 0.984 & 0.985 & 7.48 & 10.90 \\
\midrule
AoA $=12^\circ$ & GeoTransolver & \textbf{3.00} & \textbf{8.17} & \textbf{0.992} & \textbf{0.997} & \textbf{0.994} & \textbf{4.09} & \textbf{7.49} \\
AoA $=12^\circ$ & Transolver & 4.82 & 10.23 & 0.979 & 0.966 & 0.976 & 7.88 & 10.20 \\
\midrule
AoA $=22^\circ$ & GeoTransolver & \textbf{11.65} & \textbf{14.08} & \textbf{0.895} & \textbf{0.925} & 0.920 & \textbf{14.03} & \textbf{17.83} \\
AoA $=22^\circ$ & Transolver & 11.96 & 14.86 & 0.892 & 0.914 & \textbf{0.928} & 16.48 & 19.63 \\
\midrule
Full & GeoTransolver & \textbf{4.52} & \textbf{8.66} & 0.992 & 0.997 & \textbf{0.983} & \textbf{5.00} & \textbf{7.17} \\
Full & Transolver & 5.16 & 9.31 & \textbf{0.993} & 0.997 & 0.977 & 6.74 & 9.28 \\
\midrule
Scarce & GeoTransolver & 13.06 & 20.15 & 0.975 & 0.993 & 0.898 & 12.92 & 17.37 \\
\midrule
Super-scarce & GeoTransolver & 42.35 & 46.39 & 0.543 & 0.815 & -0.119 & \textbf{39.40} & \textbf{40.90} \\
Super-scarce & Transolver & \textbf{38.63} & \textbf{42.72} & \textbf{0.818} & \textbf{0.956} & \textbf{0.168} & 53.54 & 47.12 \\
\midrule
Geometry & GeoTransolver & \textbf{4.82} & \textbf{8.77} & 0.989 & 0.997 & \textbf{0.956} & \textbf{5.48} & \textbf{7.56} \\
Geometry & Transolver & 5.87 & 9.92 & 0.989 & 0.997 & 0.953 & 7.09 & 9.16 \\
\midrule
Geometry-scarce & GeoTransolver & \textbf{13.54} & 17.93 & 0.956 & 0.989 & 0.667 & 19.18 & 22.77 \\
Geometry-scarce & Transolver & 13.69 & \textbf{17.50} & \textbf{0.958} & \textbf{0.990} & \textbf{0.672} & \textbf{15.68} & \textbf{17.82} \\
\midrule
Geometry-super-scarce & GeoTransolver & 48.98 & 49.06 & 0.317 & 0.840 & -4.735 & 39.87 & 43.07 \\
Geometry-super-scarce & Transolver & \textbf{30.43} & \textbf{32.58} & \textbf{0.800} & \textbf{0.951} & \textbf{0.368} & \textbf{35.30} & \textbf{34.33} \\
\midrule
OOD AoA & GeoTransolver & \textbf{48.81} & \textbf{38.40} & \textbf{-2.865} & \textbf{0.808} & \textbf{0.295} & \textbf{52.56} & \textbf{42.31} \\
OOD AoA & Transolver & 66.41 & 60.79 & -18.289 & -7.835 & -6.144 & 57.20 & 61.15 \\
\midrule
OOD deflection & GeoTransolver & \textbf{4.60} & \textbf{9.05} & \textbf{0.992} & \textbf{0.996} & \textbf{0.978} & \textbf{4.63} & \textbf{6.83} \\
OOD deflection & Transolver & 5.79 & 10.16 & 0.987 & 0.995 & 0.970 & 7.28 & 9.50 \\
\midrule
OOD stall & GeoTransolver & 40.32 & 32.13 & -0.098 & 0.659 & 0.625 & 56.23 & 38.52 \\
OOD stall & Transolver & \textbf{38.96} & \textbf{31.65} & \textbf{-0.058} & \textbf{0.699} & \textbf{0.660} & \textbf{54.51} & \textbf{38.06} \\
\bottomrule

\end{tabular}
\end{table}

Every field error in this paper uses one reduction, the \emph{equal-case macro} of Equation~\eqref{eq:macro} in Appendix~\ref{app:detailedml}: the complete-support error is formed for each case and then averaged over cases with equal weight, so each held-out case counts once regardless of how many native entities it contains. This is the reduction the benchmark ranks on; the relative errors in Table~\ref{tab:ml_native_benchmark} match the corresponding leaderboard records. Within a case the spatial weighting is unchanged: surface fields use the published nodal dual area and volume fields equal weight per valid point.

At \(4^\circ\) and \(12^\circ\) GeoTransolver has the lower error on every field quantity in both modalities, and its surface-pressure error at \(12^\circ\), \natAoatwelveGeotSurfCpRelLtwo\%, is the lowest split-level surface-pressure error in Table~\ref{tab:ml_native_benchmark}. Both architectures then deteriorate sharply at \(22^\circ\): surface-pressure error rises by a factor of between two and a half and four against \(12^\circ\), and volume errors by a factor of between about two and three and a half, and the margin between the two architectures nearly closes. The separated high-AoA regime is therefore substantially harder than the attached conditions for both models, and architecture choice matters less as the flow separates.

The Geometry and single-AoA splits share the same 126/18/36 train/validation/test geometry partition. On the same unseen test geometries, training GeoTransolver across all ten angles reduces all four field errors at \(4^\circ\), \(12^\circ\), and \(22^\circ\). At \(22^\circ\), the reductions are approximately 25--33\%; Appendix~\ref{app:ml_matched_aoa} gives the angle-resolved values. 

The data-efficiency ladders keep their test populations fixed while reducing the random-case training set from 1260 to 210 to 35 cases, and the geometry training set from 126 to 21 to 4 complete geometries (1260, 210, and 40 cases). All four field errors increase monotonically across the available tiers. Architecture ordering varies with budget and modality: Transolver leads on all four field errors at the smallest geometry tier, while the smallest random tier favours Transolver on the surface and GeoTransolver in the volume. Pitching-moment accuracy is particularly sensitive to data scarcity: its \(R^2\) is negative for GeoTransolver at the smallest tier of either ladder, while drag \(R^2\) remains above 0.8 for both models.

The OOD tests separate extrapolation along a geometric design parameter from a change of flow regime, and the separation is stark. On high-deflection geometry both architectures stay close to in-distribution accuracy, with lift and drag \(R^2\) above 0.98. Low-to-high-AoA extrapolation over all 900 held-out cases is far worse: surface-pressure error rises roughly tenfold and lift \(R^2\) goes strongly negative for both, with Transolver degrading further than GeoTransolver. The 723-case OOD-stall population behaves similarly, except that Transolver is numerically ahead on every listed stall aggregate in this checkpoint-matched pair, consistent with its lower stall-volume errors. Both models nonetheless retain roughly 39--40\% surface-pressure error and non-positive lift \(R^2\) at stall. Strong interpolation and unseen-geometry accuracy therefore does not establish reliable extrapolation across separated-flow regimes.

\begin{figure}[!t]
\centering
\includegraphics[width=\textwidth]{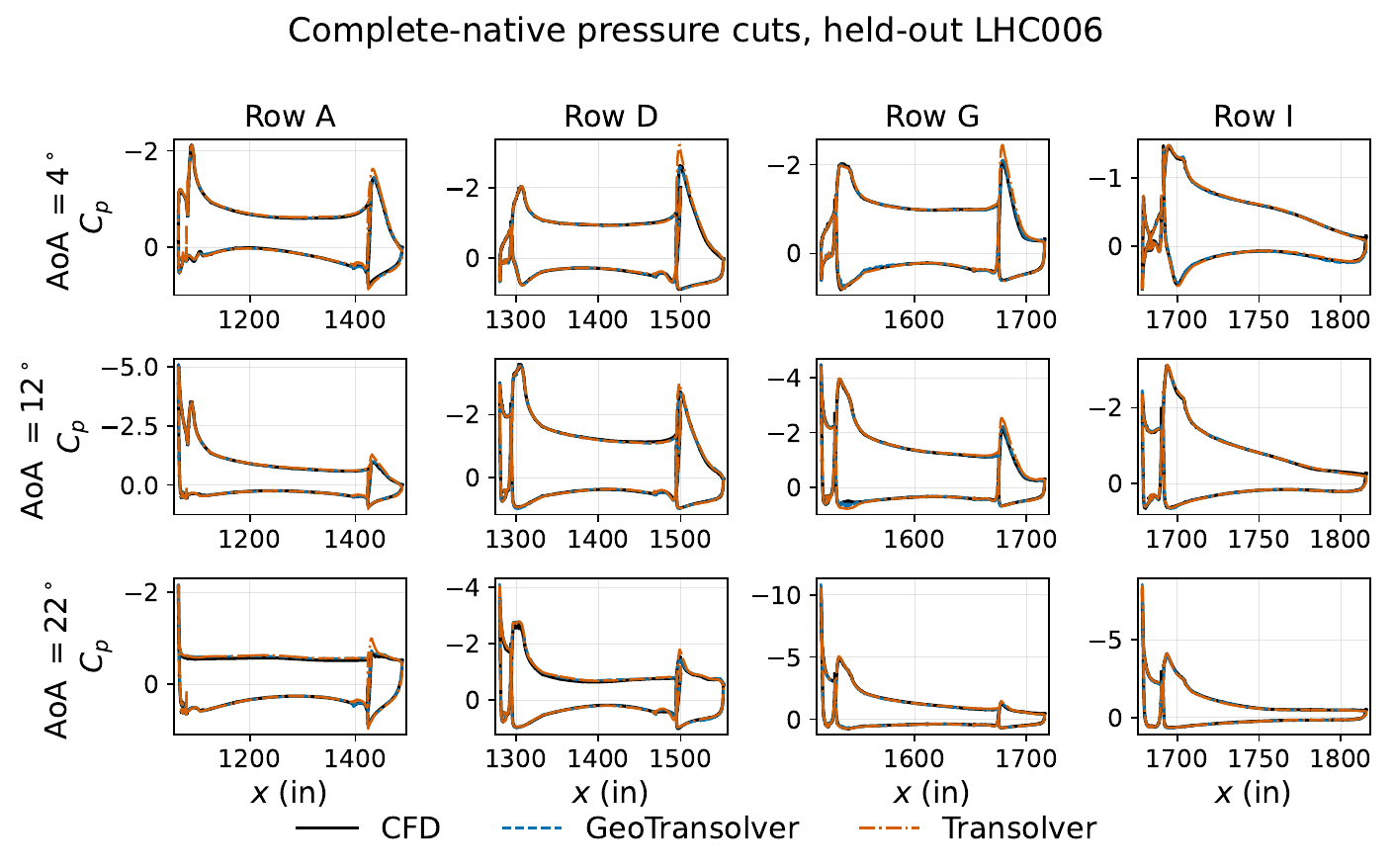}
\caption{Complete-native pressure cuts for the held-out geometry LHC006 at
\(4^\circ\), \(12^\circ\), and \(22^\circ\), on official HLPW-5 pressure rows A,
D, G, and I. Curves are drawn directly from the complete-native evaluation on
the same support used for scoring; every disconnected physical cut graph is
plotted as its own polyline, so no edge is created across a topology gap. The
\(C_p\) axis is inverted and each panel's limits are computed from every
displayed point, so no curve is clipped. The complete A--J atlas is given in
Appendix~\ref{app:detailedml}.}
\label{fig:ml_native_cp_cuts_main}
\end{figure}

Figure~\ref{fig:ml_native_cp_cuts_main} illustrates the pressure predictions for held-out geometry LHC006 at \(4^\circ\), \(12^\circ\), and \(22^\circ\). The attached cases reproduce the sectional structure closely, while the \(22^\circ\) case shows errors in suction-peak magnitude. The corresponding velocity profiles in Appendix~\ref{app:detailedml} (Fig.~\ref{fig:ml_native_velocity_profiles}) show wake errors consistent with the lower profile scores in separated flow. The appendix also provides the full pressure and velocity atlases, error tails, and regional diagnostics.

\section{Conclusions}
HiLiftAeroML provides 1,800 WMLES simulations of 180 CRM-HL variants across ten angles of attack. The CC-BY-4.0 release includes geometries, time-averaged surface and volume fields, integrated loads, validation and convergence material, deterministic splits, and native-point quadrature weights. Representative experimental comparisons support its use for studying high-lift aerodynamics while documenting the limitations of the generating simulations.

The GeoTransolver and Transolver baselines demonstrate useful interpolation and unseen-geometry prediction, and expose strong degradation under data scarcity and flow-regime shifts. High-deflection extrapolation remains close to in-distribution accuracy, whereas AoA and stall extrapolation produce negative lift \(R^2\). These results identify concrete challenges for future methods without establishing a general architecture ranking. 

\subsection{Limitations}
The geometries vary one CRM-HL topology at a wind-tunnel-scale Reynolds number of \(1.6\times10^6\). Neither other aircraft families nor flight-scale conditions are represented. Transition treatment, the equilibrium wall model, finite averaging windows, and uncertainty in separated flow limit simulation fidelity. The released fields are time averages and statistical moments, rather than high-frequency flow histories.

The ML evaluation covers only two related transformer-based architectures, GeoTransolver and Transolver. These initial baselines illustrate the dataset's utility and identify challenges for future methods. The main contribution is the dataset; a comprehensive comparison across model families, hyperparameter settings, and training strategies remains future work. 

Generalisation to new aircraft, reliable flow-regime extrapolation, calibrated uncertainty, and coupled surface--volume prediction remain open tasks.

\paragraph{Outlook.}
HiLiftAeroML gives the aerospace community a shared resource for developing and testing surrogates of high-lift flow fields and aerodynamic loads. Its open data and reproducible evaluation splits support comparisons across methods and a clearer understanding of where predictions are reliable. Such surrogates could help engineers explore more design options and focus high-fidelity simulation on the configurations and flow regimes where it is most needed.

\label{paper:main_text_end}
\FloatBarrier

\section*{Ethics Statement}
This paper presents work whose goal is to advance the field of Machine Learning in Computational Fluid Dynamics. There are many potential societal consequences of our work, primarily in improving aircraft efficiency and safety, none which we feel must be specifically highlighted here as negative ethical concerns. The work involves no human subjects, no personally identifying data, and no sensitive data: every case in the release is a computational simulation of a publicly documented reference geometry, distributed under CC-BY-4.0.

\section*{Reproducibility Statement}
The simulation campaign is specified in Section~\ref{sec:flow_solver} and Appendix~\ref{app:numMeth}, with reference conditions in Table~\ref{table:dataset_parameters} and the coefficient definitions in Appendix~\ref{app:postPro-aerospace}. Validation against wind-tunnel measurements is given in Appendix~\ref{app:validation}. For the machine-learning results, Appendix~\ref{app:detailedml} records the split definitions, native supports and quadrature weights, metric equations and reduction, integrated-load conventions, evaluated checkpoints, and per-record provenance; Appendix~\ref{app:hpc} reports the measured compute for data generation, training, and inference. The released dataset, including the deterministic split assignments, is public under CC-BY-4.0, and every reported metric is re-derived from stored sufficient statistics rather than transcribed.
\ifdefined\HiLiftAnonymous\else
The companion artifact release~\citep{HiLiftAeroMLArtifacts2026} provides CPU reproduction of retained score reductions for all 23 benchmark records and the 46 corresponding model exports, with resolved configurations, normalization records, split selectors, and integrity manifests. The verified replay reconstructs aggregate metrics from retained evaluation outputs; fresh native inference and profile scoring against evaluator-owned truth are outside its scope. The checkpoint weights are distributed under CC-BY-4.0, with existing source-code licenses retained.
\fi

\section*{AI Use Statement}
Large language models were used as general-purpose assistants throughout this work. Their use covered editing and condensing the manuscript text; writing and refactoring the supporting analysis, evaluation, and figure-generation code; and reviewing the manuscript for internal consistency and for errors, including feedback on how results were presented and interpreted. The simulation data themselves were produced by the flow solver described in Section~\ref{sec:flow_solver}, and the reported metrics by the evaluation pipeline of Appendix~\ref{app:detailedml}; no result in this paper was generated by a language model. Every reported metric is re-derived from stored sufficient statistics, and all AI-assisted text, code, and analysis were reviewed and verified by the authors, who take full responsibility for the content of this paper.

\ificlrfinal
\section*{Acknowledgements}
The authors acknowledge computing resources provided by Cadence, NVIDIA, the Texas Advanced Computing Center, CSCS, and the Oak Ridge Leadership Computing Facility, a DOE Office of Science User Facility supported under Contract DE-AC05-00OR22725.
\fi

\bibliography{referencesextra,mendeley}
\bibliographystyle{iclr2027_conference}
\appendix
\newpage
\section{Numerical methodology}
\label{app:numMeth}

The simulations solve the compressible Navier-Stokes equations using a large-eddy simulation (LES) formulation. Following the resolved-energy formulation described by \citet{goc2021large}, the equations for conservation of mass, momentum, and energy are expressed in modeled LES form, using Einstein summation notation, as:

\begin{align}
    \frac{\partial \bar{\rho}}{\partial t} + \frac{\partial (\bar{\rho} \tilde{u}_j)}{\partial x_j} &= 0, \label{eq:NSeqs-cont} \\
    \frac{\partial (\bar{\rho} \tilde{u}_i)}{\partial t} + \frac{\partial (\bar{\rho} \tilde{u}_i \tilde{u}_j)}{\partial x_j} &= -\frac{\partial \bar{p}}{\partial x_i} + \frac{\partial \tilde{\sigma}_{ij}}{\partial x_j} - \frac{\partial \tau_{ij}^{\mathrm{SGS}}}{\partial x_j}, \label{eq:NSeqs-mom} \\
    \frac{\partial (\bar{\rho} E_r)}{\partial t} + \frac{\partial [(\bar{\rho} E_r + \bar{p})\tilde{u}_j]}{\partial x_j} &= \frac{\partial}{\partial x_j}\left[\tilde{u}_i\left(\tilde{\sigma}_{ij}-\tau_{ij}^{\mathrm{SGS}}\right)-\tilde{q}_j-Q_j^{\mathrm{SGS}}\right]. \label{eq:NSeqs-energy}
\end{align}

Here, $\bar{\rho}$ and $\bar{p}$ denote filtered density and pressure, and $\tilde{u}_i$ denotes the Favre-filtered velocity. The resolved specific total energy is
\begin{equation}
    E_r = \tilde{e} + \frac{1}{2}\tilde{u}_i\tilde{u}_i,
\end{equation}
where $\tilde{e}$ is the Favre-filtered specific internal energy. The quantities $\tilde{\sigma}_{ij}$ and $\tau_{ij}^{\mathrm{SGS}}$ denote the molecular and modeled subgrid-scale stresses, respectively, while $\tilde{q}_j$ and $Q_j^{\mathrm{SGS}}$ denote the molecular and modeled subgrid-scale heat fluxes. The molecular stress is that of a Newtonian fluid, and the molecular heat flux follows Fourier's law,
\begin{equation}
    \tilde{q}_j = -k_{\mathrm{th}} \frac{\partial \tilde{T}}{\partial x_j},
\end{equation}
with $k_{\mathrm{th}}$ the thermal conductivity and $\tilde{T}$ the Favre-filtered temperature. The modeled SGS heat flux represents unresolved heat transport. SGS stress work is retained explicitly in the energy flux in Equation~(\ref{eq:NSeqs-energy}). At the filtered-equation level, the SGS stress is defined by
\begin{equation}
    \tau_{ij}^{\mathrm{SGS}} = \bar{\rho} (\widetilde{u_i u_j} - \tilde{u}_i \tilde{u}_j).
\end{equation}
Its deviatoric part is approximated using the turbulence model described below.

A fundamental scaling parameter in fluid dynamics is the dimensionless Reynolds number $Re$, representing the ratio between inertial and viscous forces:

\begin{equation}
    \mathrm{Re} = \frac{U_{\text{ref}} \, C_{\text{ref}}}{\nu}
    \label{eq:Re}
\end{equation}

For the aircraft configurations in this dataset, the simulations are performed at a chord-based Reynolds number of $Re_{MAC} = 1.6 \times 10^6$, where the reference length is the Mean Aerodynamic Chord (MAC). The Mach number is set to $M=0.2$.

\subsection{Turbulence modelling}
\label{app:numMeth-turbModel}

To capture complex, unsteady flow phenomena such as flow separation, a Wall-Modeled Large Eddy Simulation (WMLES) approach is employed.

\subsubsection{Subgrid-Scale Model}
Informed by investigations highlighting the importance of dynamic modeling for flow separation, the Dynamic Smagorinsky subgrid-scale (SGS) model is utilized. The deviatoric part of the SGS stress tensor is modeled using the Boussinesq hypothesis:

\begin{equation}
    \tau_{ij}^{\mathrm{SGS}} - \frac{1}{3}\tau_{kk}^{\mathrm{SGS}}\delta_{ij} = -2 \mu_t \left(\tilde{S}_{ij} - \frac{1}{3}\tilde{S}_{kk}\delta_{ij}\right)
\end{equation}

where $\tilde{S}_{ij} = \frac{1}{2} (\frac{\partial \tilde{u}_i}{\partial x_j} + \frac{\partial \tilde{u}_j}{\partial x_i})$ is the resolved strain rate tensor. Subtracting $\tilde{S}_{kk}\delta_{ij}/3$ gives its deviatoric part, so both sides remain trace-free when the resolved velocity has nonzero divergence. The SGS eddy viscosity, $\mu_t$, is defined as:

\begin{equation}
    \mu_t = \bar{\rho} (C_S \Delta)^2 |\tilde{S}| \;\text{, with } |\tilde{S}| = \sqrt{2 \tilde{S}_{ij} \tilde{S}_{ij}}
\end{equation}

Here, $\Delta$ is the filter width proportional to the grid size. Unlike constant coefficient models, the coefficient $C_S$ is computed dynamically in space and time based on the energy content of the smallest resolved scales. This dynamic procedure allows the model to vanish in laminar flow regions and adjust to the local turbulence structure, providing improved accuracy in transitional and separated flow regimes.

\subsubsection{Wall Modeling}
Wall shear stress is supplied by an equilibrium wall model that couples the near-wall velocity profile to the resolved LES velocity at an off-wall matching location~\citep{lehmkuhl2018large,goc2021large}. The model approximates the wall-model layer as having constant total shear stress in the wall-normal direction,

\begin{equation}
    \frac{d}{d\eta} \left[ (\mu + \mu_{t,\mathrm{wm}}) \frac{d u_{\parallel}}{d\eta} \right] = 0,
\end{equation}

where $\eta$ is the wall-normal distance, $u_{\parallel}$ is the wall-parallel velocity, and $\mu_{t,\mathrm{wm}}$ is the wall-model eddy viscosity. The wall condition $u_{\parallel}(0)=0$ and matching condition $u_{\parallel}(h_{\mathrm{wm}})=u_{\mathrm{LES}}$ relate the wall stress to the LES velocity. The resulting stress provides the momentum boundary condition for the LES. This equilibrium approximation concerns the tangential momentum balance across the wall-model layer; it does not imply equality of turbulent production and dissipation throughout the boundary layer.

\subsection{Flow solver and discretisation}
\label{app:numMeth-solver}

\begin{figure}[t] %
    \centering
    \includegraphics[width=\textwidth]{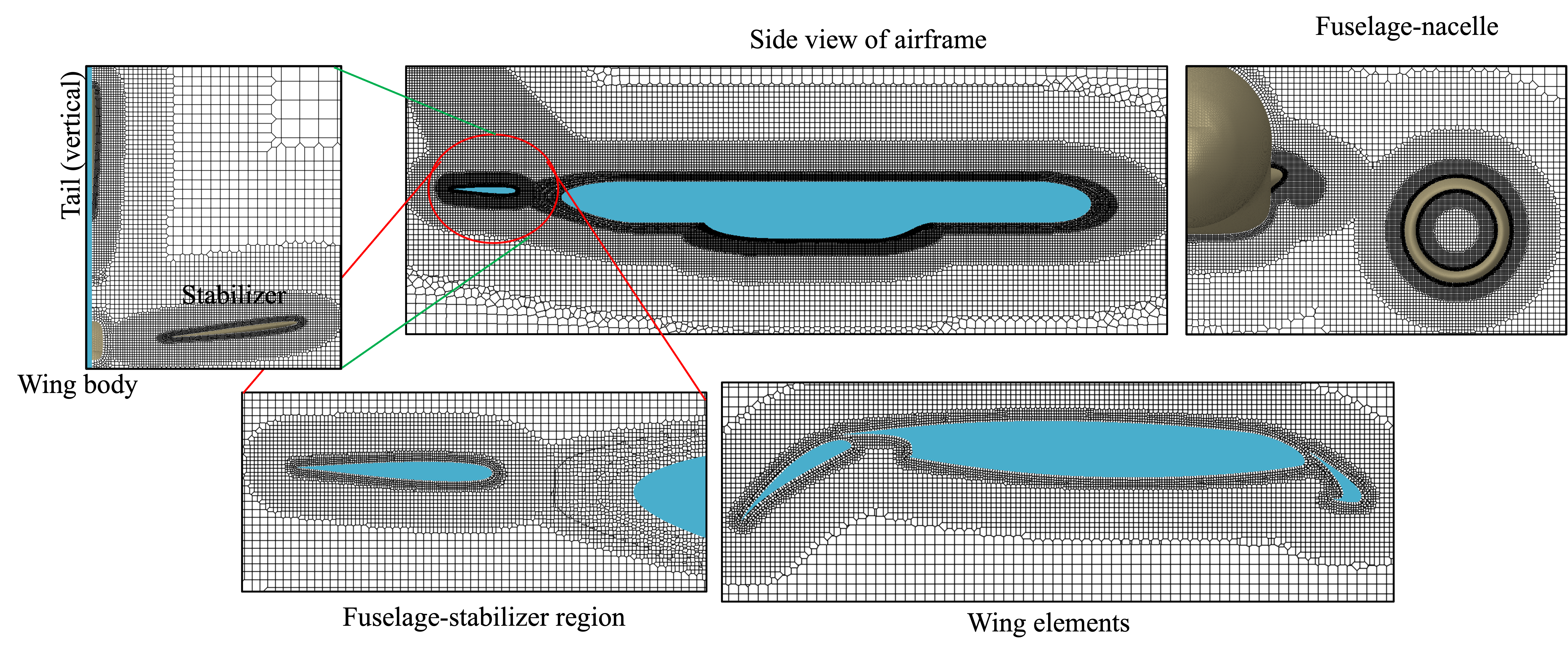}
    \caption{Grid distribution (from a side-view) on the airframe, with specific details of the three-element airfoil slice (taken at mid-span of the main wing element), around the fuselage and the nacelle, the vertical tail and the horizontal stabilizers respectively. Note that these images represent a grid four times coarser than the baseline grid for visual clarity.}
    \label{fig:baselinemesh_main}
\end{figure}

Section~\ref{sec:flow_solver} describes the Fidelity Charles discretisation and Fidelity Stitch grid construction. The baseline mesh is shown in Fig.~\ref{fig:baselinemesh_main}. The filtered equations and turbulence closures are given above; the boundary conditions and averaging procedure are detailed below.

\subsubsection{Boundary Conditions}
\label{app:boundary_conditions}
\begin{figure}[!htb]
    \centering
    \includegraphics[width=0.68\linewidth]{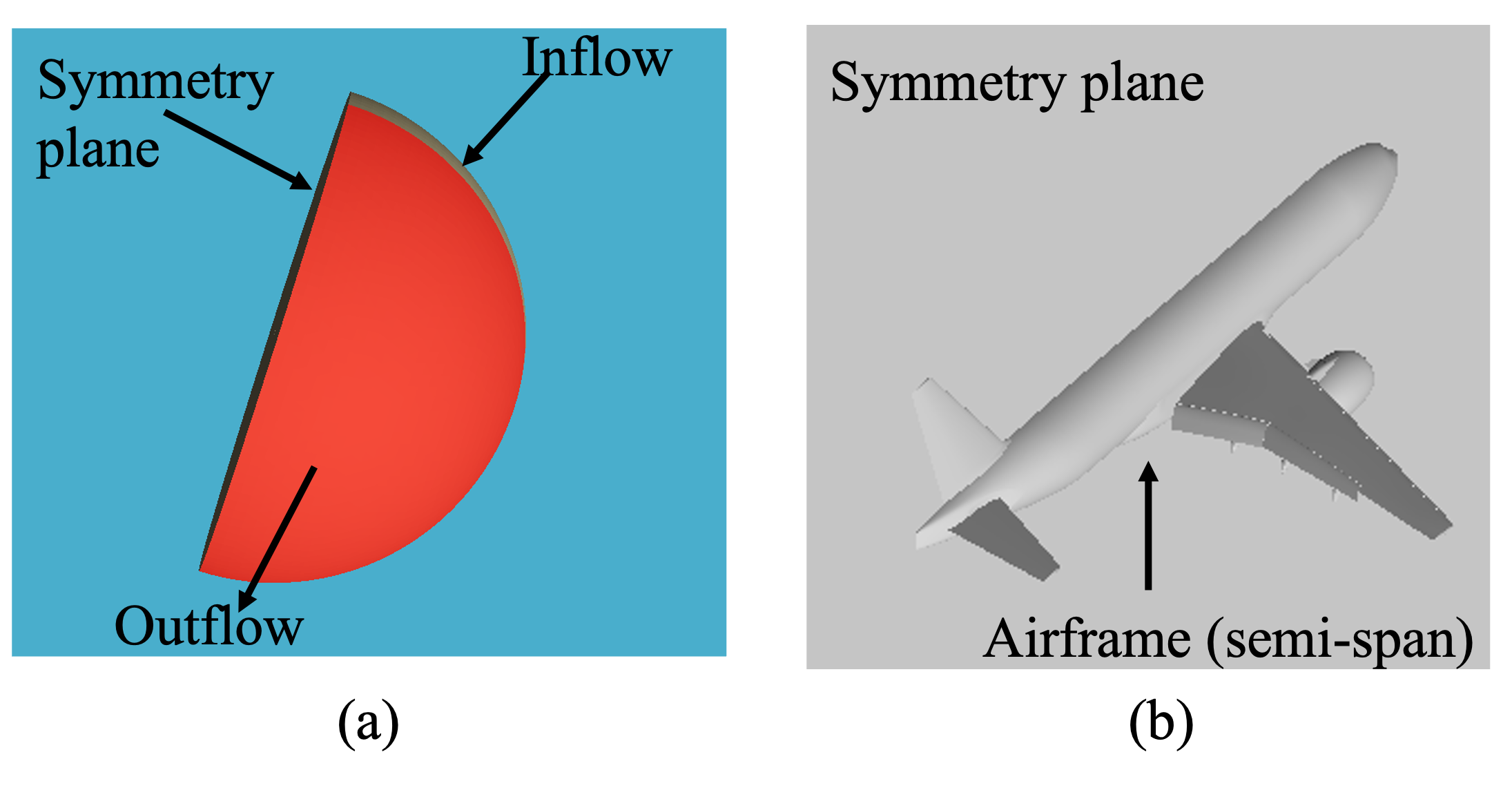}
    \caption{Schematic of the baseline case setup showing the inflow and outflow regions and the symmetry plane along which the semi-span aircraft model is mounted.}
    \label{fig:baselinesetup}
\end{figure}

The simulation domain is a hemisphere with a radius of approximately $75 \times MAC$. Figure~\ref{fig:baselinesetup} shows the semi-span aircraft mounted along the symmetry plane and the inflow and outflow regions. The boundary conditions are defined as follows:
\begin{itemize}
    \item \textbf{Inflow:} A uniform plug flow is specified at the inlet (the forward half of the hemisphere).
    \item \textbf{Outflow:} A characteristic non-reflecting boundary condition with a specified outlet pressure is applied to the rear half.
    \item \textbf{Symmetry:} A no-stress boundary condition is applied at the symmetry plane.
    \item \textbf{Walls:} All solid surfaces of the aircraft are treated viscously using the equilibrium wall model.
\end{itemize}

Reference dimensions, flow conditions, the moment centre, and derived normalization quantities are collected in Table~\ref{tab:dataset-refQuantCoeffs}.

\subsection{Time-averaging procedure}
\label{app:postPro-flowStats}

WMLES generates a time-dependent solution that requires statistical averaging. The time-averaging strategy is based on Convective Time Units (CTU), defined by the flow transit time over the geometry.

The procedure is as follows:
\begin{enumerate}
    \item \textbf{Initial Transient:} The adapted-grid solution is advanced to flush out non-physical transients, and the running statistics are reset after 10 CTU in every case. The usable averaging window is therefore the total adapted-grid duration less this 10 CTU, and is shorter than the total run.
    \item \textbf{Data Collection:} Averaging then continues to the end of the case.
    \begin{itemize}
        \item For $AoA \le 12^{\circ}$ the adapted-grid run is nominally 15 CTU, giving a usable averaging window of about 5 CTU.
        \item For $AoA > 12^{\circ}$, the nominal run is extended to 30--200 CTU depending on how slowly the wake statistics converge, giving a usable window of 20--190 CTU.
    \end{itemize}
\end{enumerate}
Individual cases depart from these nominal durations.

The released per-case load files report the standard deviation $\sigma$ of the instantaneous signal in \texttt{*\_stdev}. The \texttt{*\_stderr} and \texttt{*\_ci95} columns use the nominal CTU-based convention
\begin{equation*}
N_{\mathrm{CTU}}=\frac{T_{\mathrm{avg}}}{\mathrm{CTU}},\qquad
SE_{\mathrm{CTU}}=\frac{\sigma}{\sqrt{N_{\mathrm{CTU}}}},\qquad
h_{\mathrm{CTU}}=1.96\,SE_{\mathrm{CTU}},
\end{equation*}
where $T_{\mathrm{avg}}$ is the usable averaging duration and $h_{\mathrm{CTU}}$ is the stored half-width. This convention assumes one independent sample per CTU; the resulting bands are approximate indicators of averaging uncertainty.

\begin{figure}[H]
    \centering
    \includegraphics[width=0.48\textwidth]{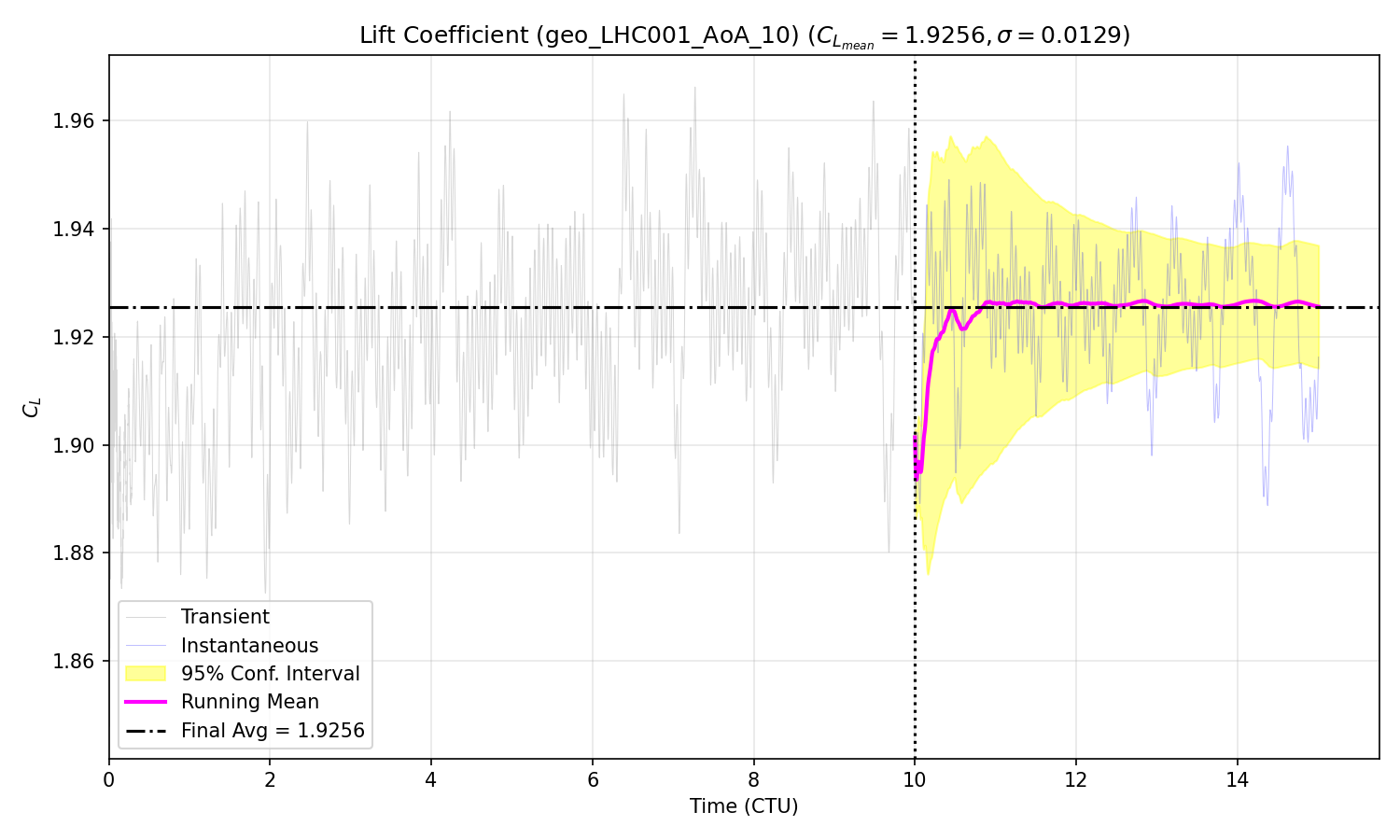}
    \hfill
    \includegraphics[width=0.48\textwidth]{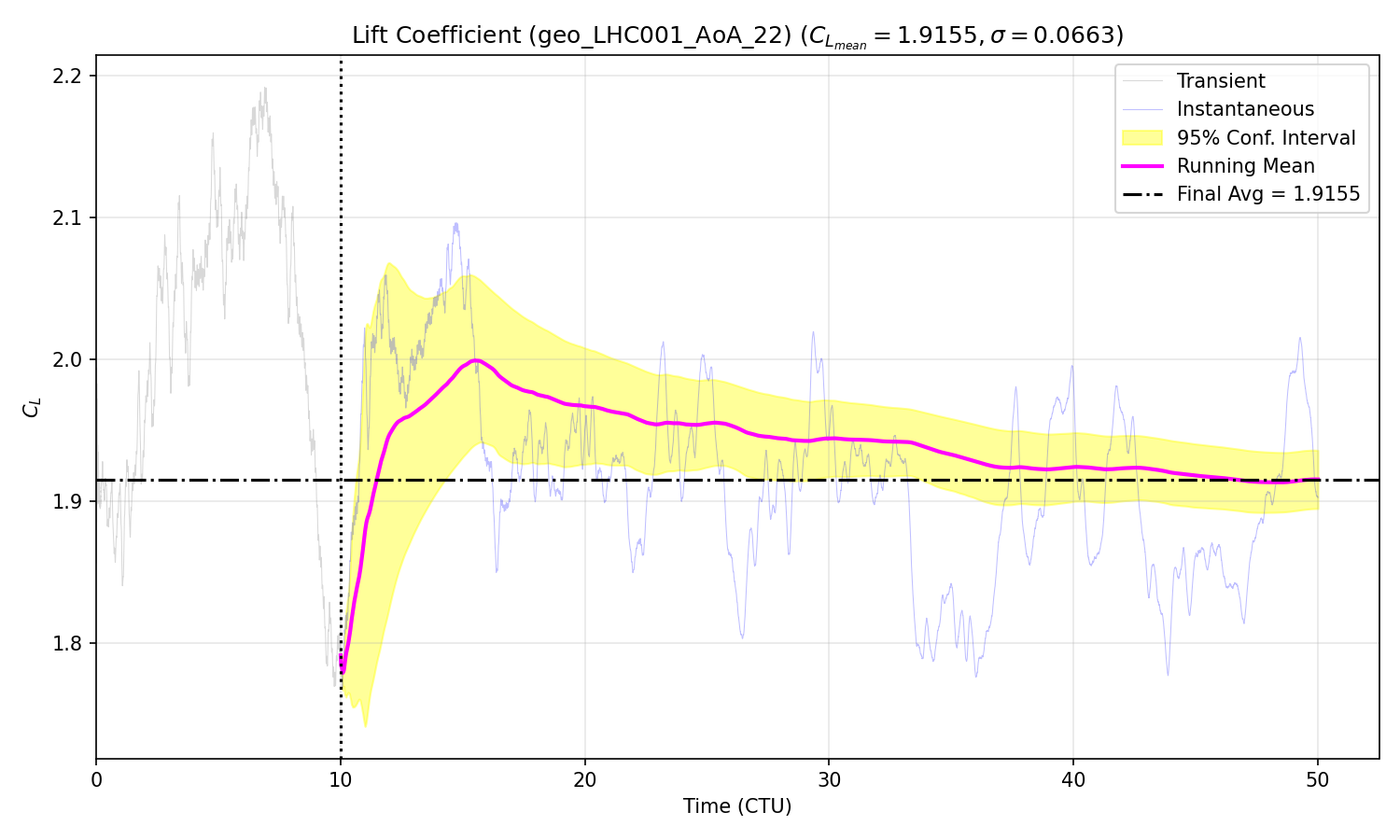}
    
    \vspace{0.3cm} %
    
    \includegraphics[width=0.48\textwidth]{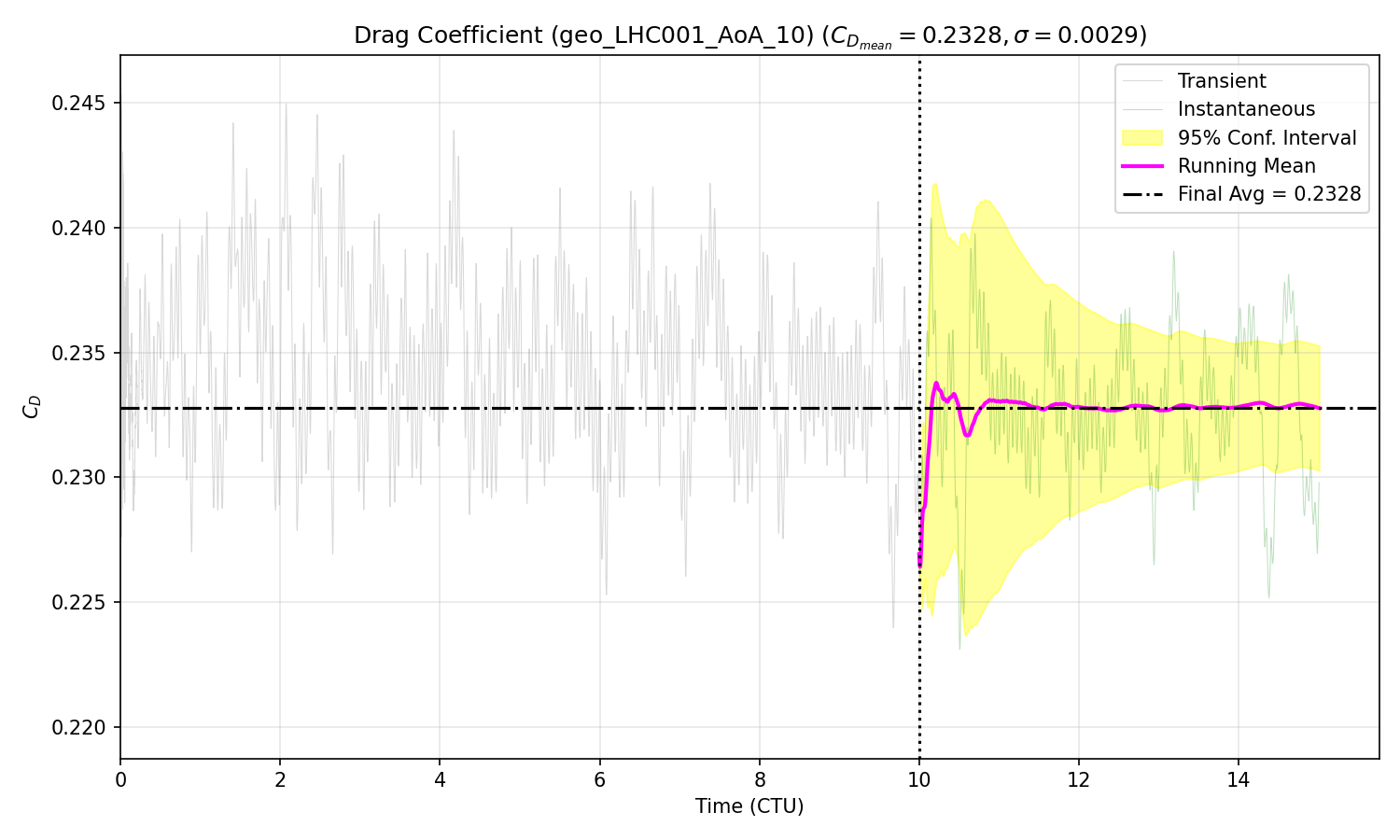}
    \hfill
    \includegraphics[width=0.48\textwidth]{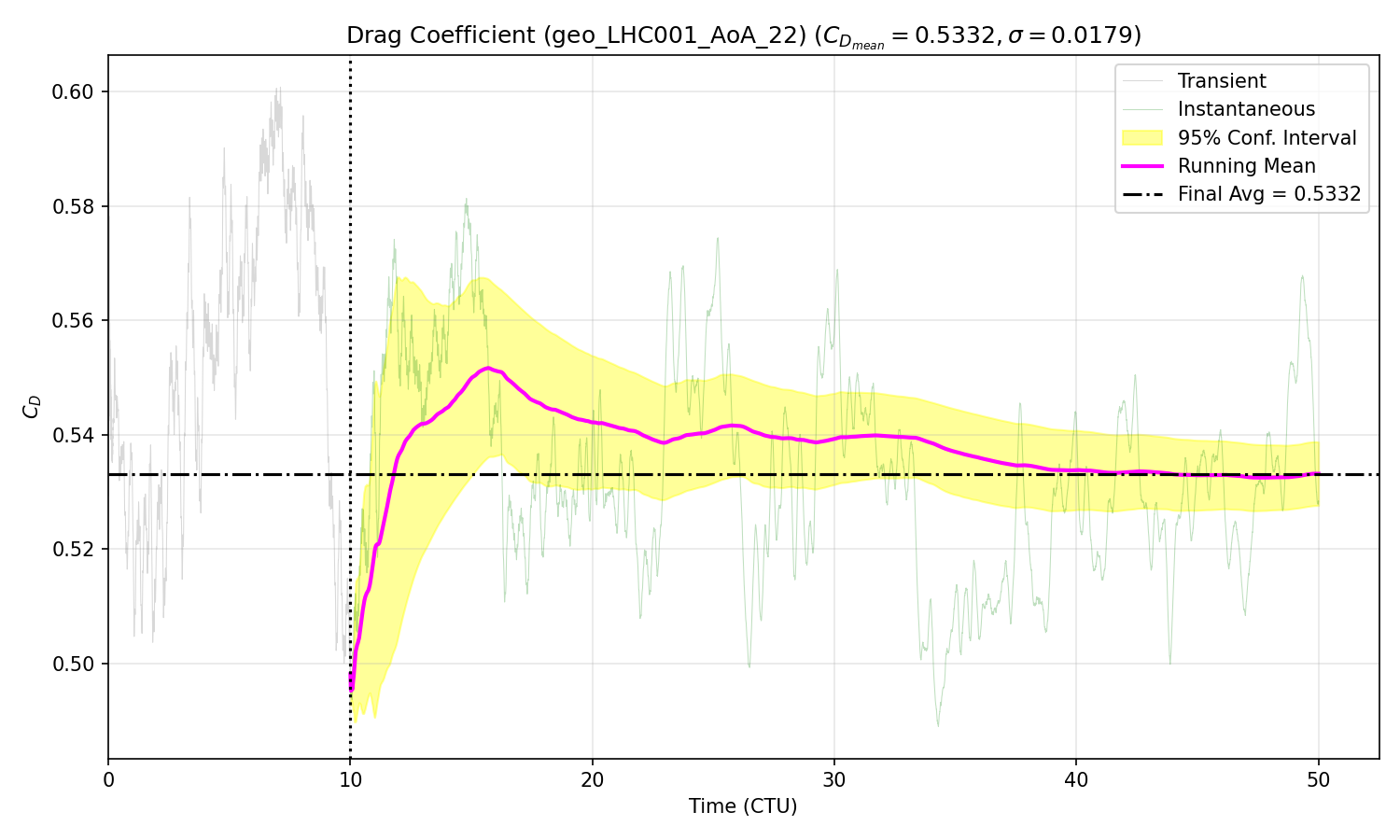}
    
    \vspace{0.3cm} %
    
    \includegraphics[width=0.48\textwidth]{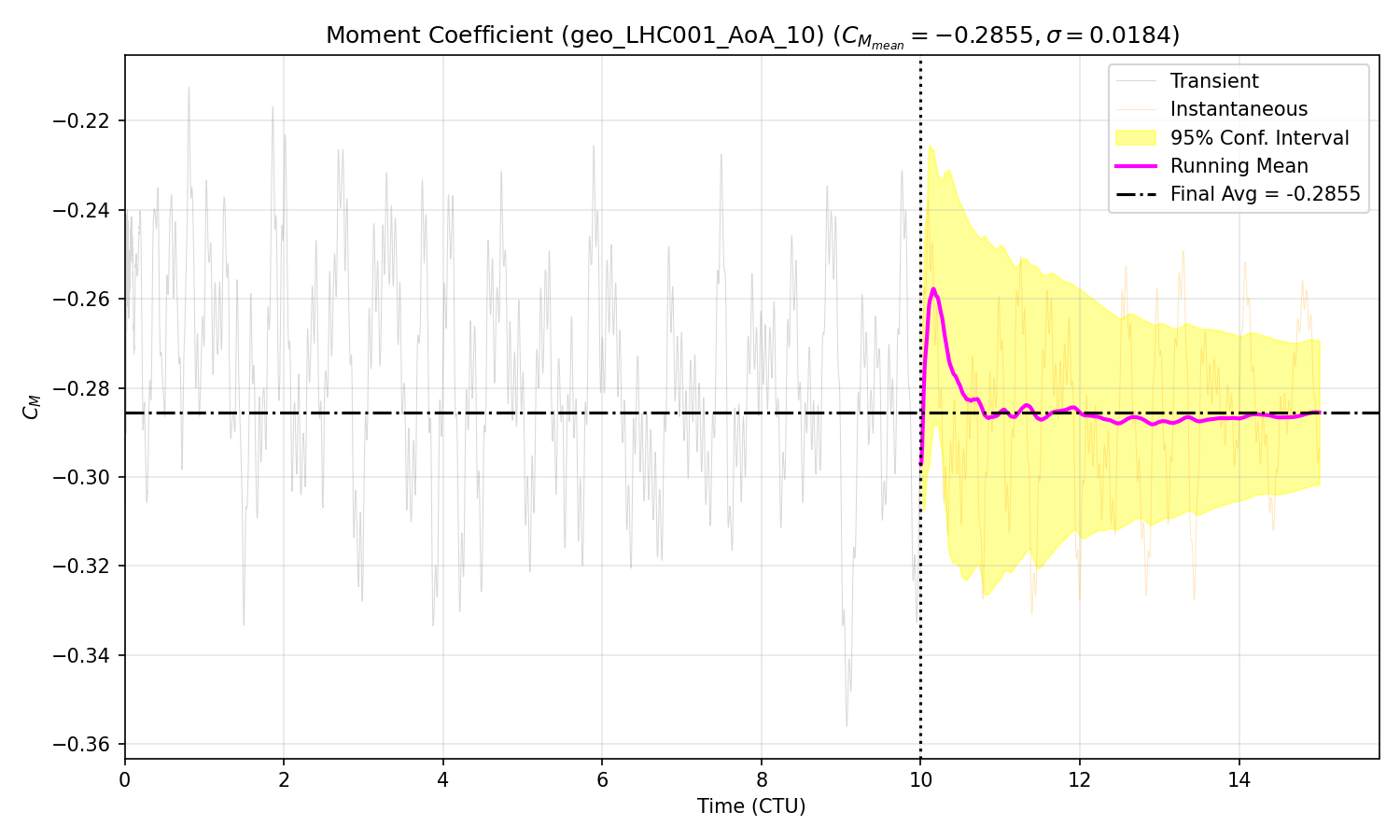}
    \hfill
    \includegraphics[width=0.48\textwidth]{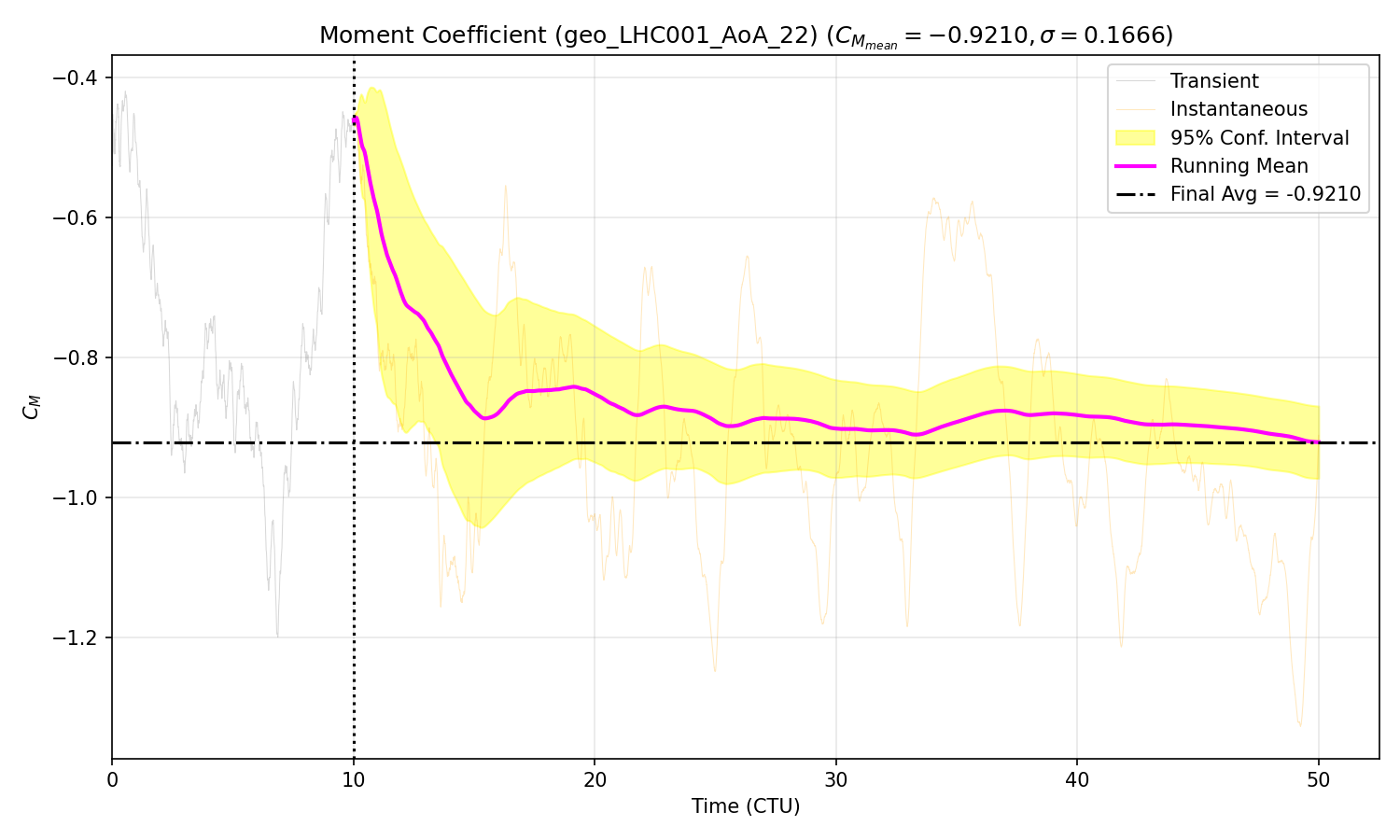}
    
    \caption{Example statistical convergence plots for Lift ($C_L$), Drag ($C_D$), and Pitching Moment ($C_M$) coefficients for geometry LHC001 at 10$^\circ$ AoA (left column) and 22$^\circ$ AoA (right column). The plots display the instantaneous signal (scatter), the running mean (solid line), and nominal CTU-based bands of the running mean $\pm 1.96\,SE_{\mathrm{CTU}}$ (shaded/dashed lines), plotted against Convective Time Units.}
    \label{fig:combined_convergence_plots}
\end{figure}

Figure~\ref{fig:combined_convergence_plots} shows examples of the instantaneous lift, drag, and pitching-moment signals, their running means, and the corresponding nominal CTU-based bands.

\begin{figure}[H]
    \centering
    \includegraphics[width=0.6\textwidth]{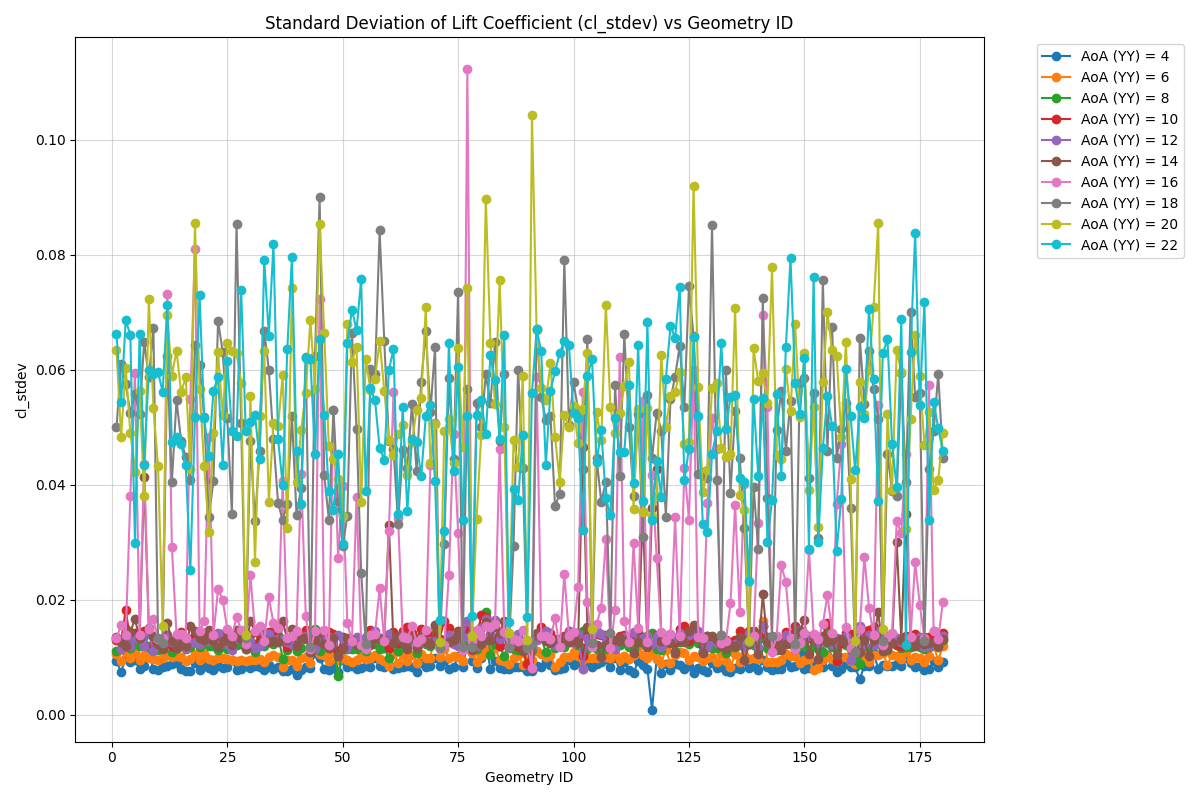}
    
    \vspace{0.1cm} %
    
    \includegraphics[width=0.6\textwidth]{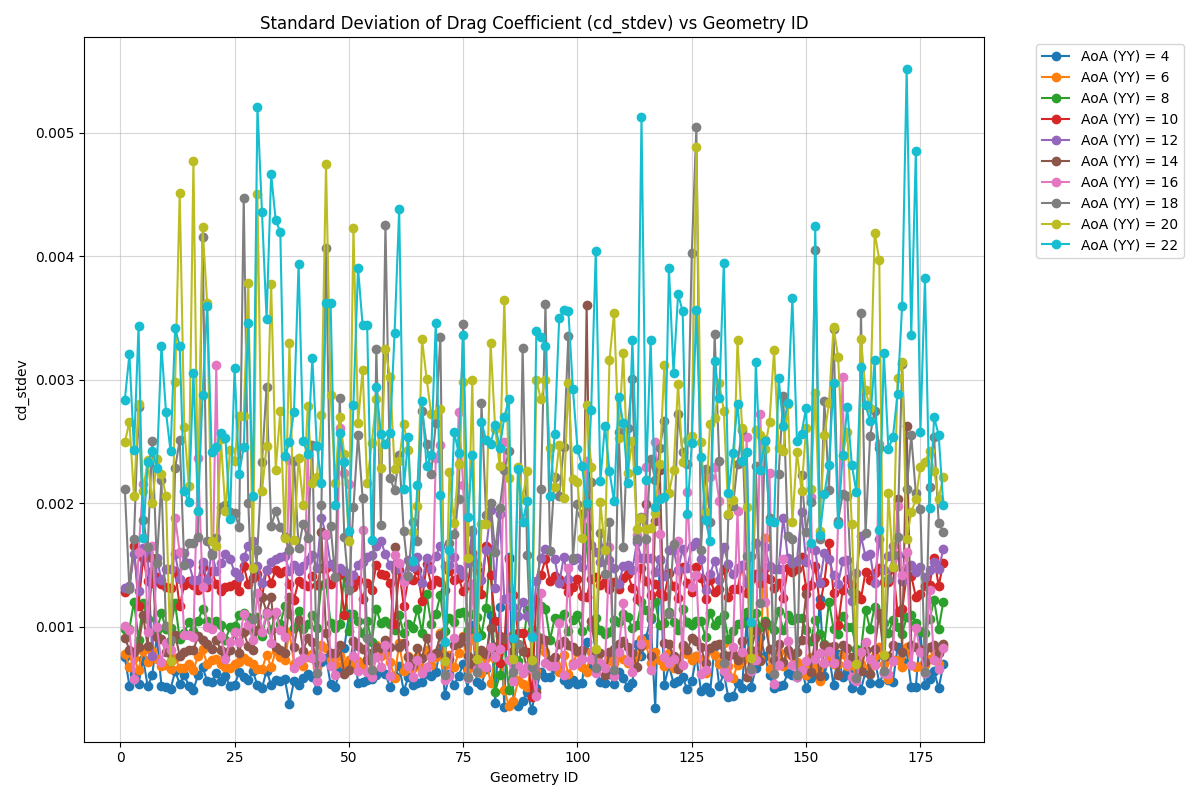}
    
    \vspace{0.1cm} %
    
    \includegraphics[width=0.6\textwidth]{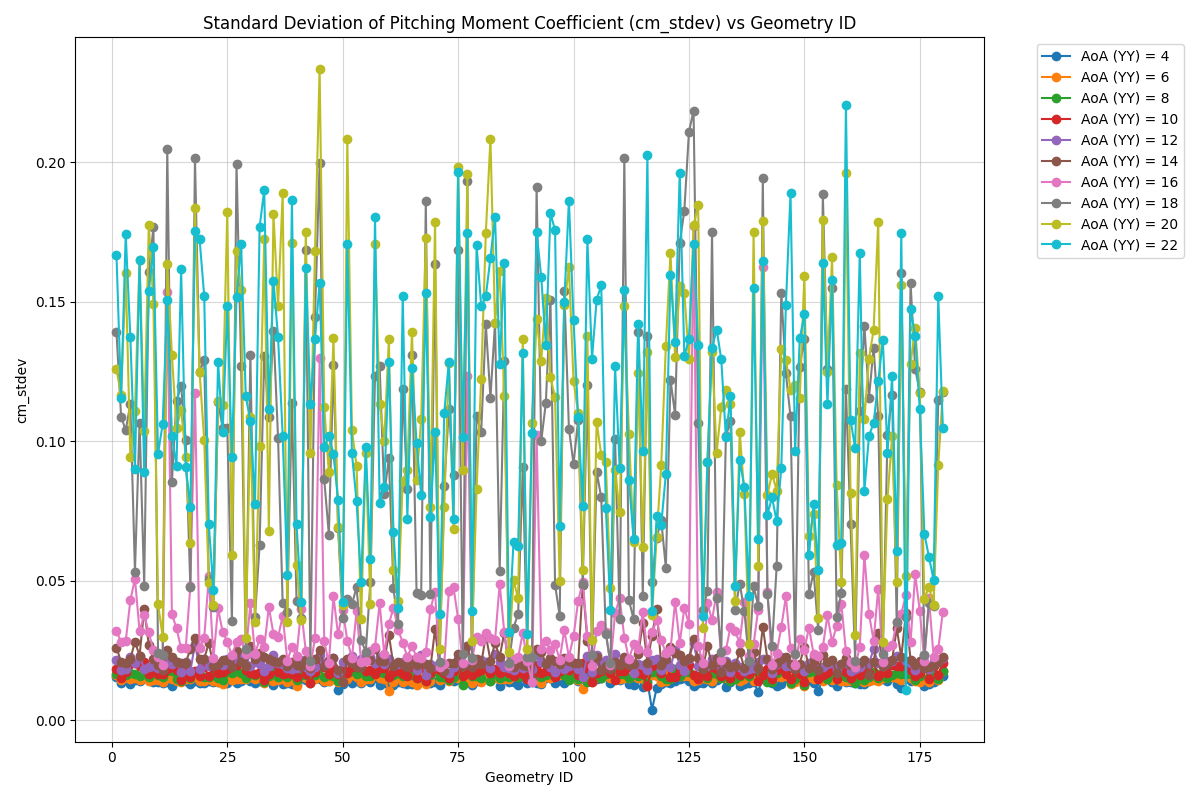}
    
    \caption{Standard deviation of Lift ($C_L$), Drag ($C_D$), and Pitching Moment ($C_M$) coefficients plotted against Geometry ID for various Angles of Attack.}
    \label{fig:stdev_plots}
\end{figure}

Figure~\ref{fig:stdev_plots} shows how the standard deviation of the instantaneous load signals varies with geometry, Angle of Attack (AoA), and aerodynamic coefficient ($C_L$, $C_D$, and $C_M$). Individual convergence plots (e.g., \texttt{plot\_CD\_geo\_LHCi\_AoA\_j.png}, \texttt{plot\_CL\_geo\_LHCi\_AoA\_j.png}, and \texttt{plot\_CM\_geo\_LHCi\_AoA\_j.png}) are provided in each released run directory.

\subsection{Volume export}
\noindent
The aggregate amount of volumetric data is particularly large due to the combination of the number of cases and the size of the individual grids.  While the underlying Voronoi grid has many favorable properties for the solution process, it exacerbates the size of the discrete mesh export due to the arbitrary face complexity of the cells and polyhedral topology.  These factors are not necessarily conducive for the purposes of training downstream machine learned models. To ameliorate some of these challenges, an oct-tree data structure is constructed from the cell centroids of the Voronoi grid.  The resultant oct-tree volumetric grid is discretely edge walkable.  The depth of the oct-tree is such that the size of the leaves is proportional to the local cell size, where the proportionality constant is a user defined parameter.  A proportionality constant of one would nominally result in an oct-tree leaf that contains a single centroid, while a larger proportionality constant allows for the volumetric grid to be coarsened. For this dataset the constant was set so that the leaf size tracks the local Voronoi length scale, so the released export carries essentially the native solution resolution rather than a coarsened one: for a representative case the oct-tree holds \(375\times10^{6}\) terminal leaves against \(314\times10^{6}\) solver control volumes. The reduction from that leaf count to the released cell count is the removal of leaves failing an inside/outside test against the geometry, not a change of resolution. Raising the constant coarsens the export further should a smaller representation be required \citep{ashton2025fluidintelligenceforwardlook}. This procedure resulted in surface meshes of approximately 30M cells (140M points) and volume mesh of 150-300M cells.

\section{Definition of aerodynamic quantities used in the dataset}
\label{app:postPro-aerospace}

\subsection{Solver Reference Definitions}
The flow solver utilizes a consistent set of definitions to derive the reference flow conditions from the input parameters (Mach number, Reynolds number, Angle of Attack, and atmospheric conditions). The simulations are performed using English units (slugs, inches, seconds, Rankine). The specific derivations used in the solver setup are defined as follows:

\textbf{Thermodynamic Constants:}
The gas constant $R_{gas}$ is converted to consistent units ($\mathrm{in}^2/(\mathrm{s}^2\,{}^\circ\mathrm{R})$) from the standard English value:
\begin{equation}
    R_{gas} = R_{gas,English} \times \left(12 \frac{in}{ft}\right)^2 \quad [\mathrm{in}^2/(\mathrm{s}^2\,{}^\circ\mathrm{R})]
\end{equation}
where $R_{gas,English} = 1716.594 \, \mathrm{ft}^2/(\mathrm{s}^2\,{}^\circ\mathrm{R})$ and $\gamma = 1.4$.

\textbf{Freestream Conditions:}
The speed of sound ($a_{\infty}$) and freestream velocity magnitude ($U_{\infty}$) are calculated as:
\begin{equation}
    a_{\infty} = \sqrt{\gamma R_{gas} T_{ref}} \;, \quad U_{\infty} = M \cdot a_{\infty} \quad [in/s]
\end{equation}

To match the specified Reynolds number ($Re$) and Mach number ($M$), the reference density ($\rho_{\infty}$) and viscosity ($\mu_{ref}$) are derived as:
\begin{eqnarray}
    && p_{ref} = 12 \times p_{atm} \quad [\mathrm{slug}/(\mathrm{in}\,\mathrm{s}^2)] \\
    && \rho_{\infty} = \frac{p_{ref}}{R_{gas} T_{ref}} \quad [slugs/in^3] \\
    && \mu_{ref} = \frac{\rho_{\infty} U_{\infty} C_{ref}}{Re} \quad [\mathrm{slug}/(\mathrm{in}\,\mathrm{s})]
\end{eqnarray}
where $p_{atm}$ is the atmospheric pressure in psi ($lbf/in^2$), $T_{ref}$ is the reference temperature in Rankine, and $C_{ref}$ is the reference chord length ($C_{MAC}$).

\textbf{Velocity Components and Time Scale:}
The inflow velocity components are rotated based on the angle of attack ($\alpha$). Note that $z$ is the body-vertical direction; it coincides with the lift direction only at $\alpha=0$, since lift is defined in the wind axes obtained by rotating this frame through $\alpha$:
\begin{eqnarray}
    && U_x = U_{\infty} \cos\left(\alpha \cdot \frac{\pi}{180}\right) \\
    && U_y = 0 \\
    && U_z = U_{\infty} \sin\left(\alpha \cdot \frac{\pi}{180}\right)
    \label{eq:postPro-uinf}
\end{eqnarray}
The convective time scale, defined as the Convective Time Unit (CTU), is given by:
\begin{equation}
    CTU = \frac{C_{ref}}{U_{\infty}}
\end{equation}

\subsection{Force and moment coefficients}
The dynamic pressure ($q_{\infty}$) used for normalization is defined consistent with the solver inputs:
\begin{equation}
    q_{\infty} = \frac{1}{2} \rho_{\infty} U_{\infty}^2 \quad [\mathrm{slug}/(\mathrm{in}\,\mathrm{s}^2)]
\end{equation}

The aerodynamic loads are obtained by integrating the surface traction over the wetted surface $S$. The traction is the sum of a pressure part and a viscous part,
\begin{equation}
    t_i = t^p_i + t^v_i \;, \quad \text{where} \quad t^p_i = (p - p_{ref})\, n_i \quad \text{and} \quad t^v_i = \tau_{w,i} \;,
    \label{eq:postPro-traction}
\end{equation}
Here $n_i$ is the unit boundary-face normal directed out of the fluid domain, that is into the body, so that a surface pressure above $p_{ref}$ gives a traction pushing on the wall and the positive sign in Eq.~(\ref{eq:postPro-traction}) is the force exerted by the fluid on the body; this is the orientation carried by the released \texttt{N\_BF} field and used by the load evaluator. The viscous part $\tau_{w,i}$ is the wall-shear-stress vector exerted by the fluid on the wall, taken in the same body frame. The solver reports the force and moment components in the body (geometry) frame, in which the model is fixed and the freestream is inclined by $\alpha$ as in Eq.~(\ref{eq:postPro-uinf}):
\begin{equation}
    F_i = \int_S t_i \, dS \;, \qquad M_i = \int_S \left[ (\mathbf{x} - \mathbf{r}_{\text{MRP}}) \times \mathbf{t} \right]_i dS \;, \qquad i \in \{x,y,z\} \;.
    \label{eq:postPro-bodyForces}
\end{equation}

Lift and drag are defined along the wind axes, so the released coefficients apply a rotation of $\alpha$ about the $y$-axis before normalisation:
\begin{eqnarray}
    && C_L = \frac{F_z\cos\alpha - F_x\sin\alpha}{q_{\infty}S_{\text{ref}}} \;\text{,}\hspace{0.03\textwidth}
    C_D = \frac{F_x\cos\alpha + F_z\sin\alpha}{q_{\infty}S_{\text{ref}}} \;\text{,}\hspace{0.03\textwidth}
    C_Y = \frac{F_y}{q_{\infty}S_{\text{ref}}} \;\text{.}
    \label{eq:postPro-forceCoeffs}
\end{eqnarray}
The body-axis axial force can be negative at high incidence and must not be interpreted as drag.

It is often useful to decompose these coefficients into their pressure and viscous components. Writing $F^k_i = \int_S t^k_i \, dS$ for $k \in \{p,v\}$ and applying the same rotation to each part separately gives
\begin{eqnarray}
    && C_{L,k} = \frac{F^k_z\cos\alpha - F^k_x\sin\alpha}{q_{\infty}S_{\text{ref}}} \;\text{,}\hspace{0.05\textwidth}
    C_{D,k} = \frac{F^k_x\cos\alpha + F^k_z\sin\alpha}{q_{\infty}S_{\text{ref}}} \;\text{,}
    \label{eq:postPro-liftDragComponents}
\end{eqnarray}
so that $C_L = C_{L,p} + C_{L,v}$ and $C_D = C_{D,p} + C_{D,v}$. Because the viscous traction acts predominantly along the freestream, $C_{L,v}$ is small and negative at incidence, whereas $C_{D,v}$ carries almost all of the skin-friction contribution.

The pitching moment is taken about the $y$-axis through the moment reference point $\mathbf{r}_{\text{MRP}}$ and is therefore unchanged by the rotation in Eq.~(\ref{eq:postPro-forceCoeffs}):
\begin{equation}
    C_M = \frac{M_y}{q_{\infty}S_{\text{ref}} C_{\text{ref}}} \;\text{,} \label{eq:postPro-momCoeffs}
\end{equation}

\subsection{Field storage and variable catalogue}

Table~\ref{tab:dataset-output-quantities} is the catalogue of released volume and surface fields, including their names, units, and averaging conventions. Volume fields are stored as \texttt{PointData} on an octree representation derived from the Voronoi grid; \texttt{CellID} is stored as \texttt{CellData}. The catalogue also lists the surface wall-shear and wall-distance statistics, boundary normals, and projected volume quantities. File formats and native-point surface quadrature weights are described in Appendix~\ref{app:dataset-dataDesc}.

\clearpage
\section{Extended validation results of numerical methodology}
\label{app:validation}

This section provides the experimental comparisons reported in the data-generation study of \citet{ashton2026hiliftgeneration}, supporting the CFD validation summarized in the main paper. Validation of the CFD workflow is carried out for the Landing configuration with Vertical and Horizontal stabilizers, experimentally identified as LDG-HV \citep{mouton2024testing}. In the following, we demonstrate that the Wall-Modeled LES (WMLES) methodology presented in the paper provides a reliable correlation to high-quality experimental wind tunnel data, which builds confidence that the approach can reasonably predict the wide range of geometric variations present in the HiLiftAeroML dataset.

\subsection{Integrated Forces and Moments}

To establish the baseline accuracy of the simulations, we compare the integrated forces from the present wall-modeled LES with the experiments of \citet{mouton2024testing} conducted at the ONERA F1 Pressurized Low-Speed Wind Tunnel. The validation focuses on the LDG-HV configuration at a Reynolds number of $Re_{MAC} = 1.6 \times 10^6$. Comparison is made between the baseline grid (approx. 100M control volumes) and the final adapted grid (approx. 300-500M control volumes), highlighting the efficacy of the solution-based adaptation approach \citep{agrawal2023reynolds}.

Fig.~\ref{fig:forces} confirms that the integrated forces are reasonably predicted relative to the experiments. While the lift coefficient ($C_L$) is not greatly changed due to the grid adaptation process, significant improvements in the drag coefficient ($C_D$) and pitching moment ($C_M$) are observed on the adapted grid. Specifically, the adaptation process corrects the stall behavior and post-stall predictions, aligning them more closely with experimental observations. Some non-smoothness observed on the coarse baseline grid at high angles of attack can be attributed to the relatively short time history used to drive the adaptation process.

\subsection{Grid Adaptation and Resolution}

\begin{figure}[htb]
    \centering
    \begin{subfigure}[b]{0.49\textwidth}
        \centering
        \includegraphics[width=\textwidth]{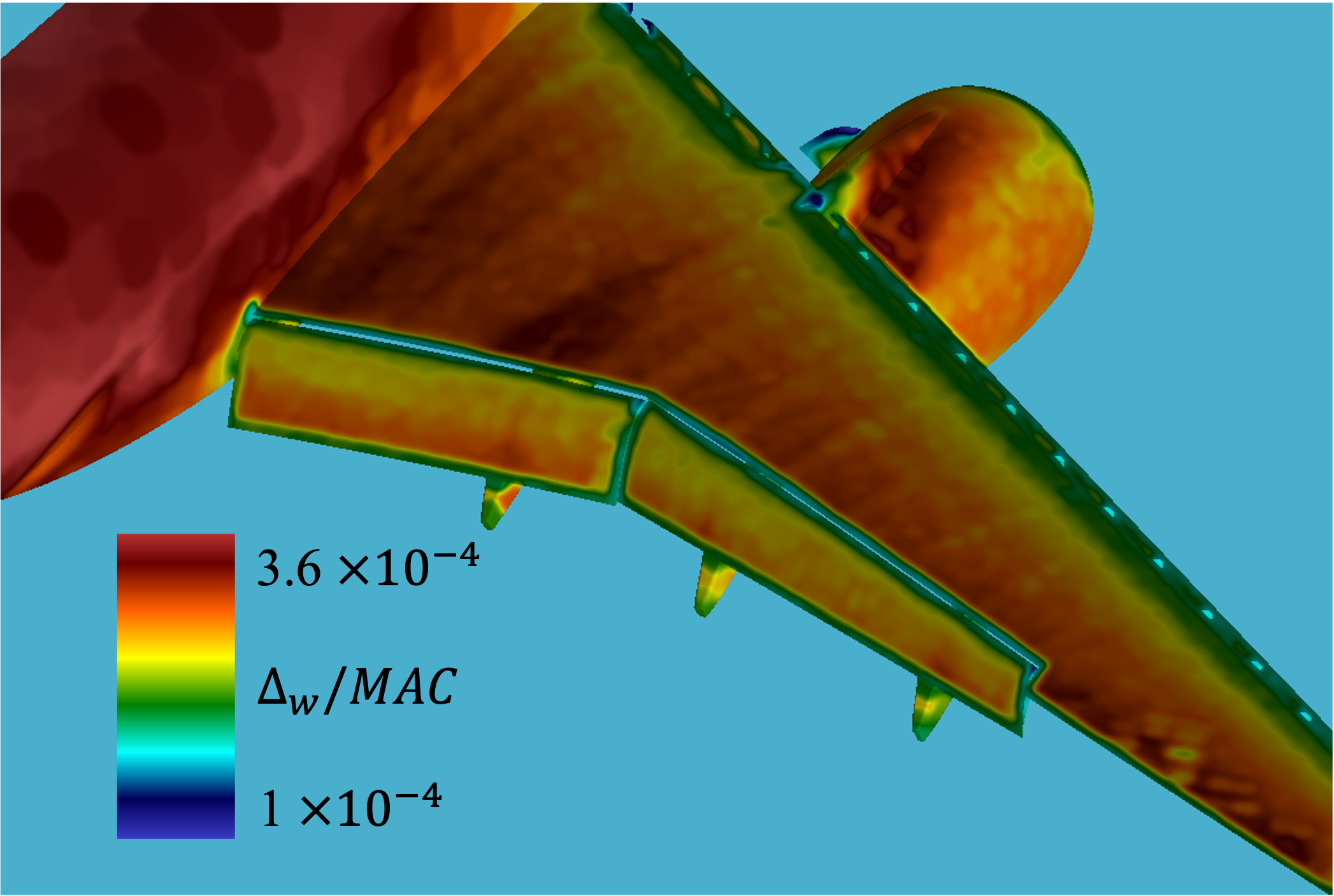}
        \caption{Airframe resolution}
        \label{fig:val-dxost-airframe}
    \end{subfigure}
    \hfill
    \begin{subfigure}[b]{0.49\textwidth}
        \centering
        \includegraphics[width=\textwidth]{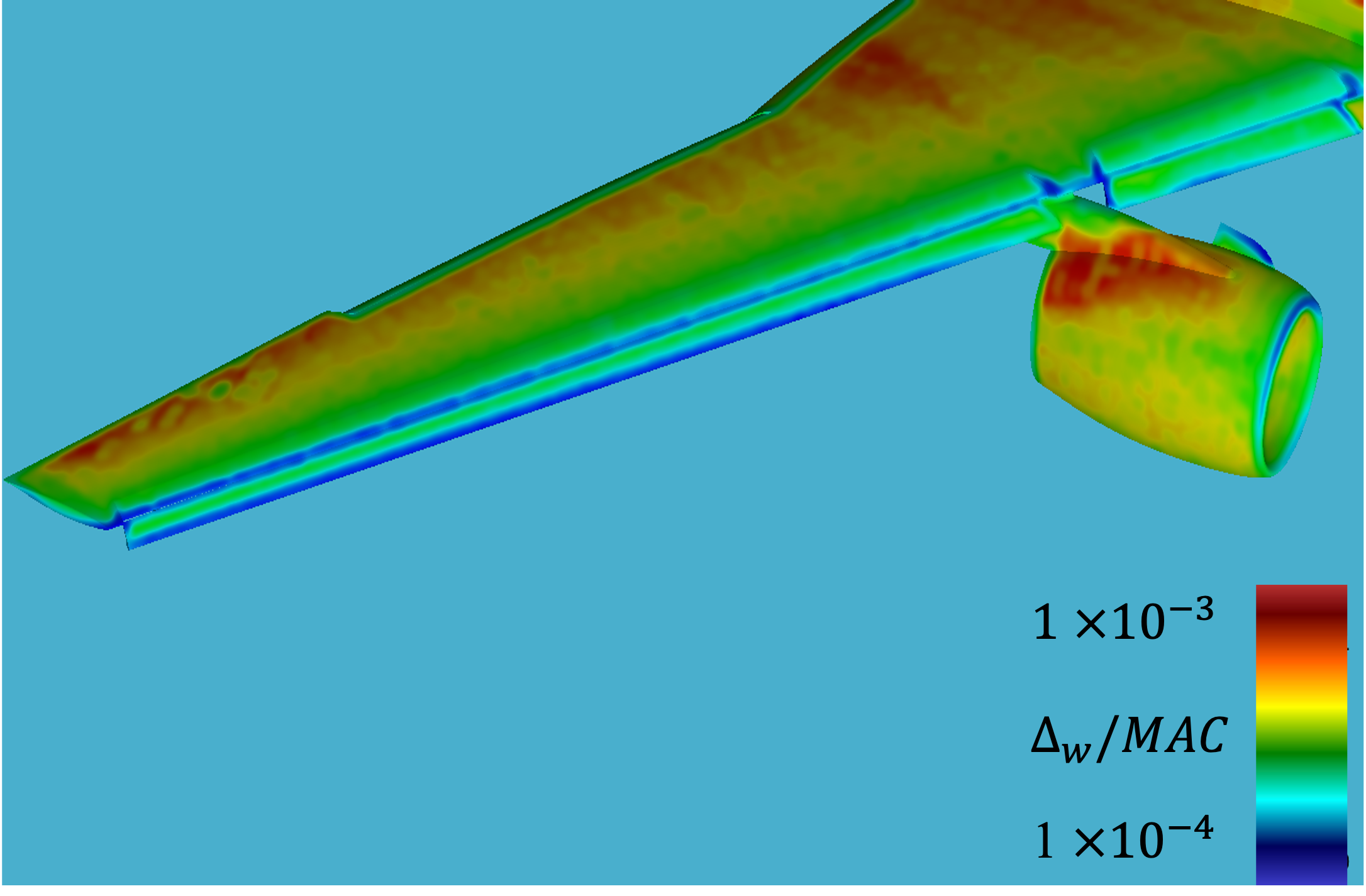}
        \caption{Slat zoom-in}
        \label{fig:val-dxost-slat}
    \end{subfigure}
    \caption{Contours of near-wall resolution from the adaptation approach \citep{agrawal2023reynolds} at $\alpha=18^\circ$.}
    \label{fig:dxostimage-val}
\end{figure}

A critical component of the methodology is the use of dynamic grid adaptation. Fig.~\ref{fig:dxostimage-val} presents the resulting near-wall grid resolutions suggested by the solution-based adaptation approach for the $\alpha=18^\circ$ condition.

It is apparent that the leading edges of the wing element necessitate more grid refinement than the rest of the wing. Similarly, finer grids are dynamically allocated to the slat element where the strong inviscid acceleration of the flow imposes strict pressure-gradient based resolution requirements. This targeted refinement allows the solver to capture complex flow physics, such as slat wakes and confluence, without the prohibitive cost of a uniformly fine mesh.

\subsection{Surface Pressure and Flow Physics}

To validate the local flow physics, we compare surface pressure distributions and near-wall flow patterns. Although the specific LDG-HV geometry used for force validation lacked pressure taps, comparisons are made against the similar LDG (no tail) configuration for which experimental pressure data exists.

\begin{figure}[htb]
    \centering
    \includegraphics[width=0.8\textwidth]{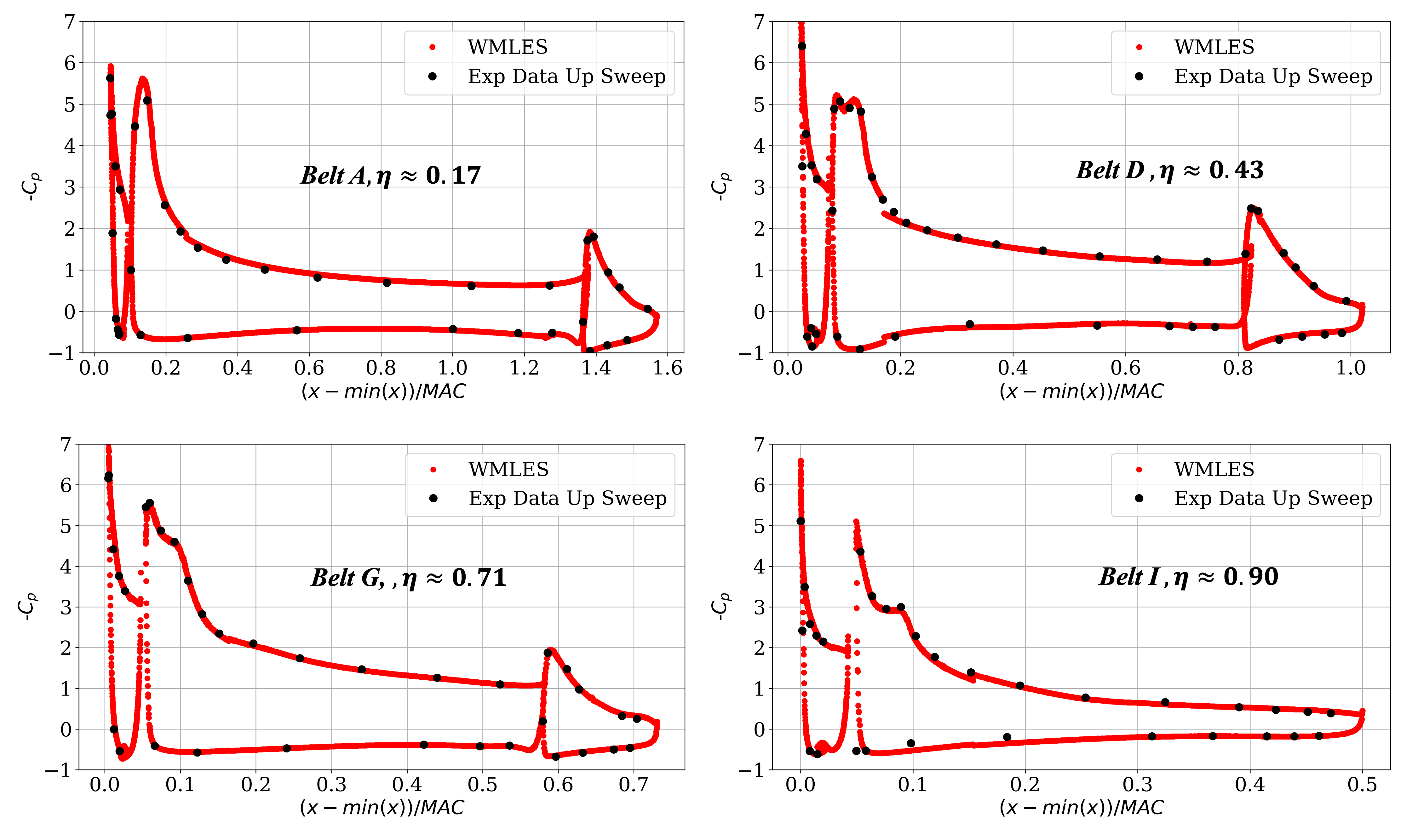}
    \caption{Comparisons of surface pressure belts from current wall-modeled LES (adapted grid) with experiments \citep{mouton2024testing} at $\alpha = 18^\circ$ for the LDG configuration.}
    \label{fig:wmles_18_cp-val}
\end{figure}

Fig.~\ref{fig:wmles_18_cp-val} shows the comparison of four surface pressure belts (A, D, G, and I) at $\alpha = 18^\circ$. Overall, favorable agreement between the simulation and experiments is obtained. Notably, the suction peaks on the main wing and flap elements (see Belts A, D, and G) are well captured. This region has historically been challenging for RANS simulation campaigns, and the accurate prediction here underscores the fidelity of the WMLES approach.

\begin{figure}
    \centering
     \includegraphics[width=\linewidth]{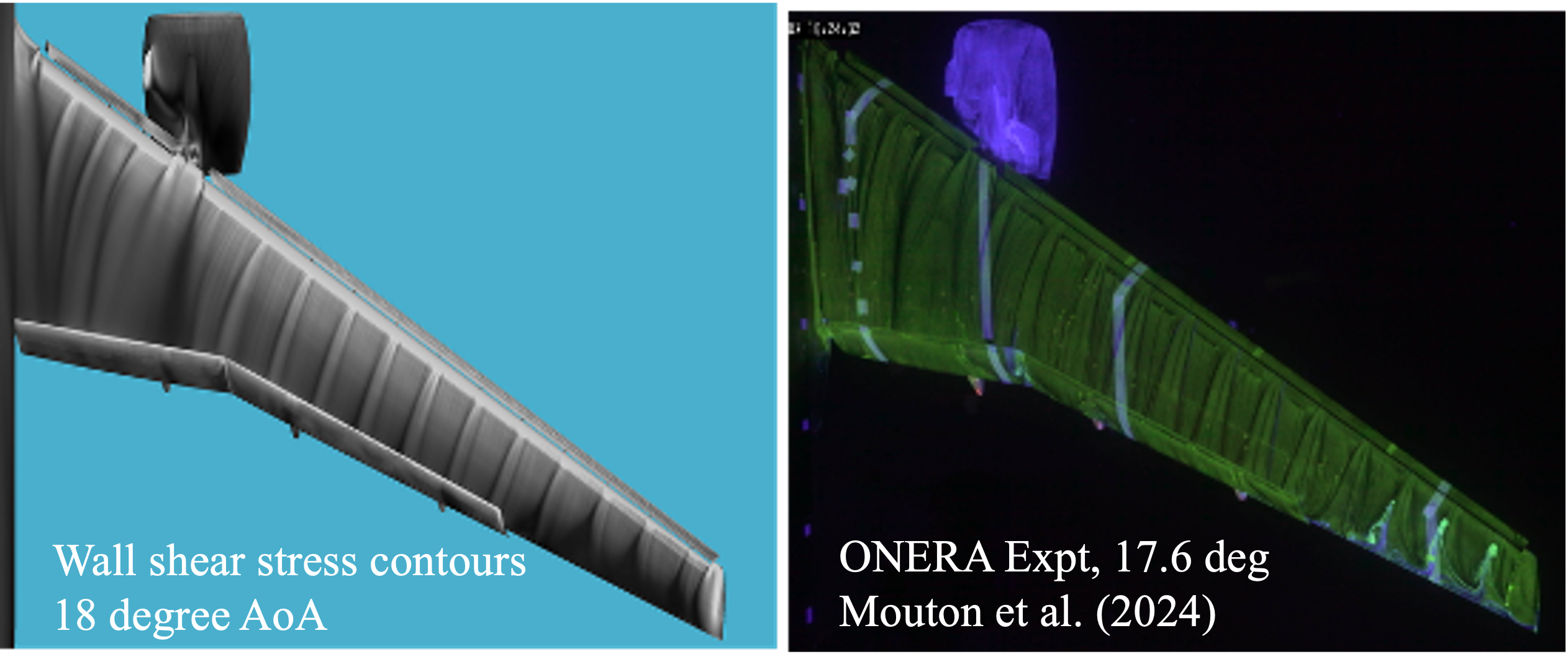}
    \caption{Comparison of experimental oil-film visualization at $\alpha=17.6^\circ$ \citep{mouton2024testing} with averaged wall-stress contours from the adapted-grid simulation at $\alpha=18^\circ$ on the suction side of the LDG-HV configuration at $Re_{MAC} = 1.6 \times 10^6$.}
    \label{fig:tau18}
\end{figure}

Finally, Fig.~\ref{fig:tau18} presents a qualitative comparison of near-surface flow patterns. On the adapted grid, the simulation correctly predicts the outboard wedge-shaped separation patterns near the wing tip, consistent with the experimental oil-flow visualization. Both the simulations and experiment also show evidence of flow separation on the flap near the Yehudi break.

In summary, the employed numerical methodology delivers an encouraging agreement with experiments for both integrated forces and local flow features. The ability to capture complex separation patterns and suction peaks at high angles of attack provides confidence in the dataset's utility for training machine learning models for high-lift aerodynamics.

\FloatBarrier
\section{Details of ML Evaluation}
\label{app:detailedml}

This section expands the machine-learning evaluation in the main paper. It documents the split methodology, the complete-native-support evaluation protocol, the exact field, force, and profile calculations, the aggregation and scoring rules, case-level and regional diagnostics, and run provenance. Results are from single training runs unless explicitly stated otherwise.

\subsection{Dataset Splits Methodology}

The HiLiftAeroML dataset features 14 deterministic train, validation, and test split families (Table~\ref{tab:splits_summary2}) curated to measure specific machine-learning capabilities, from basic data efficiency to out-of-distribution (OOD) generalization. The splits cover the 1800 complete LHC cases: 180 geometry variants evaluated at 10 angles of attack from \(4^\circ\) to \(22^\circ\). The split manifest file is included in the dataset repository and stores case IDs using the on-disk naming convention, sorted by geometry number and then AoA. Several principles guided the construction of these splits:
\begin{itemize}
    \item \textbf{In-distribution validation:} validation cases are drawn from the same statistical distribution as the corresponding training data, so hyperparameter tuning does not use OOD data.
    \item \textbf{Consistent ratios:} splits based on random or ranked geometry selection use an 80/20 test fraction, corresponding to an approximate $70/10/20$ train/validation/test partition. Physics-driven splits scale naturally based on their regime boundaries.
    \item \textbf{Shared geometries:} the geometry data-efficiency and single-AoA splits retain the same 18 validation and 36 testing geometries, enabling one-to-one comparisons across training-set sizes and flow regimes.
    \item \textbf{Exploratory benchmark status:} train, validation, and test memberships are disjoint, but test artifacts were inspected while this initial baseline campaign and its reporting protocol were developed. The results are therefore reproducible exploratory baselines, not an untouched competition leaderboard. Future method selection should use train/validation data only and evaluate the fixed test set once.
\end{itemize}

\begin{table}[t]
\centering
\caption{Summary of HiLiftAeroML dataset splits and their evaluation targets.}
\label{tab:splits_summary2}
\scriptsize
\resizebox{\textwidth}{!}{%
\begin{tabular}{llcccl}
\toprule
\textbf{Split} & \textbf{Type} & \textbf{Train} & \textbf{Val} & \textbf{Test} & \textbf{What it tests} \\
\midrule
\texttt{full} & In-dist & 1260 & 180 & 360 & Baseline random case-level split \\
\texttt{medium} & In-dist & 510 & 180 & 360 & Intermediate case-level data efficiency \\
\texttt{scarce} & In-dist & 210 & 180 & 360 & Data efficiency, $1/6$ of full training data \\
\texttt{super\_scarce} & In-dist & 35 & 180 & 360 & Extreme data efficiency, $1/36$ of full training data \\
\texttt{geometry} & In-dist & 1260 & 180 & 360 & Generalization to unseen geometries \\
\texttt{geometry\_medium} & In-dist & 510 & 180 & 360 & Unseen-geometry learning from 51 training geometries \\
\texttt{geometry\_scarce} & In-dist & 210 & 180 & 360 & Unseen-geometry learning from 21 training geometries \\
\texttt{geometry\_super\_scarce} & In-dist & 40 & 180 & 360 & Unseen-geometry learning from 4 training geometries \\
\texttt{single\_aoa\_4} & In-dist & 126 & 18 & 36 & Geometry generalization at $4^\circ$ lower AoA \\
\texttt{single\_aoa\_12} & In-dist & 126 & 18 & 36 & Geometry generalization at $12^\circ$ mid-range \\
\texttt{single\_aoa\_22} & In-dist & 126 & 18 & 36 & Geometry generalization at $22^\circ$ post-stall \\
\texttt{aoa} & OOD & 788 & 112 & 900 & Low-AoA training to high-AoA extrapolation \\
\texttt{deflection} & OOD & 1260 & 180 & 360 & Low-deflection training to high-deflection extrapolation \\
\texttt{stall} & OOD & 942 & 135 & 723 & Pre-stall training to post-stall extrapolation \\
\bottomrule
\end{tabular}%
}
\end{table}

\paragraph{Recommended comparisons.}
The splits are intended to form four complementary ladders. For case-level data efficiency, \texttt{full}, \texttt{medium}, \texttt{scarce}, and \texttt{super\_scarce} share the same validation and test sets while reducing the number of training cases from 1260 to 510, 210, and 35. For geometry data efficiency, \texttt{geometry}, \texttt{geometry\_medium}, \texttt{geometry\_scarce}, and \texttt{geometry\_super\_scarce} retain the same held-out geometries while reducing the training population from 126 to 51, 21, and 4 complete geometries. To assess whether operating-condition coverage helps unseen-geometry prediction, each \texttt{single\_aoa} split is compared with the matching angle slice of \texttt{geometry}: the train, validation, and test geometry IDs are identical, and each matched test contains the same 36 geometry--AoA cases. The random case-level \texttt{full} split remains useful for interpolation and data-efficiency benchmarks, but it is not used for this cross-AoA conclusion. For flow-regime extrapolation, \texttt{aoa} and \texttt{stall} test whether a model trained in lower-AoA or pre-stall conditions can extrapolate to high-AoA separated flow, while \texttt{deflection} withholds an intentionally harder high-deflection tail of the design space.

\paragraph{In-distribution splits.}
The \texttt{full} split is a random case-level $70/10/20$ partition of all 1800 cases. Because this partition is at the case level, a geometry may appear in training at one AoA and in validation or test at another AoA. The \texttt{medium}, \texttt{scarce}, and \texttt{super\_scarce} splits use the same validation and test sets as \texttt{full}, but replace the training set with nested random subsamples of \texttt{full\_train}. \texttt{medium} contains 510 cases, near the geometric midpoint between the 1260-case \texttt{full} and 210-case \texttt{scarce} training sets; \texttt{scarce} uses \(1/6\) of the full training set and \texttt{super\_scarce} uses \(1/36\). Thus,
\begin{equation}
\text{\texttt{super\_scarce\_train}} \subset
\text{\texttt{scarce\_train}} \subset
\text{\texttt{medium\_train}} \subset
\text{\texttt{full\_train}},
\end{equation}
where the added midpoint resolves the previous six-fold jump from \texttt{full} to \texttt{scarce} without changing any published assignment.

\paragraph{Geometry and single-AoA splits.}
The \texttt{geometry} split partitions the 180 geometries into 126 training, 18 validation, and 36 test geometries, retaining all 10 AoA values for each selected geometry. Unlike \texttt{full}, this setup prevents the model from seeing a test geometry at any AoA during training. Its nested \texttt{geometry\_medium}, \texttt{geometry\_scarce}, and \texttt{geometry\_super\_scarce} variants reuse the same validation and test geometries but retain 51, 21, and 4 complete training geometries, respectively. Keeping all 10 AoAs per selected geometry gives 510, 210, and 40 training cases and isolates geometry coverage from operating-condition coverage:
\begin{equation}
\begin{split}
\text{\texttt{geometry\_super\_scarce\_train}} &\subset
\text{\texttt{geometry\_scarce\_train}} \\ &\subset
\text{\texttt{geometry\_medium\_train}} \\ &\subset
\text{\texttt{geometry\_train}}.
\end{split}
\end{equation}
The \texttt{single\_aoa\_4}, \texttt{single\_aoa\_12}, and \texttt{single\_aoa\_22} splits reuse the same 126/18/36 geometry partition but restrict the cases to a single AoA. Their train, validation, and test lists are exactly the corresponding angle slices of the \texttt{geometry} lists. These variants isolate geometry generalization in pre-stall, mid-range, and post-stall regimes, while comparison with the matching \texttt{geometry} slice measures the observed benefit of adding the other nine AoAs for the same training geometries.

\paragraph{OOD AoA and deflection splits.}
The \texttt{aoa} split trains and validates on cases with \(\alpha \leq 12^\circ\) and tests on cases with \(\alpha \geq 14^\circ\); all 180 geometries appear in the train-side and test-side populations. The \(12^\circ/14^\circ\) cutoff is physics motivated: near \(14^\circ\)--\(16^\circ\), the dominant aerodynamic sensitivity shifts from flap deflection toward slat deflection, and many geometries begin to stall. The validation set is a random \(1/8\) subset of the lower-AoA pool. The \texttt{deflection} split ranks geometries by the mean of the inboard/outboard flap and slat deflection angles, trains and validates on the lower 80\% of this ranked list, and tests on the upper 20\%, representing the most aggressive high-lift settings. It provides a distribution-shifted geometry test: both splits hold out complete geometries, while \texttt{deflection} selects the upper tail of the prescribed deflection score rather than a random geometry subset.

\paragraph{Stall split.}
The \texttt{stall} split is defined from the aerodynamic response rather than from geometry parameters or a fixed AoA threshold. For each geometry, a cubic spline is fit to \(C_L(\alpha)\), and the stall onset is identified as the first AoA where
\begin{equation}
\frac{d C_L}{d \alpha} \leq 0.
\end{equation}
Cases at and beyond that AoA are labeled post-stall and used for testing; earlier cases are pre-stall and are divided into train and validation sets, with validation again selected as a random \(1/8\) subset of the train-side pool. All 180 geometries exhibit a detected stall onset in this criterion. Most stall at \(16^\circ\) (100 geometries), with a secondary cluster at \(14^\circ\) (44 geometries), yielding 1077 pre-stall cases and 723 post-stall cases.

\begin{figure}[t]
\centering
\includegraphics[width=\textwidth]{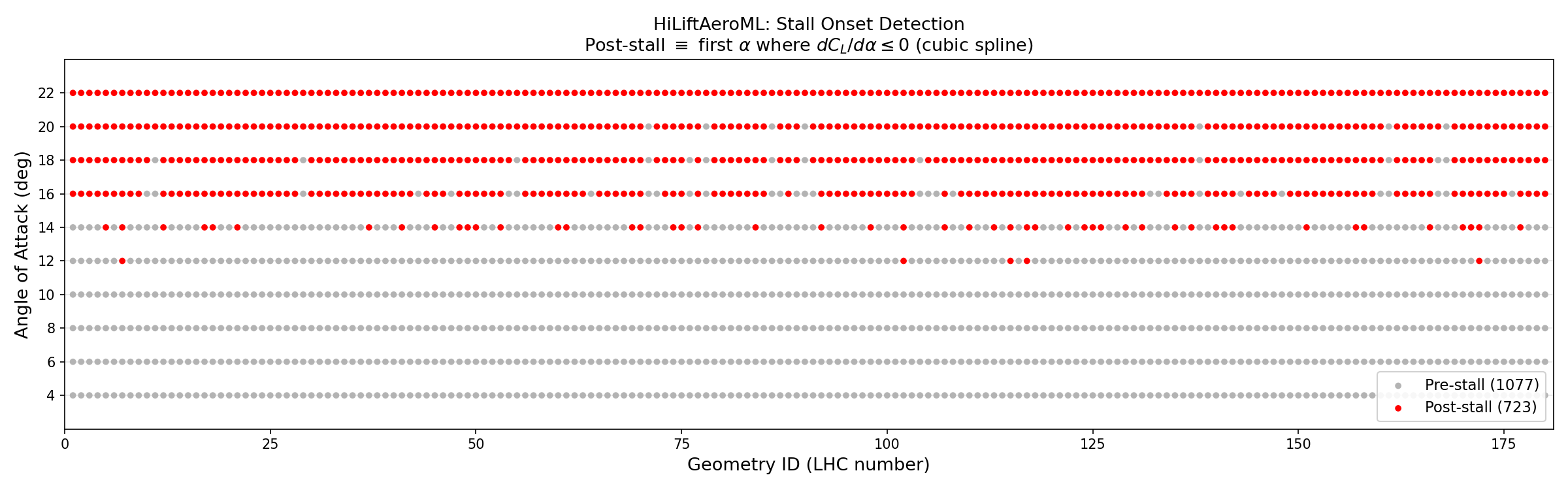}
\caption{Stall-onset labeling used for the \texttt{stall} split. Each dot is one geometry--AoA case; gray cases are pre-stall and red cases are post-stall according to the first AoA where a cubic-spline estimate satisfies \(dC_L/d\alpha \leq 0\).}
\label{fig:app_stall_onset_split}
\end{figure}

\paragraph{Reproducibility.}
Splits are produced deterministically with a fixed seed by \path{splits/generate_splits.py}. The manifest stores case-ID lists for every train, validation, and test population, including \path{medium_train}, \path{geometry_medium_train}, and the previously published keys, making membership independent of downstream model code.

\paragraph{Completed result coverage.}
Twelve of the 14 split families have completed surface and volume records under the complete-native-support protocol, giving the 23 evaluations in Table~\ref{tab:ml_native_benchmark}: 12 for GeoTransolver and 11 for Transolver, forming 11 checkpoint-comparable model pairs. The \texttt{scarce} split has a GeoTransolver record only. The \texttt{medium} and \texttt{geometry\_medium} families were added to the published manifest after this baseline campaign and have no completed records. Across all 23 records, the protocol produced \(8{,}142\) case--model--split evaluations involving \(1{,}355\) distinct CFD cases.

\subsection{Model Architecture Details}

\paragraph{Transolver.}
Transolver~\citep{wu2024transolver} addresses the cost and noise of applying transformer attention directly to high-resolution simulation meshes. The key operation is Physics-Attention: point or cell features are projected onto a fixed number of learned physical-state slices; attention is applied between the resulting slice tokens; and the updated slice information is mapped back to the original mesh tokens. Because the number of slices is fixed, the attention cost scales linearly with the number of presented entities for a fixed slice count, rather than quadratically in the entity count. In the PhysicsNeMo implementation used here, the unstructured surface model instantiates \texttt{PhysicsAttentionIrregularMesh} blocks, preceded by an MLP projection and followed by residual MLP updates. All reported rows use sixteen Transolver blocks, hidden width 320, MLP ratio four (1280-wide feed-forward layers), eight heads, and 256 physical-state slices, for 18.54M trainable parameters. Centered surface-cell centroids and cell normals form the six-dimensional geometry embedding, while the three-component freestream vector \(U_\infty\) is broadcast to every presented cell as the functional input. Thus, AoA enters through freestream conditioning, whereas local surface geometry enters through the centroid--normal embedding. This configuration was selected to approximately capacity match GeoTransolver, so that the two headline architectures are compared at similar parameter counts rather than at each architecture's individually tuned optimum.

\paragraph{GeoTransolver.}
GeoTransolver~\citep{adams2025geotransolverlearningphysicsirregular} extends Transolver for irregular CAE domains by adding Geometry-Aware Latent Embeddings (GALE). The arXiv paper motivates GALE as a way to keep every transformer layer anchored to geometry, boundary conditions, and operating state: physical-state self-attention is retained, but each GALE block also cross-attends to shared context tokens built from geometry and global parameters. In the PhysicsNeMo implementation, \texttt{ContextProjector} tokenizes geometry and global embeddings into the same slice-token form used by Physics-Attention, and each \texttt{GALE\_block} mixes self-attention and context cross-attention with a learnable gate. Every reported configuration uses twelve GALE blocks, base hidden width 256, eight heads, 256 slices, and weighted state mixing. The selected configuration enables the multi-scale ball-query branch (\texttt{include\_local\_features=true}) at radii \(0.1/0.5/2.0\), neighbour caps \(16/32/64\), and local width 32. Concatenating the three local outputs increases the effective GALE width from 256 to 352 and the active capacity from 9.05M to 17.06M. The branch was enabled on the basis of preliminary single-AoA experiments that are not reported here; because it also changes capacity, those experiments did not isolate the contribution of neighbourhood information, and we make no claim about it.

\subsection{Preprocessing and Training Representation}
\label{app:ml_preprocessing_decision}

All ML rows use the PhysicsNeMo \texttt{.pdmsh} representation rather than the original solver VTU files at training time. The surface path reads the boundary pmesh, where each surface cell has a unique centroid, outward normal, area, pressure, wall shear, and attached freestream metadata.
For surface GeoTransolver and Transolver, each case is represented by centered, sampled surface-cell centroids and cell normals. The targets are pressure, temperature, density, velocity, and wall shear in a common nondimensional layout followed by a surface-only training z-score, with the reported surface metrics focused on pressure and wall shear after that z-score has been reversed. Freestream velocity is broadcast as the Transolver functional input and is projected as GeoTransolver global context. In the selected GeoTransolver rows, three ball-query coordinate features are appended to the initial embedding and tokenized local context; this is an architectural change, not merely a preprocessing flag. For volume GeoTransolver and Transolver, each case is represented by centered, sampled interior volume points, STL-derived signed distance and SDF normals, and the same freestream metadata. All volume models predict pressure, temperature, density, and velocity.

\paragraph{Training sampling.}
The reader subsamples existing mesh entries with a seeded generator, and training reseeds that generator by epoch, so a sampled subset is reproducible but changes between epochs. The 200k training setting denotes the number of existing surface cells or volume points presented per case in one model pass, not a newly generated point cloud. This is a property of \emph{training} only. It is important not to conflate it with evaluation: the models are trained on stochastic 200k-entity samples but are evaluated on the complete native support, as defined next.

Note that the surface entity differs between the two stages: training presents surface-cell centroids, whereas evaluation scores the native boundary points with their nodal dual areas. Both architectures map coordinates to fields and so can be queried at any location on the body, and the training and evaluation coordinates both lie on the same wetted surface, so evaluating at nodes rather than at cell centroids requires no interpolation of either the prediction or the truth: each is read at the coordinate where it is stored. The volume stage uses the same entity, the native volume point, at both stages.

\subsection{Complete-Native-Support Evaluation}
\label{app:ml_native_protocol}

\paragraph{Bounded exhaustive inference.}
Evaluation covers every retained native entity of every held-out case exactly once. Because a single attention pass over a 100M-entity boundary or a 200M-point volume is not feasible, the evaluator partitions each case into bounded, globally mixed chunks of at most 200k entities and runs one forward pass per chunk. Mixing is global rather than spatially blocked: the surface uses a case-keyed PCG64 permutation of the native point order, and the volume uses balanced ranges of the seed-42 randomized PDMsh order. Each chunk therefore presents an approximately unbiased sample of the whole body to the model, matching the statistics of the training context while still visiting every entity exactly once across the full sweep.

Three properties of this protocol should be stated explicitly. First, it is an exhaustive traversal, not a subsample: coverage and weight-coverage fractions are exactly 1.0 for every case and support in every reported record. Second, it is \emph{not} one whole-mesh attention pass; the model never sees the entire case in a single context, so the result should not be read as the limit of a global-attention model. Third, it is not a mesh-convergence study: the native mesh is fixed, and no refinement sequence is involved.

\paragraph{Surface support and weights.}
The surface support is all retained native boundary \texttt{PointData} points. The scored quantities are the scalar pressure coefficient and the three-component wall-shear coefficient. The primary spatial measure is the published native nodal dual area, so the surface metric approximates a physical area integral rather than a count over an adaptively refined mesh. For the vector quantity, a point's weight multiplies the complete squared three-component error norm at that point rather than being applied per component. An equal-native-node surface reduction is retained in the released artifacts as a secondary diagnostic but is not the primary surface measure.

\paragraph{Volume support and validity.}
The volume support is every valid native volume \texttt{PointData} point. Validity is decided by the raw Float32 rule \(\mathrm{avg}(P)\neq 0.0\) before normalization, which excludes the documented invalid all-zero rows (Appendix~\ref{app:dataset-quality}). The scored quantities are the scalar pressure coefficient and the three-component velocity. The primary spatial measure is one equal weight per valid native point. No volume or cell-volume weighting is introduced into the primary volume result: the native volume discretization is strongly graded, and cell-volume weighting would let a small number of large far-field cells dominate a metric that is intended to describe the resolved flow.

\paragraph{Physical variables.}
All relative metrics use equivalent nondimensional physical variables,
\begin{equation}
C_p=\frac{p-p_\infty}{q_\infty},\qquad
\frac{\boldsymbol{\tau}_w}{q_\infty},\qquad
\frac{\mathbf{u}}{\lvert\mathbf{U}_\infty\rvert},
\end{equation}
evaluated after the training normalization has been reversed. Every evaluated case shares the same nominal reference condition. Within a case, conversion to gauge pressure, dimensional wall shear, or dimensional velocity multiplies truth and prediction by the same scale and leaves relative \(L_2\) and \(R^2\) unchanged. Dimensional errors retain the evaluator's per-case inverse normalization, including the small difference between its field scale \(q_\infty\) and the published load reference \(q_{\mathrm{ref}}\). Absolute pressure is avoided for relative \(L_2\), because adding \(p_\infty\) would make its denominator dominated by the freestream offset and reduce the apparent relative error. Centered \(R^2\) is unchanged by a common additive offset to truth and prediction.

\subsection{Field Metric Definitions}
\label{app:ml_native_metrics}

\paragraph{Per-case complete-support metrics.}
Consider one case with complete support \(i=1,\dots,N\), nonnegative spatial weights \(w_i\), truth \(\mathbf y_i\), and prediction \(\widehat{\mathbf y}_i\). For a scalar field the norm is the absolute value; for a vector field it is the full Euclidean norm over components. The weighted relative \(L_2\) error in percent is
\begin{equation}
E_{\mathrm{rel},2}
=100\sqrt{\frac{\sum_i w_i\lVert\widehat{\mathbf y}_i-\mathbf y_i\rVert^2}
{\sum_i w_i\lVert\mathbf y_i\rVert^2}}\;\%.
\label{eq:native_rel_l2}
\end{equation}
The companion weighted quantities are
\begin{align}
E_{\mathrm{rel},1}&=100\,\frac{\sum_i w_i\lVert\widehat{\mathbf y}_i-\mathbf y_i\rVert_1}{\sum_i w_i\lVert\mathbf y_i\rVert_1}\;\%,\\
\operatorname{MAE}&=\frac{\sum_i w_i\lVert\widehat{\mathbf y}_i-\mathbf y_i\rVert_1}{d\sum_i w_i},\\
\operatorname{RMSE}&=\sqrt{\frac{\sum_i w_i\lVert\widehat{\mathbf y}_i-\mathbf y_i\rVert^2}{d\sum_i w_i}},
\end{align}
where \(d\) is the number of components. The coefficient of determination uses the same complete support and the weighted truth mean \(\bar{\mathbf y}=\sum_i w_i\mathbf y_i/\sum_i w_i\),
\begin{equation}
R^2=1-\frac{\sum_i w_i\lVert\widehat{\mathbf y}_i-\mathbf y_i\rVert^2}
{\sum_i w_i\lVert\mathbf y_i-\bar{\mathbf y}\rVert^2}.
\label{eq:native_r2}
\end{equation}
These definitions require positive denominators. Relative errors are undefined for zero truth norm, and \(R^2\) is undefined for zero truth variance. The benchmark asset builder rejects non-positive field-error denominators and non-positive truth variance; no additive denominator offset is used for the reported relative \(L_2\) or \(R^2\) values.

\paragraph{Sufficient statistics.}
Each case stores the numerator \(S_e=\sum_i w_i\lVert\widehat{\mathbf y}_i-\mathbf y_i\rVert^2\) and denominator \(S_t=\sum_i w_i\lVert\mathbf y_i\rVert^2\) of Equation~\eqref{eq:native_rel_l2}, together with \(\sum_i w_i\mathbf y_i\) and \(\sum_i w_i\) for the centered form. These statistics are additive across cases and chunks. For fixed predictions, accumulating them over chunks gives the same complete-support reduction up to floating-point summation order. The predictions themselves depend on the bounded contexts used by the model.

\paragraph{Case aggregation.}
Every field error reported in this paper uses the \emph{equal-case macro} reduction: the complete-support metric is formed for each case from that case's sufficient statistics, and the \(K\) case values are then averaged with equal weight,
\begin{equation}
E^{\mathrm{macro}}_{\mathrm{rel},2}
=\frac{1}{K}\sum_{c=1}^{K}
100\sqrt{\frac{S_e^{(c)}}{S_t^{(c)}}}\;\%.
\label{eq:macro}
\end{equation}
Each held-out case therefore counts once, independently of how many native entities it contains. A pooled field error would instead combine case numerators and denominators before taking their ratio. With equal-point volume weights, this gives greater influence to cases with more points or larger truth magnitudes; with surface dual-area weights, influence depends on physical area and the truth field, rather than point count alone. Equal-case averaging makes the per-case distribution of Section~\ref{app:ml_case_error_analysis} the distribution of the reported mean. It is the reduction the benchmark ranks on, so the relative field errors match the corresponding leaderboard records.

Note that \emph{within} a case, sufficient statistics are still accumulated over the complete support with the spatial weights of Section~\ref{app:ml_native_metrics}: dual area on the surface, equal weight per valid point in the volume. The equal-case choice concerns aggregation across cases only.

Coefficient \(R^2\) involves no spatial reduction at all: \(C_L\), \(C_D\), and \(C_M\) are one scalar per case, so their \(R^2\) is the ordinary case-set statistic of Equation~\eqref{eq:coefficient_r2}.

\subsection{Operating-condition coverage on matched geometries}
\label{app:ml_matched_aoa}

The Geometry split permits a direct case-ID-matched test of operating-condition coverage: it uses the same 126/18/36 train/validation/test geometry partition as each single-AoA split, so its angle-specific test subsets contain exactly the same 36 unseen cases, scored on exactly the same complete native support. Training GeoTransolver on all ten AoAs rather than only the target angle lowers complete-support surface pressure/wall-shear errors from \natAoafourGeotSurfCpRelLtwo\%/\natAoafourGeotSurfTauRelLtwo\% to \natangleGeometryAoafourGeotSurfCpRelLtwo\%/\natangleGeometryAoafourGeotSurfTauRelLtwo\% at \(4^\circ\), from \natAoatwelveGeotSurfCpRelLtwo\%/\natAoatwelveGeotSurfTauRelLtwo\% to \natangleGeometryAoatwelveGeotSurfCpRelLtwo\%/\natangleGeometryAoatwelveGeotSurfTauRelLtwo\% at \(12^\circ\), and from \natAoatwentytwoGeotSurfCpRelLtwo\%/\natAoatwentytwoGeotSurfTauRelLtwo\% to \natangleGeometryAoatwentytwoGeotSurfCpRelLtwo\%/\natangleGeometryAoatwentytwoGeotSurfTauRelLtwo\% at \(22^\circ\). Volume pressure/velocity errors fall correspondingly from \natAoafourGeotVolPRelLtwo\%/\natAoafourGeotVolVRelLtwo\% to \natangleGeometryAoafourGeotVolPRelLtwo\%/\natangleGeometryAoafourGeotVolVRelLtwo\%, from \natAoatwelveGeotVolPRelLtwo\%/\natAoatwelveGeotVolVRelLtwo\% to \natangleGeometryAoatwelveGeotVolPRelLtwo\%/\natangleGeometryAoatwelveGeotVolVRelLtwo\%, and from \natAoatwentytwoGeotVolPRelLtwo\%/\natAoatwentytwoGeotVolVRelLtwo\% to \natangleGeometryAoatwentytwoGeotVolPRelLtwo\%/\natangleGeometryAoatwentytwoGeotVolVRelLtwo\%. Every quantity improves at every angle, and at \(22^\circ\) the reductions are both the largest and the most uniform, at roughly 25--33\% across all four, against spreads of 14--27\% at \(4^\circ\) and 6--28\% at \(12^\circ\). Broader AoA coverage therefore helps on the same unseen geometries without removing the difficulty of separated high-AoA flow, which remains several times harder than \(4^\circ\) or \(12^\circ\).

The multi-angle model also uses ten times as many training cases, and the comparison is not compute matched; the observed reductions therefore do not isolate the causal effect of AoA coverage.

\subsection{Integrated Loads}
\label{app:metric_force_convention}

Lift, drag, and pitching moment are derived from the same complete native surface prediction that produces the field metrics. No separately fitted scalar force head is used. Spatial errors can cancel during integration, so integrated-load and pointwise-field accuracy need not have the same ordering, particularly when assessed with different metrics.

\paragraph{Integration of the surface interpolant.}
Pressure and viscous tractions are represented by a piecewise-linear interpolant on the planar triangles of each native polygon's ordered fan. Forces use degree-one integration of that interpolant; the pitching moment uses degree-two integration of \((\mathbf r-\mathbf{r}_{\text{MRP}})\times\boldsymbol{t}\), where \(\boldsymbol t\) is the interpolated traction. The quadrature is exact for these polynomial integrands, and the same rule is applied to predicted and reference fields. Its exactness is relative to the surface interpolant; geometric and field-discretization errors remain.

\paragraph{Reference quantities.}
The published references are \(S_{\text{ref}}=191.8447776\,\mathrm{m}^2\), \(C_{\text{ref}}=7.00532\,\mathrm{m}\), and \(\mathbf{r}_{\text{MRP}}=(33.67786,0,4.51993)\,\mathrm{m}\); equivalently, the native dataset metadata stores \(297360.0\,\mathrm{in}^2\), \(275.80\,\mathrm{in}\), and \((1325.9,0,177.95)\,\mathrm{in}\). Forces are integrated in body axes and projected to wind axes at the case AoA. Coefficients are normalized with the published \(q_\mathrm{ref}\), \(S_{\text{ref}}\), and \(C_{\text{ref}}\), and the \(q_\infty/q_\mathrm{ref}\) conversion is applied exactly once. The pitching coefficient follows the dataset's positive-body-\(y\) convention,
\begin{equation}
C_M=\frac{M_y}{q_{\mathrm{ref}} S_{\text{ref}}C_{\text{ref}}}.
\end{equation}

\paragraph{Coefficient scoring.}
For each coefficient, \(R^2\) is evaluated across the predicted/truth pair from every case of the split, using the ordinary case-set definition
\begin{equation}
R^2(C)=1-\frac{\sum_{k}(\widehat C_k-C_k)^2}{\sum_{k}(C_k-\bar C)^2}.
\label{eq:coefficient_r2}
\end{equation}
Drag and lift \(R^2\) are the headline integrated-load quantities. Pitching-moment \(R^2\) is reported alongside them as a diagnostic, but it is not read as independent evidence, because \(C_M\) is already strongly determined by the same surface pressure field that the \(C_p\) and \(C_L\) results reflect. Coefficient MAE is the casewise mean \(K^{-1}\sum_k|\widehat C_k-C_k|\) in dimensionless coefficient points; the corresponding dimensional load differences are \(q_{\mathrm{ref}} S_{\text{ref}}\operatorname{MAE}(C_L)\), \(q_{\mathrm{ref}} S_{\text{ref}}\operatorname{MAE}(C_D)\), and \(q_{\mathrm{ref}} S_{\text{ref}}C_{\text{ref}}\operatorname{MAE}(C_M)\).

\subsection{Absolute errors in engineering units}
\label{app:ml_absolute_errors}

Table~\ref{tab:ml_native_rmse} complements the relative field errors and coefficient \(R^2\) with absolute root-mean-square errors (RMSE) for the same 23 records and checkpoints. Field RMSE is calculated over each complete case, using surface dual-area weights or equal valid-volume-point weights, and then averaged equally over cases. Wall shear and velocity use the componentwise definition above, with \(d=3\). Pressure and shear errors are converted from \(\mathrm{slug}/(\mathrm{in}\,\mathrm{s}^2)\) to Pa using a factor of 574.5631; velocity errors are converted from in/s to m/s using 0.0254. Coefficient RMSE uses one predicted/truth coefficient pair per test case,
\begin{equation}
\operatorname{RMSE}(C)=\sqrt{\frac{1}{K}\sum_{k=1}^{K}(\widehat C_k-C_k)^2},
\qquad C\in\{C_L,C_D,C_M\}.
\label{eq:coefficient_rmse}
\end{equation}
This differs from averaging per-case absolute coefficient errors, which gives MAE.

\begin{table}[!htb]
\centering
\caption{Absolute RMSE on complete native support. GeoT denotes GeoTransolver and Trans denotes Transolver; Geom. denotes Geometry. Field values are equal-case means of per-case RMSE; coefficient values use Equation~\eqref{eq:coefficient_rmse}. Lift and moment errors are in coefficient points; drag errors are in drag counts, where one count is \(10^{-4}\) in \(C_D\). Lower is better. The records, test cases, and checkpoints match the relative-error tables.}
\label{tab:ml_native_rmse}
\small
\setlength{\tabcolsep}{3pt}
\renewcommand{\arraystretch}{1.07}
\begin{tabular}{@{}llrrrrrrr@{}}
\toprule
& & \multicolumn{2}{c}{Surface} & \multicolumn{2}{c}{Volume} & \multicolumn{3}{c}{Integrated loads} \\
\cmidrule(lr){3-4}\cmidrule(lr){5-6}\cmidrule(l){7-9}
Split & Model & \(p\) & \(\boldsymbol{\tau}_w\) & \(p\) & \(\mathbf u\) & \(C_L\) & \(C_D\) & \(C_M\) \\
& & (Pa) & (Pa) & (Pa) & (m/s) & (points) & (counts) & (points) \\
\midrule
AoA $=4^\circ$ & GeoT & 63.9 & 0.65 & 90.5 & 2.23 & 0.0148 & 9.1 & 0.0086 \\
 & Trans & 81.4 & 0.70 & 157.5 & 4.05 & 0.0209 & 25.3 & 0.0085 \\
\addlinespace[2pt]
AoA $=12^\circ$ & GeoT & 65.3 & 0.67 & 161.6 & 3.06 & 0.0107 & 13.6 & 0.0055 \\
 & Trans & 105.2 & 0.84 & 315.2 & 4.17 & 0.0176 & 49.6 & 0.0107 \\
\addlinespace[2pt]
AoA $=22^\circ$ & GeoT & 215.6 & 1.05 & 415.0 & 5.56 & 0.0487 & 141.6 & 0.0676 \\
 & Trans & 224.0 & 1.11 & 484.1 & 6.06 & 0.0492 & 152.1 & 0.0644 \\
\addlinespace[2pt]
Full & GeoT & 87.3 & 0.68 & 165.9 & 2.66 & 0.0287 & 75.2 & 0.0262 \\
 & Trans & 99.6 & 0.73 & 224.5 & 3.46 & 0.0276 & 78.1 & 0.0302 \\
\addlinespace[2pt]
Scarce & GeoT & 248.4 & 1.59 & 414.7 & 6.48 & 0.0500 & 123.6 & 0.0641 \\
\addlinespace[2pt]
Super-scarce & GeoT & 795.3 & 3.67 & 1258.0 & 15.21 & 0.2153 & 615.5 & 0.2124 \\
 & Trans & 725.2 & 3.38 & 1749.9 & 17.57 & 0.1357 & 300.6 & 0.1832 \\
\addlinespace[2pt]
Geometry & GeoT & 94.0 & 0.68 & 189.2 & 2.76 & 0.0285 & 83.3 & 0.0364 \\
 & Trans & 112.9 & 0.77 & 243.0 & 3.35 & 0.0294 & 80.9 & 0.0375 \\
\addlinespace[2pt]
Geom.-scarce & GeoT & 258.3 & 1.38 & 631.8 & 8.28 & 0.0582 & 165.0 & 0.1003 \\
 & Trans & 263.4 & 1.35 & 529.0 & 6.46 & 0.0565 & 156.5 & 0.0995 \\
\addlinespace[2pt]
Geom.-super-scarce & GeoT & 926.2 & 3.78 & 1302.2 & 15.72 & 0.2290 & 617.5 & 0.4161 \\
 & Trans & 583.7 & 2.52 & 1188.3 & 12.49 & 0.1240 & 340.9 & 0.1381 \\
\addlinespace[2pt]
OOD AoA & GeoT & 969.2 & 3.06 & 1717.2 & 14.52 & 0.4132 & 497.6 & 0.1931 \\
 & Trans & 1409.8 & 4.95 & 2140.1 & 21.71 & 0.9232 & 3376.4 & 0.6149 \\
\addlinespace[2pt]
OOD deflection & GeoT & 98.5 & 0.76 & 166.4 & 2.66 & 0.0273 & 78.4 & 0.0316 \\
 & Trans & 124.1 & 0.85 & 265.5 & 3.70 & 0.0342 & 92.5 & 0.0370 \\
\addlinespace[2pt]
OOD stall & GeoT & 776.6 & 2.48 & 1754.3 & 12.86 & 0.2174 & 601.4 & 0.1538 \\
 & Trans & 753.4 & 2.45 & 1721.8 & 12.75 & 0.2134 & 565.0 & 0.1466 \\
\bottomrule

\end{tabular}
\end{table}

\paragraph{Converting coefficient errors to loads.}
For the published half-model reference area and chord, \(q_{\mathrm{ref}}=2837.11\,\mathrm{Pa}\), \(S_{\mathrm{ref}}=191.8448\,\mathrm{m}^2\), and \(C_{\mathrm{ref}}=7.00532\,\mathrm{m}\). Multiplying the \(C_L\) or \(C_D\) RMSE by \(q_{\mathrm{ref}}S_{\mathrm{ref}}\) gives lift or drag RMSE in N; multiplying the \(C_M\) RMSE by \(q_{\mathrm{ref}}S_{\mathrm{ref}}C_{\mathrm{ref}}\) gives pitching-moment RMSE in N\,m. Thus, 0.01 lift-coefficient points correspond to 5.44~kN, one drag count to 54.4~N, and 0.01 moment-coefficient points to 38.1~kN\,m at the dataset's reference conditions. These are loads for the simulated half-model and its published reference scale.

\subsection{Sectional Pressure Cuts}
\label{app:ml_cp_cuts}

\begin{figure}[t]
\centering
\includegraphics[width=\columnwidth]{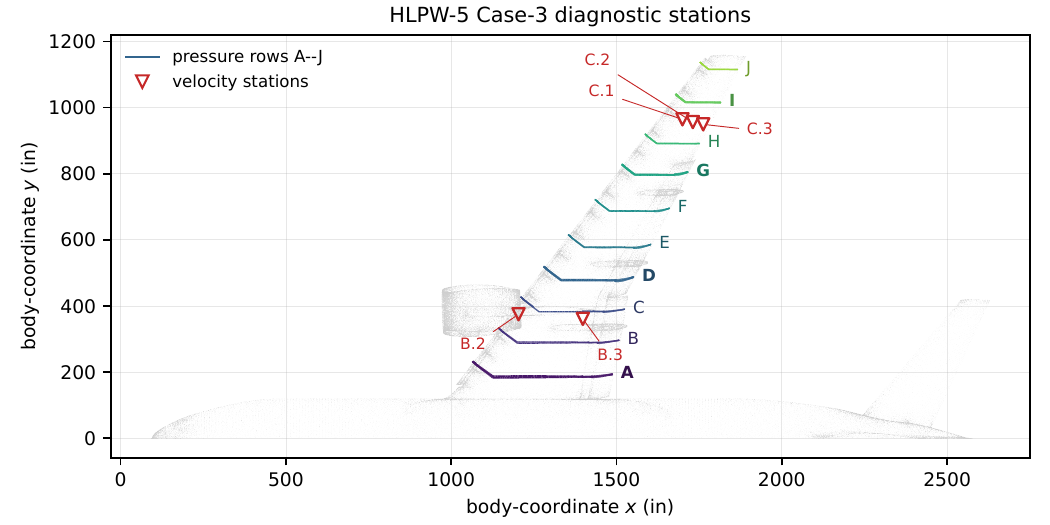}
\caption{Planform location of the HLPW-5 Case-3 diagnostic stations on the released geometry. Grey shows the aircraft surface; coloured lines show the official pressure planes intersected with their corresponding main-wing, slat or flap surfaces. The bend near the main-wing leading edge is part of the station definition. Rows used in the main paper are labelled in bold; red markers locate the five retained velocity stations. Component restrictions apply to this locator figure only, not to the scored extraction.}
\label{fig:app_hlpw5_station_map}
\end{figure}

\paragraph{Where the stations are.}
The diagnostic coordinates follow the published HLPW-5 Case-3 station definitions~\citep{clark2025HLPW5}; the subset of velocity stations used here is author-selected. The workshop's pressure-station tables and extraction macros define rows A--J by planes \(Ax+By+Cz=D\) in body-coordinate inches. Each main-wing row has a short segment approximately normal to the leading edge, followed by a nearly streamwise aft segment, with the switch at the published streamwise coordinate. Separate planes define the slat and, for rows A--G, the deployed flap; these segments need not align across the gaps between deployed elements. The five retained velocity stations B.2, B.3, C.1, C.2 and C.3 use the published nominal-grid velocity-template coordinates, which fix each station at a single \((x,y)\) and vary only in \(z\). Figure~\ref{fig:app_hlpw5_station_map} locates both sets on the aircraft.

\paragraph{Construction.}
All ten rows are scored; no row is dropped. Within a row the cut through the high-lift geometry is generally not a single curve, and every disconnected physical cut graph is preserved as its own object. At most 128 samples are retained per physical connected graph, placed uniformly in physical graph arc length, so that a long main-element cut and a short slat cut are each resolved without either dominating the sample budget. No edge is ever created between disconnected branches or across a topology gap.

\paragraph{Scoring.}
Truth is centered independently within each physical graph, which removes a constant offset per branch that would otherwise inflate a shared \(R^2\). The piecewise-linear squared error and the corresponding centered squared truth are integrated against physical arc length, pooled over the graphs of one case, and converted to a case \(R^2\). Case scores are then averaged with equal weight in the same way as Equation~\eqref{eq:macro}.

\paragraph{Storage and decoding.}
Predictions are stored as \(\mathrm{int16}(\mathrm{round}(C_p\cdot1024))\) with branch-local delta coding, giving a quantization step of just under \(10^{-3}\) in \(C_p\). Every figure in this paper decodes that representation exactly, by cumulative summation within each branch followed by division by 1024. Figures are plotted against the physical streamwise coordinate \(x\) in inches with an inverted \(C_p\) axis, and each panel's limits are computed from every displayed finite point, so no curve is clipped.

\subsection{Velocity Profiles}
\label{app:ml_velocity_profiles}

The scored profile contract uses an author-selected subset of five of the sixteen catalogued workshop stations: B.2, B.3, C.1, C.2, and C.3. These provide local diagnostics at both inboard (B.2 and B.3) and outboard (C.1--C.3) spanwise locations, as shown in Figure~\ref{fig:app_hlpw5_station_map}. Each station requests 801 vertical rows, giving 4005 requested rows per case, of which 3991 are valid for the held-out geometry shown in this paper. The authoritative validity mask is applied, and invalid rows are retained as explicit gaps: they are never interpolated, bridged, or dropped in a way that would silently shorten a profile.

The plotted and scored profile scalar is the speed magnitude \(|U|/|U_\infty|\). It is well defined at every valid row without reference to a local edge direction, but it does not distinguish opposite velocity directions and can therefore hide flow-reversal errors. These scalar profiles complement the full velocity-vector error scored over the volume. Physical polyline arc-length weights are used along each station, truth is centered independently by station, and station sufficient statistics are pooled within a case before the case \(R^2\) values are averaged with equal weight. Velocity values are exact Float32 after the lossless storage transform is decoded.

\paragraph{Atlas over further held-out geometries.}
\begin{figure}[!t]
\centering
\includegraphics[width=\textwidth]{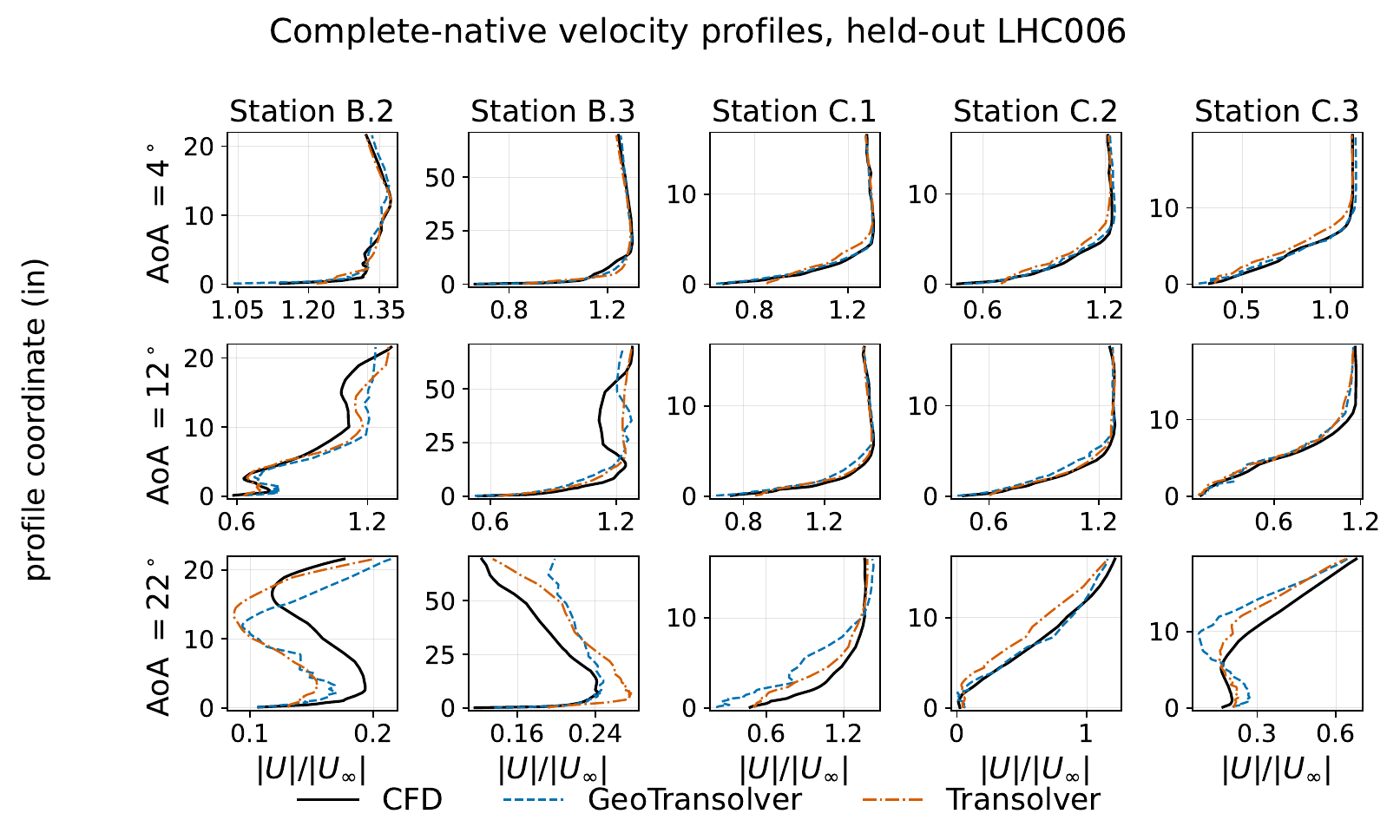}
\caption{Complete-native velocity profiles for the held-out geometry LHC006 at
\(4^\circ\), \(12^\circ\), and \(22^\circ\), at the five retained HLPW-5
stations. The plotted scalar is the speed ratio \(|U|/|U_\infty|\). Invalid
rows are retained as explicit gaps and are never interpolated or bridged. Both
models track the attached profiles closely but miss the separated wake
structure at \(22^\circ\).}
\label{fig:ml_native_velocity_profiles}
\end{figure}

\begin{figure}[p]
\centering
\includegraphics[width=\textwidth]{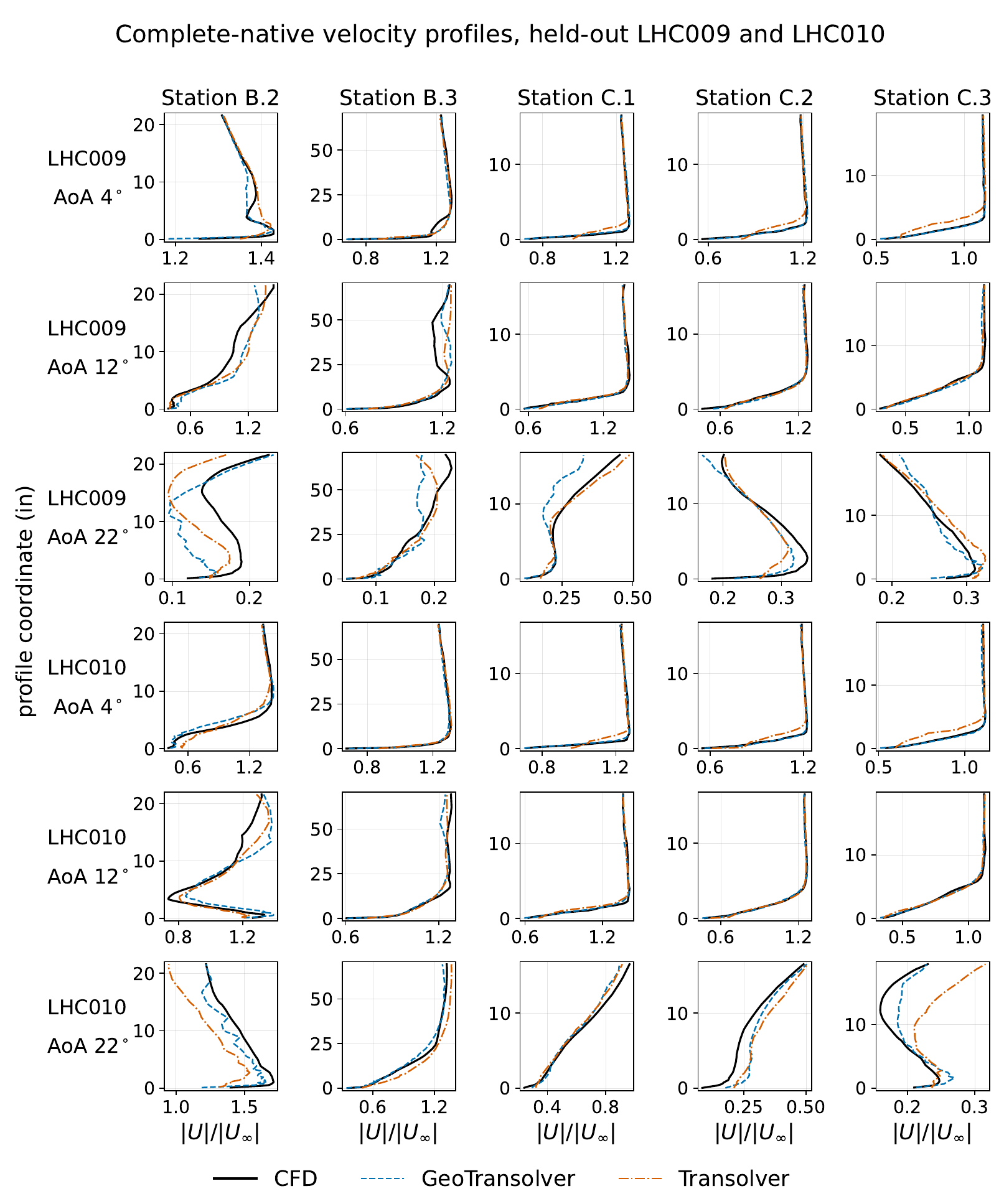}
\caption{Velocity-profile atlas, part 1: held-out geometries LHC009 and LHC010 at \(4^\circ\), \(12^\circ\), and \(22^\circ\), on the five retained stations. Truth and predictions are read at identical points from the same fixed-AoA records used for Figure~\ref{fig:ml_native_velocity_profiles}; validity gaps are retained as explicit breaks and are never interpolated or bridged, and each panel's limits are computed from every displayed finite point.}
\label{fig:app_native_velocity_atlas_p1}
\end{figure}

\begin{figure}[p]
\centering
\includegraphics[width=\textwidth]{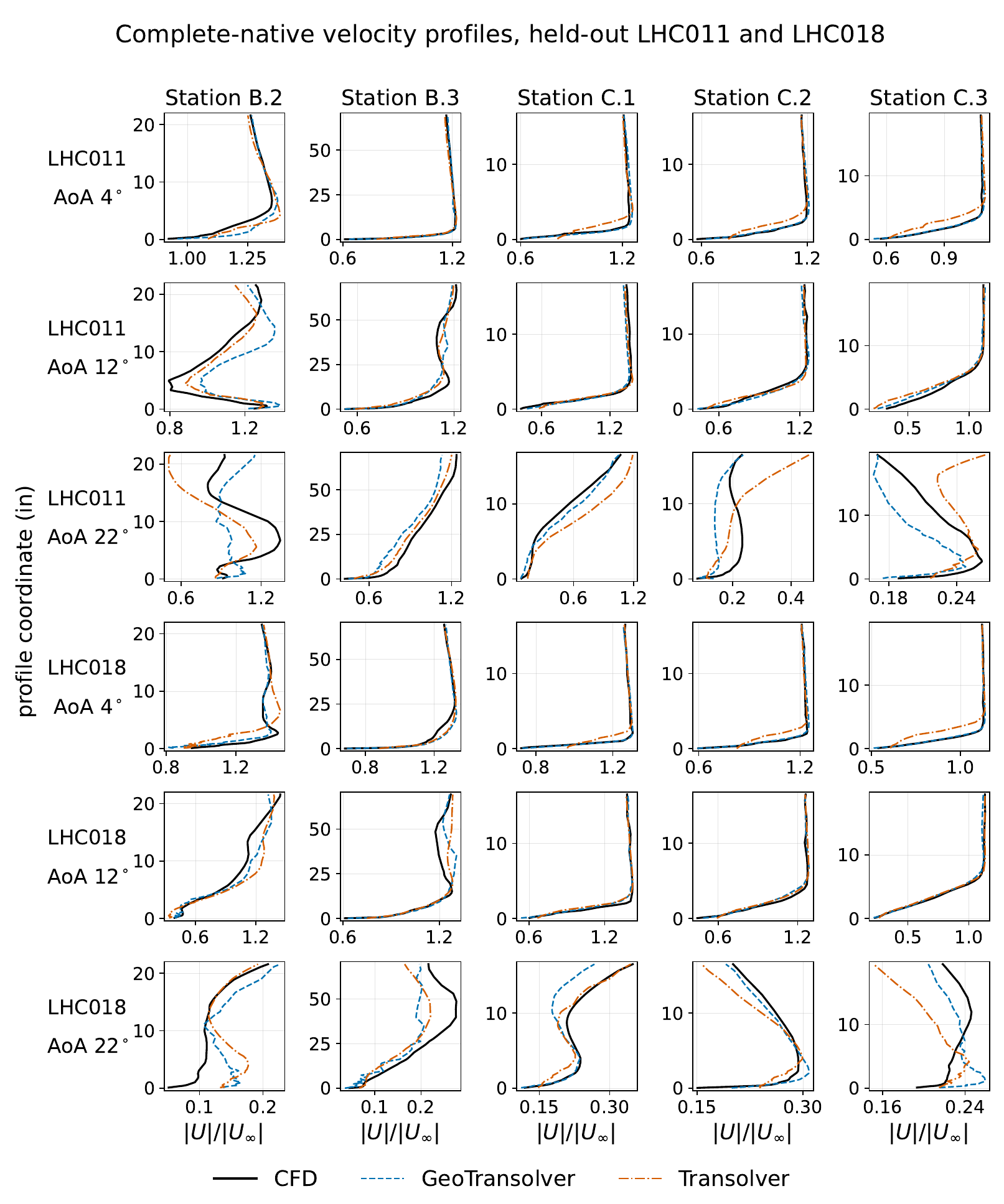}
\caption{Velocity-profile atlas, part 2: held-out geometries LHC011 and LHC018, plotted under the same conventions as Figure~\ref{fig:app_native_velocity_atlas_p1}.}
\label{fig:app_native_velocity_atlas_p2}
\end{figure}

Figures~\ref{fig:app_native_velocity_atlas_p1} and~\ref{fig:app_native_velocity_atlas_p2} extend Fig.~\ref{fig:ml_native_velocity_profiles} to four further held-out geometries at the same three angles, using the same five stations, the same fixed-AoA records, and the same plotting conventions. The geometries are the first four held-out cases by identifier other than LHC006, so they are chosen by name and not by error; across the three angles their per-case profile errors fall between roughly the tenth and the eighty-fifth percentile of the 36 held-out geometries, and LHC006 itself is likewise unexceptional. The qualitative reading is stable across all five geometries: the \(4^\circ\) profiles are tracked closely at every station, the \(12^\circ\) profiles are tracked closely except for departures in the near-wake station B.2, and the \(22^\circ\) profiles miss the separated structure on every geometry. This is the case-level counterpart of the strongly negative equal-case velocity-profile \(R^2\) that the stalled splits produce.

\begin{figure}[p]
\centering
\includegraphics[width=\textwidth]{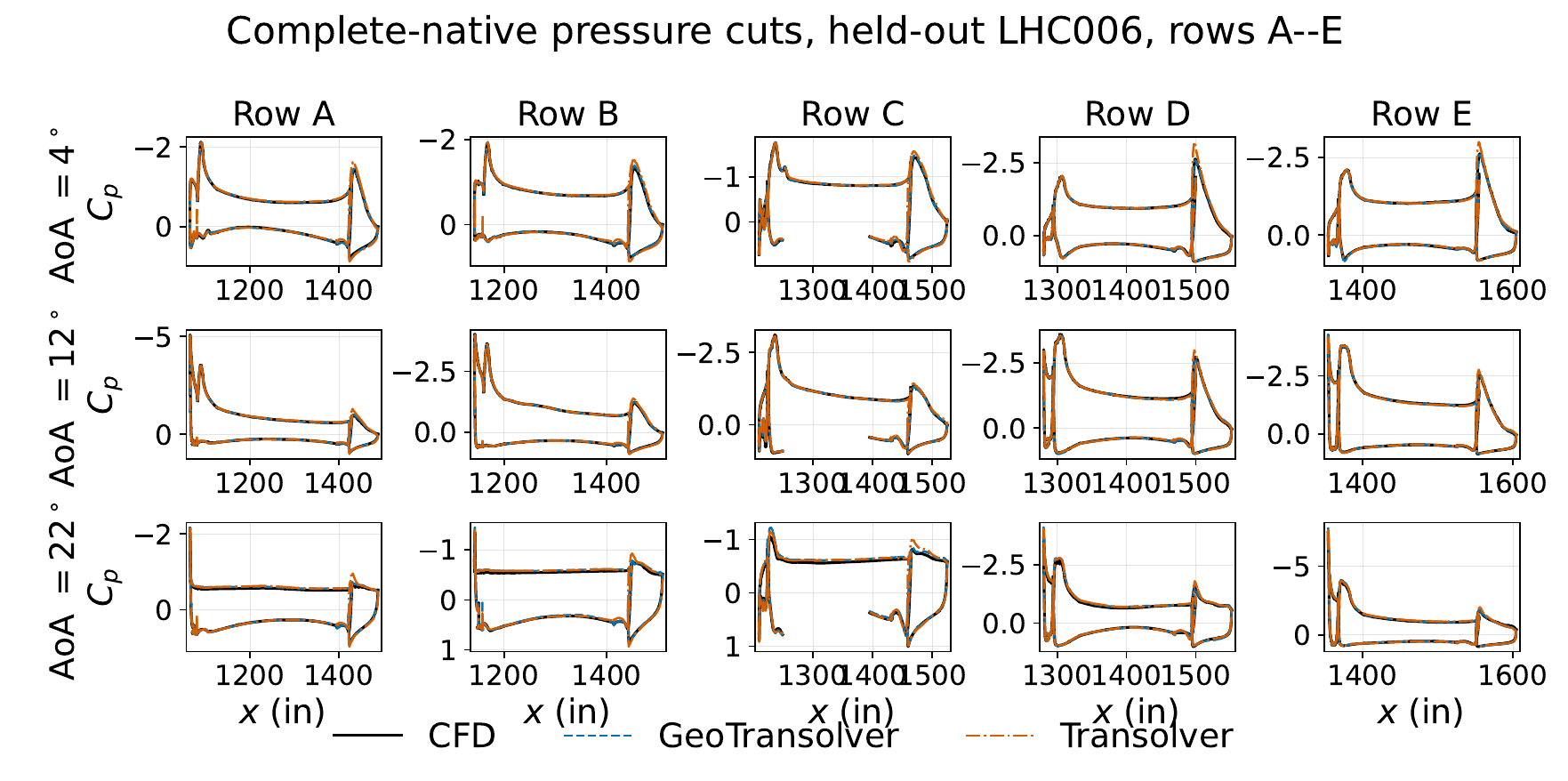}
\caption{Complete A--E pressure-cut atlas for the held-out geometry LHC006 at \(4^\circ\), \(12^\circ\), and \(22^\circ\). Each disconnected physical cut graph is drawn as its own polyline; the visible breaks are genuine topology gaps in the cut, not missing data.}
\label{fig:app_native_cp_atlas_ae}
\end{figure}

\begin{figure}[p]
\centering
\includegraphics[width=\textwidth]{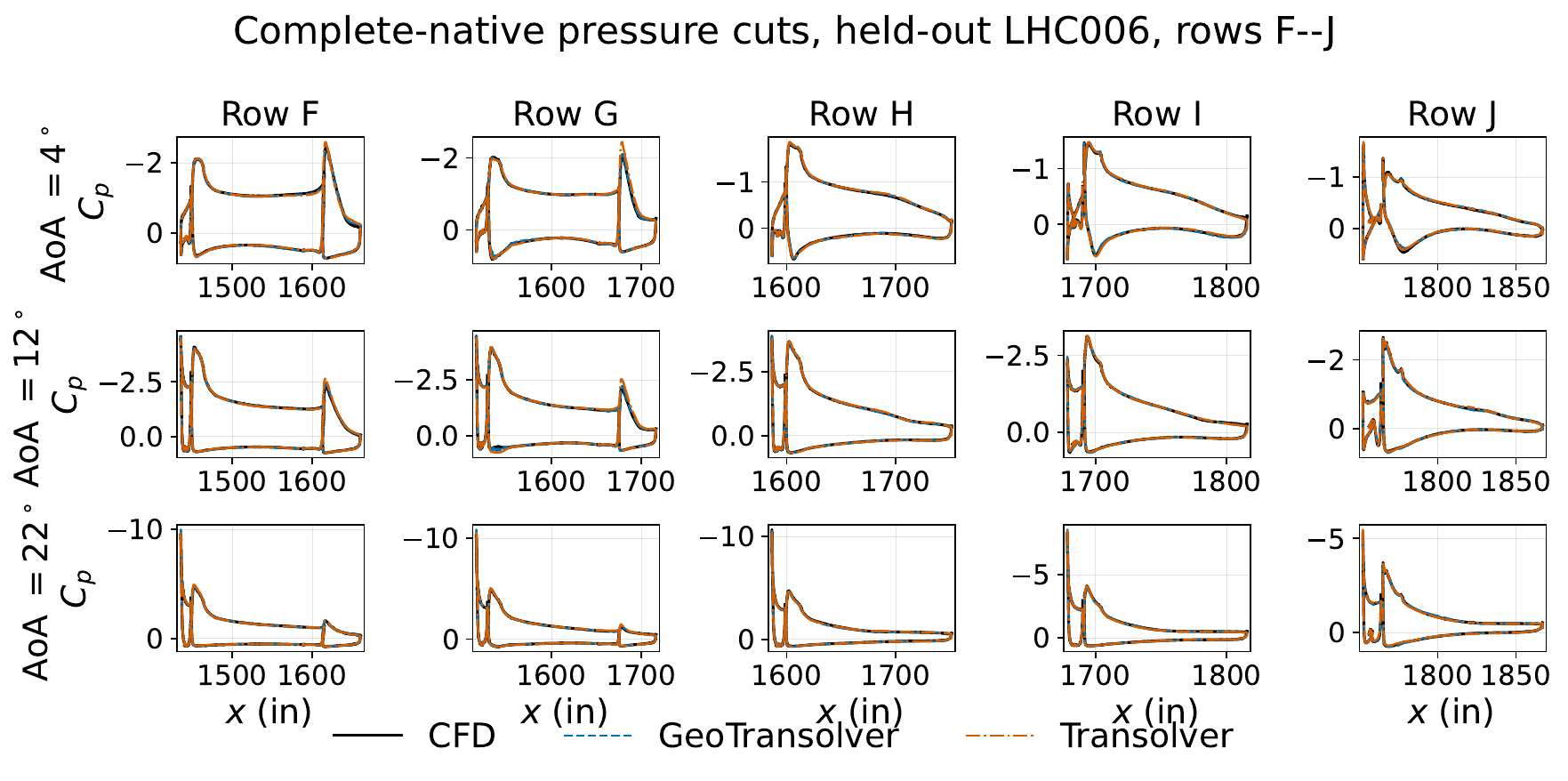}
\caption{Complete F--J pressure-cut atlas for the held-out geometry LHC006 at \(4^\circ\), \(12^\circ\), and \(22^\circ\), plotted under the same conventions as Figure~\ref{fig:app_native_cp_atlas_ae}.}
\label{fig:app_native_cp_atlas_fj}
\end{figure}

\begin{figure}[p]
\centering
\includegraphics[width=\textwidth]{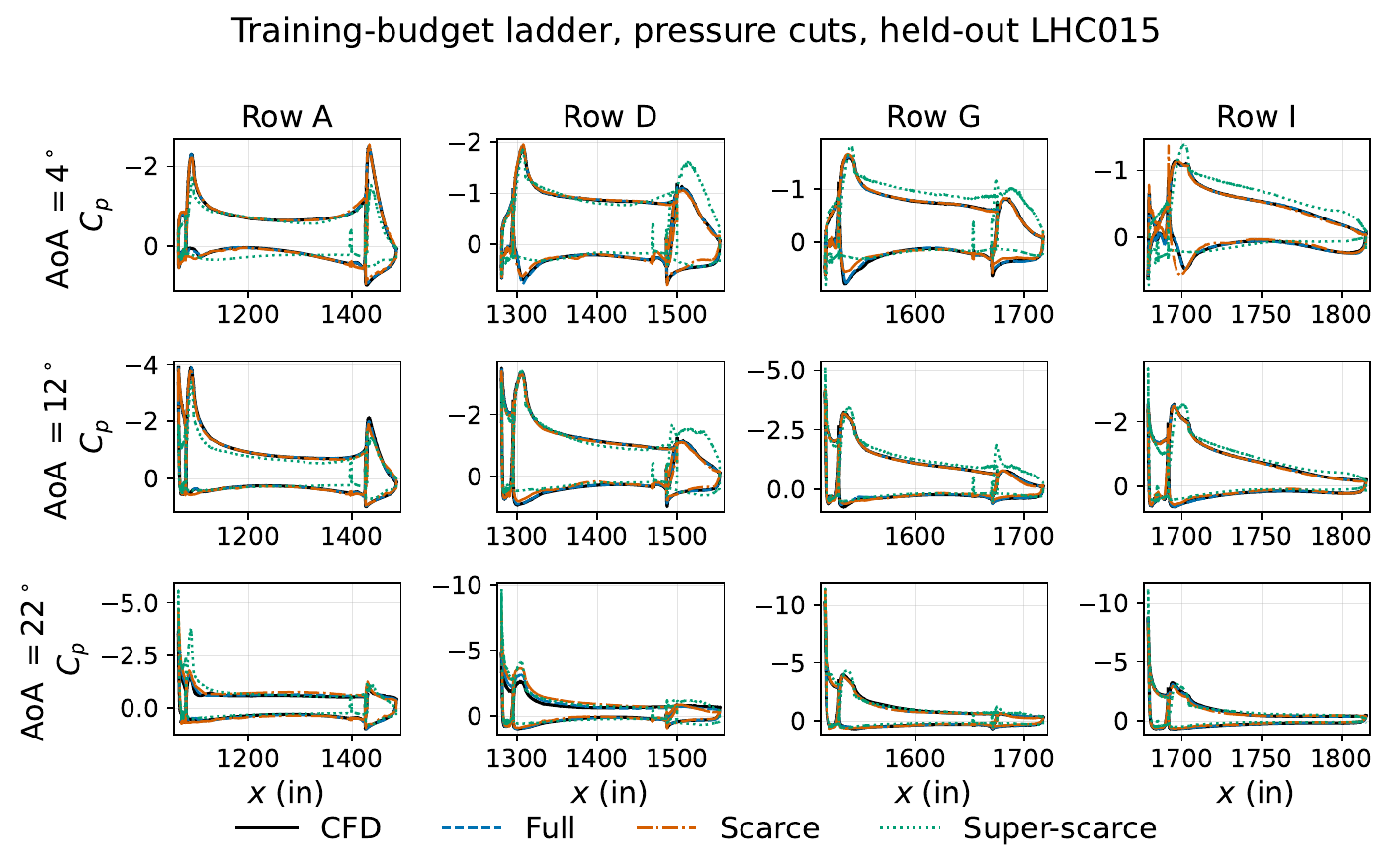}
\caption{Training-budget ladder, pressure cuts. GeoTransolver trained on \texttt{full} (1260 cases), \texttt{scarce} (210), and \texttt{super\_scarce} (35), evaluated on the held-out geometry LHC015 at \(4^\circ\), \(12^\circ\), and \(22^\circ\). The architecture, case, points, and evaluation protocol are held fixed, while training-set size, training exposure, and checkpoint horizon differ. Plotting conventions are those of Figure~\ref{fig:app_native_cp_atlas_ae}.}
\label{fig:app_native_cp_ladder}
\end{figure}

\begin{figure}[p]
\centering
\includegraphics[width=\textwidth]{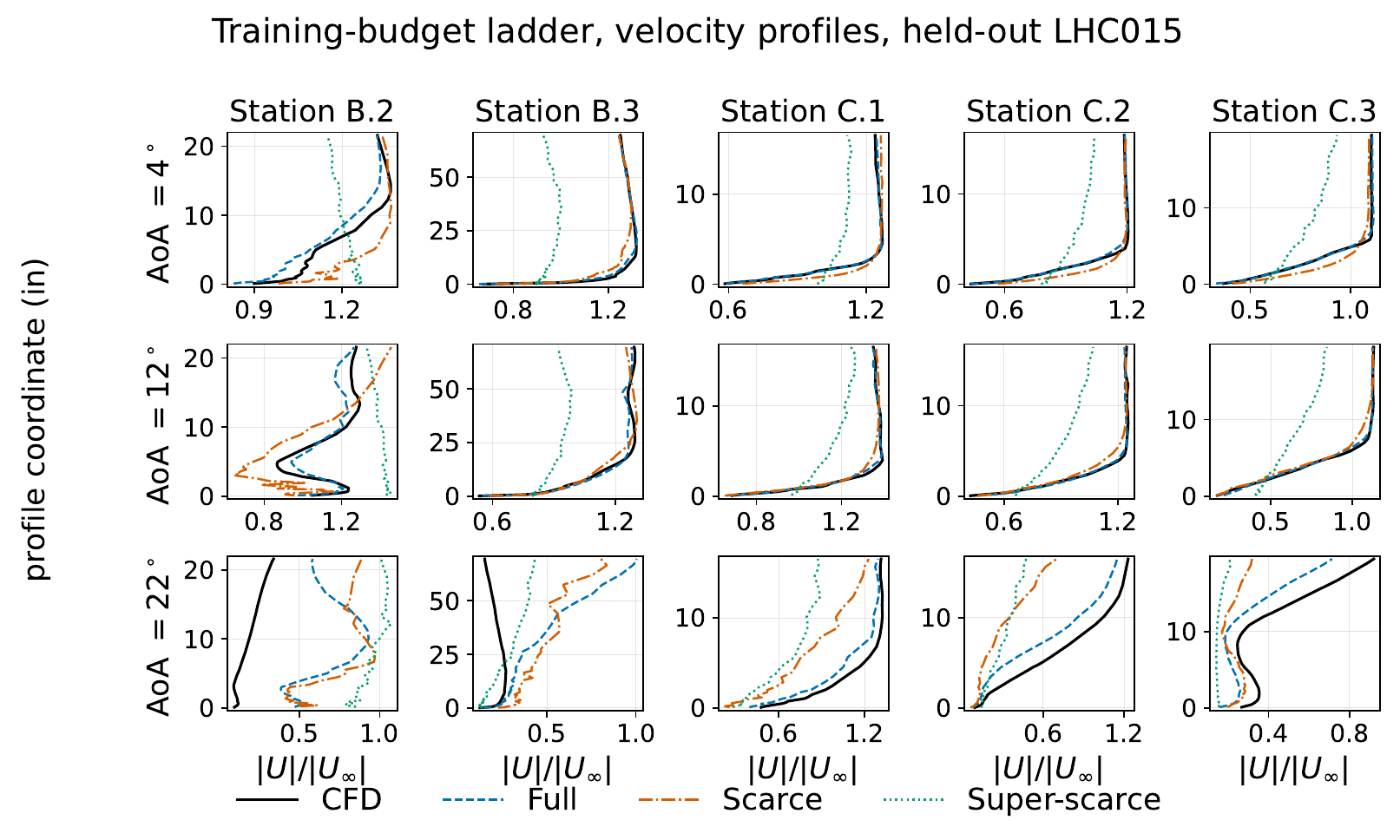}
\caption{Training-budget ladder, velocity profiles, for the same three training budgets and the same held-out geometry as Figure~\ref{fig:app_native_cp_ladder}. Validity gaps are retained as explicit breaks. At \(22^\circ\), none of these three tested runs reproduces the separated profile for this case; this comparison does not isolate the effect of data volume.}
\label{fig:app_native_velocity_ladder}
\end{figure}

\subsection{Two-Dimensional Scalar Fields}
\label{app:ml_scalar_fields}

The scored evaluators retain sufficient statistics and profile captures rather than dense prediction arrays, so field maps cannot be read out of the packaged results. They are produced instead by replaying completed checkpoints through the same evaluator with an additional plot-only capture stream that stores the per-point predicted and reference fields the chunk loop already computes. The replay changes no scored quantity, and this is verifiable rather than asserted: recomputing the case metrics from the capture reproduces the published per-case values to six decimal places for both architectures, on both supports.

The one methodological requirement for these figures is that truth and every prediction appear on identical support. That is enforced structurally. Support positions are selected once per figure and the same integer index array then gathers the reference field and every model, so no interpolation, resampling, or scalar transfer occurs anywhere in the pipeline and no model is ever drawn on a carrier derived from another model. Two display selections are used. The surface planform keeps the maximum-\(z\) native point of each \((x,y)\) bin, giving an upper-surface view in which every plotted value is a stored native value at a stored native coordinate. The volume slice keeps native points within a thin spanwise slab and draws them in \((x,z)\). Pressure error is shown as a signed difference in \(C_p\) rather than a relative error, because \(C_p\) crosses zero and a pointwise ratio is meaningless in the near-freestream field. The display window for the slices is computed from the reference field alone and then applied identically to every panel.

\begin{figure}[p]
\centering
\includegraphics[width=\textwidth]{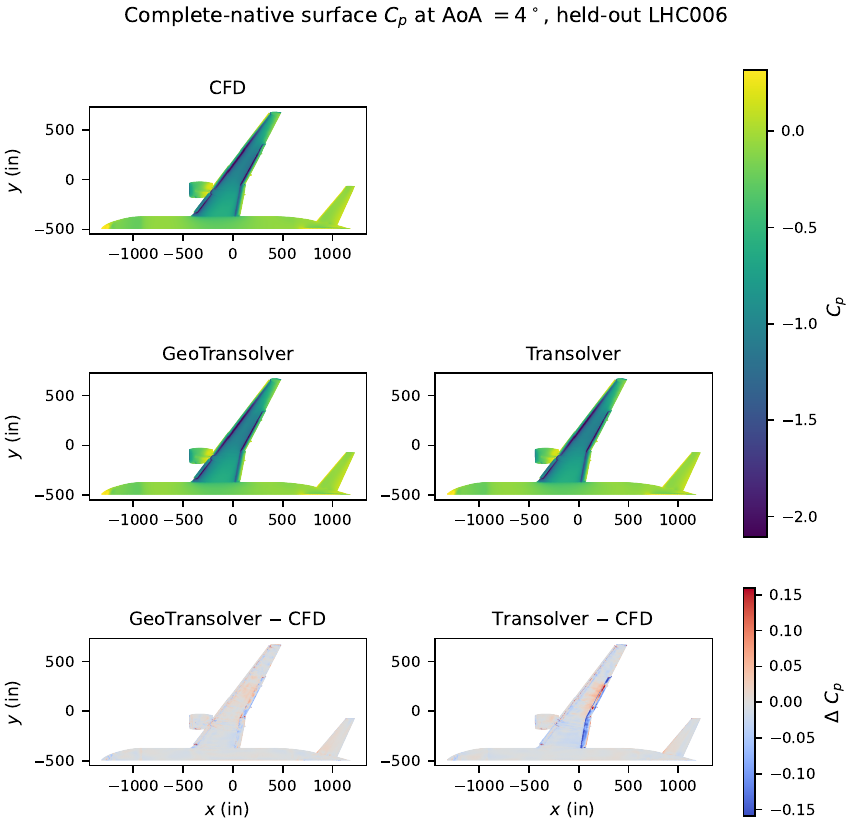}
\caption{Complete-native surface pressure at \(4^\circ\), held-out geometry LHC006, as an upper-surface planform view. Reference, both architectures, and the two signed error fields are drawn on identical native support. Error is a signed \(C_p\) difference on a symmetric scale.}
\label{fig:app_native_surface_cp_aoa4}
\end{figure}

\begin{figure}[p]
\centering
\includegraphics[width=\textwidth]{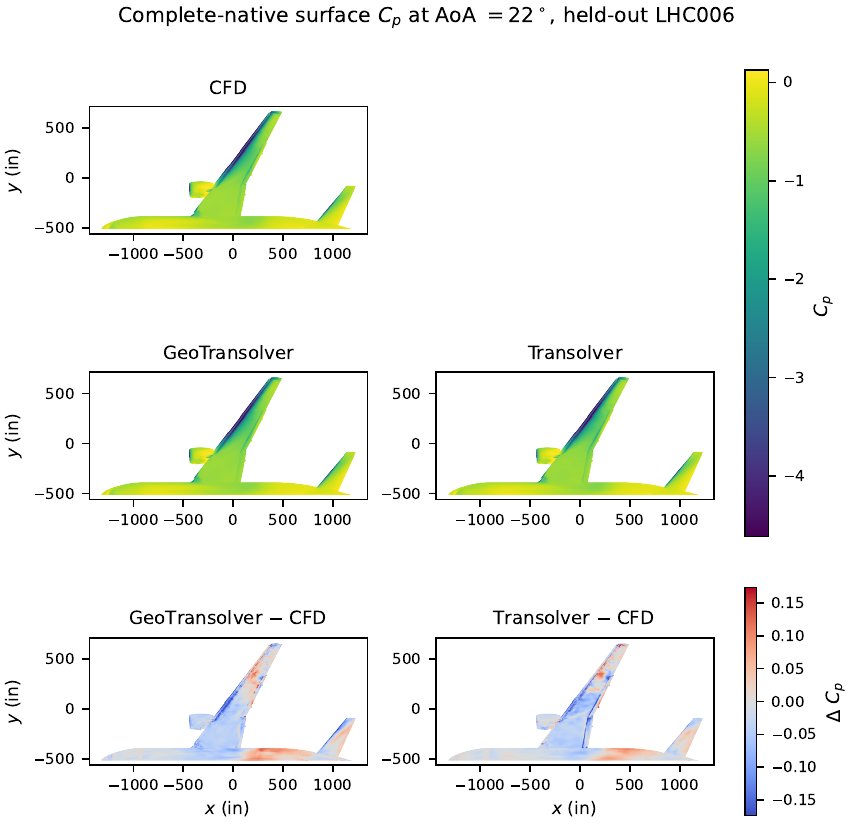}
\caption{Complete-native surface pressure at \(22^\circ\), plotted under the conventions of Figure~\ref{fig:app_native_surface_cp_aoa4}. The error is no longer confined to the suction peaks.}
\label{fig:app_native_surface_cp_aoa22}
\end{figure}

\begin{figure}[p]
\centering
\includegraphics[width=\textwidth,height=0.88\textheight,keepaspectratio,trim=0 0 0 16bp,clip]{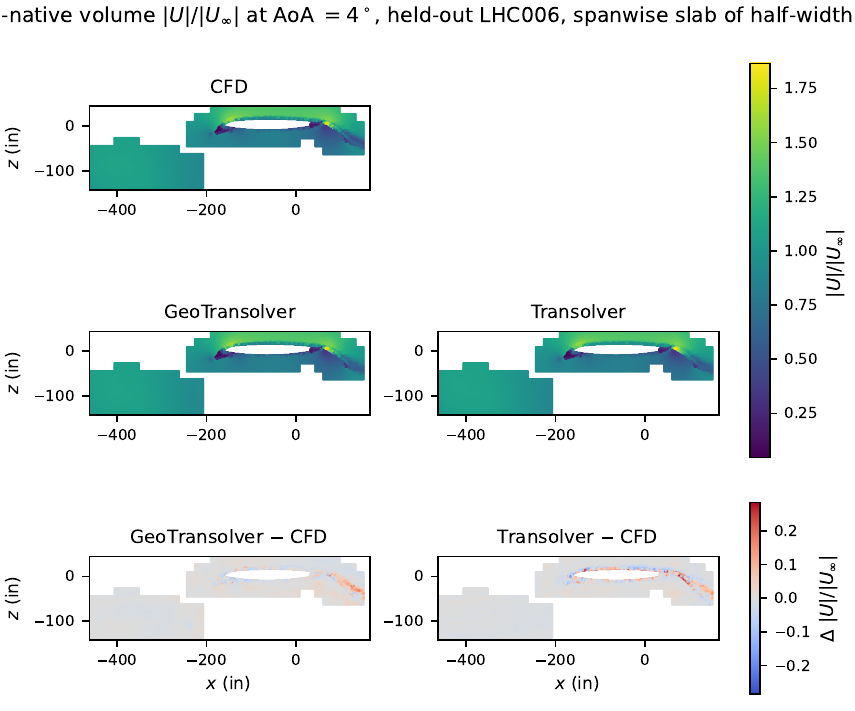}
\caption{Complete-native volume speed \(|U|/|U_\infty|\) at \(4^\circ\) on a spanwise slab, held-out geometry LHC006. Panels share a colour scale; the two error panels share a symmetric scale. The visible blocking is the native adapted grid, not a display artefact.}
\label{fig:app_native_volume_velocity_aoa4}
\end{figure}

\begin{figure}[p]
\centering
\includegraphics[width=\textwidth,height=0.88\textheight,keepaspectratio,trim=0 0 0 16bp,clip]{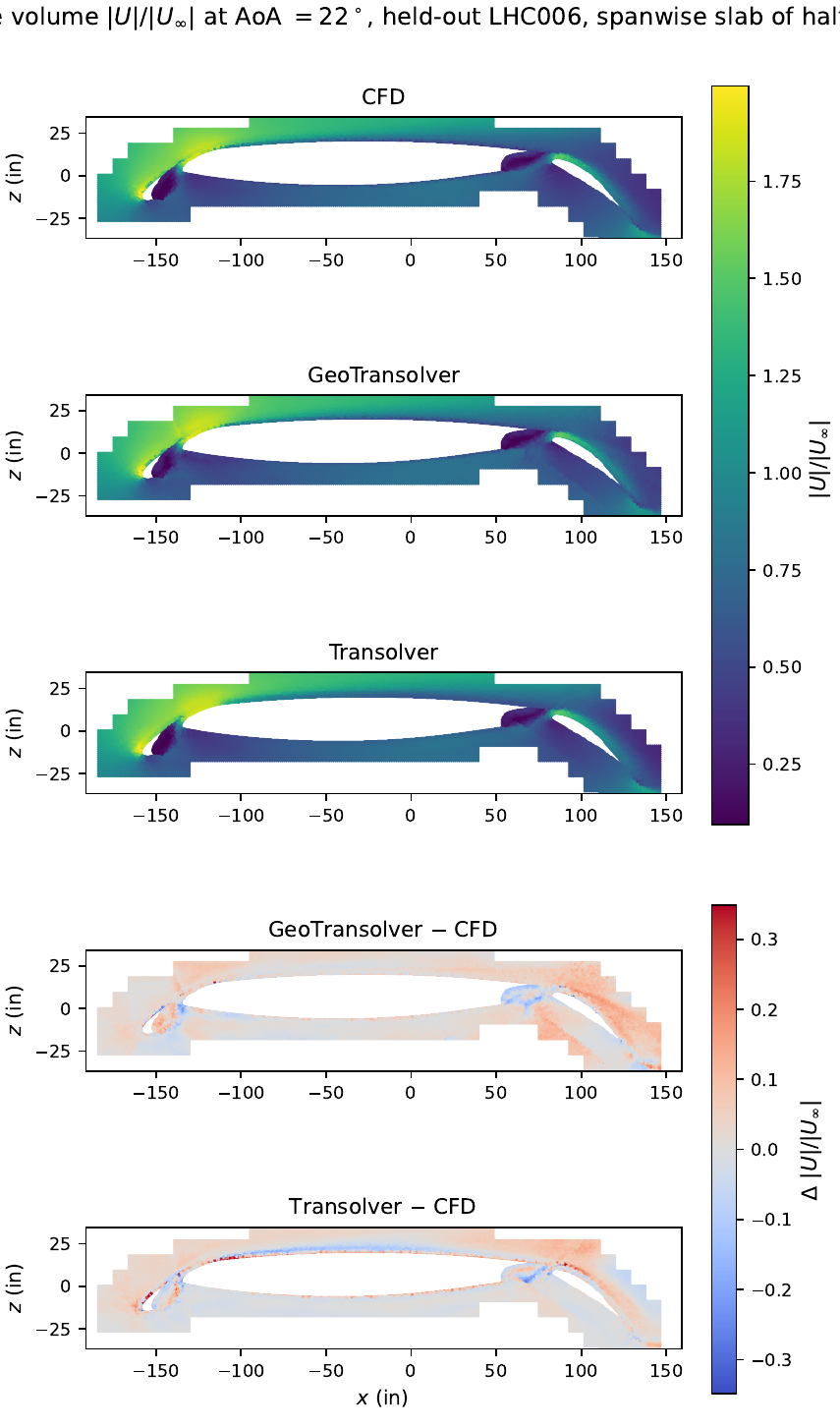}
\caption{Complete-native volume speed at \(22^\circ\), plotted under the conventions of Figure~\ref{fig:app_native_volume_velocity_aoa4}. Both architectures over-predict the speed through the separated region aft of the main element, which is the field-level form of the negative equal-case velocity-profile \(R^2\) reported for the stalled splits.}
\label{fig:app_native_volume_velocity_aoa22}
\end{figure}

Figures~\ref{fig:app_native_surface_cp_aoa4} and~\ref{fig:app_native_surface_cp_aoa22} give the surface pressure field at \(4^\circ\) and \(22^\circ\); Figures~\ref{fig:app_native_volume_velocity_aoa4} and~\ref{fig:app_native_volume_velocity_aoa22} give the volume speed field on a spanwise slice at the same two angles. The corresponding wall-shear and volume-pressure figures are produced by the same script and released with the generated assets.

These maps show where the scalar errors of Table~\ref{tab:ml_native_benchmark} live. At \(4^\circ\) the predicted and reference fields are visually difficult to separate and the error panels are dominated by thin signed bands along the slat and flap suction peaks, which is the spatial statement of a pressure error near 5\%. At \(22^\circ\) the character changes: the error is no longer confined to peaks but fills the region aft of the main element, and both architectures over-predict the speed through the separated region, producing an attached-looking field where the reference has a low-momentum wake. That is the same failure the velocity-profile diagnostics report as a strongly negative equal-case \(R^2\), shown here as a field rather than as a line.

\subsection{Case-Level and Regional Diagnostics}
\label{app:ml_case_error_analysis}

\paragraph{Case tails.}
The equal-case macro reduction makes the per-case distribution meaningful, and Table~\ref{tab:ml_native_case_tails} shows that the distribution is strongly skewed on the harder splits. On the \(22^\circ\) split, for example, the GeoTransolver case median is well below its mean while the worst case is roughly three times the mean, so the split average understates how badly the hardest geometries are predicted. The released per-case table gives complete-support relative \(L_2\), \(C_p\)-cut \(R^2\), and velocity-profile \(R^2\) for every one of the \(8{,}142\) case--model--split evaluations, so tail behaviour can be re-derived without rerunning inference.

\begin{table}[p]
\centering
\caption{Case-level distribution of complete-support surface-pressure relative \(L_2\) (\%) within each split. The mean column is the equal-case macro value of Equation~\eqref{eq:macro}; median, 90th percentile, and maximum describe its dispersion, and the final column names the worst case. A median well below the mean indicates that a small number of cases dominate the split average.}
\label{tab:ml_native_case_tails}
\scriptsize
\renewcommand{\arraystretch}{0.94}
\setlength{\tabcolsep}{4.0pt}
\begin{tabular}{llrrrrrl}
\toprule
Split & Model & Cases & Mean & Median & p90 & Max & Worst case \\
\midrule
AoA $=4^\circ$ & GeoTransolver & 36 & 4.91 & 4.73 & 6.05 & 7.96 & \texttt{\scriptsize LHC117\_AoA\_4} \\
AoA $=4^\circ$ & Transolver & 36 & 6.21 & 5.68 & 8.57 & 10.75 & \texttt{\scriptsize LHC018\_AoA\_4} \\
\midrule
AoA $=12^\circ$ & GeoTransolver & 36 & 3.00 & 2.78 & 3.51 & 6.32 & \texttt{\scriptsize LHC077\_AoA\_12} \\
AoA $=12^\circ$ & Transolver & 36 & 4.82 & 4.58 & 5.97 & 8.22 & \texttt{\scriptsize LHC117\_AoA\_12} \\
\midrule
AoA $=22^\circ$ & GeoTransolver & 36 & 11.65 & 9.44 & 19.73 & 33.47 & \texttt{\scriptsize LHC113\_AoA\_22} \\
AoA $=22^\circ$ & Transolver & 36 & 11.96 & 10.71 & 18.08 & 33.58 & \texttt{\scriptsize LHC031\_AoA\_22} \\
\midrule
Full & GeoTransolver & 360 & 4.52 & 3.34 & 6.46 & 44.98 & \texttt{\scriptsize LHC170\_AoA\_14} \\
Full & Transolver & 360 & 5.16 & 4.13 & 6.32 & 43.31 & \texttt{\scriptsize LHC170\_AoA\_14} \\
\midrule
Scarce & GeoTransolver & 360 & 13.06 & 11.78 & 18.02 & 43.70 & \texttt{\scriptsize LHC170\_AoA\_14} \\
\midrule
Super-scarce & GeoTransolver & 360 & 42.35 & 40.97 & 53.53 & 74.06 & \texttt{\scriptsize LHC091\_AoA\_4} \\
Super-scarce & Transolver & 360 & 38.63 & 36.63 & 49.37 & 72.56 & \texttt{\scriptsize LHC064\_AoA\_4} \\
\midrule
Geometry & GeoTransolver & 360 & 4.82 & 3.27 & 9.27 & 42.52 & \texttt{\scriptsize LHC115\_AoA\_14} \\
Geometry & Transolver & 360 & 5.87 & 4.70 & 8.94 & 44.22 & \texttt{\scriptsize LHC115\_AoA\_14} \\
\midrule
Geometry-scarce & GeoTransolver & 360 & 13.54 & 9.96 & 26.56 & 57.99 & \texttt{\scriptsize LHC130\_AoA\_18} \\
Geometry-scarce & Transolver & 360 & 13.69 & 10.07 & 27.10 & 63.07 & \texttt{\scriptsize LHC018\_AoA\_18} \\
\midrule
Geometry-super-scarce & GeoTransolver & 360 & 48.98 & 46.41 & 65.67 & 74.62 & \texttt{\scriptsize LHC018\_AoA\_18} \\
Geometry-super-scarce & Transolver & 360 & 30.43 & 26.79 & 49.97 & 69.58 & \texttt{\scriptsize LHC009\_AoA\_20} \\
\midrule
OOD AoA & GeoTransolver & 900 & 48.81 & 42.18 & 99.33 & 142.98 & \texttt{\scriptsize LHC124\_AoA\_22} \\
OOD AoA & Transolver & 900 & 66.41 & 80.99 & 98.25 & 114.58 & \texttt{\scriptsize LHC125\_AoA\_22} \\
\midrule
OOD deflection & GeoTransolver & 360 & 4.60 & 3.61 & 6.30 & 43.87 & \texttt{\scriptsize LHC178\_AoA\_18} \\
OOD deflection & Transolver & 360 & 5.79 & 4.39 & 8.12 & 45.88 & \texttt{\scriptsize LHC178\_AoA\_18} \\
\midrule
OOD stall & GeoTransolver & 723 & 40.32 & 44.28 & 74.98 & 104.05 & \texttt{\scriptsize LHC166\_AoA\_22} \\
OOD stall & Transolver & 723 & 38.96 & 45.41 & 71.49 & 96.25 & \texttt{\scriptsize LHC166\_AoA\_22} \\
\bottomrule

\end{tabular}
\end{table}

\paragraph{Regional reports.}
Regional diagnostics exist for all 23 records. They partition the surface into four geometric regions and the volume into four geometric/flow envelopes, using the same complete support and primary field weights as the headline metrics, and report local relative \(L_2\) and \(R^2\). For the volume, each region's RMSE is preferably normalized by that case's whole-volume truth RMS before equal-case averaging; a locally normalized relative error is unstable in the near-zero-pressure farfield, where the local denominator collapses. These partitions are geometric proxies and should be read as interpretive diagnostics: they are not exact CAD part assignments and not exact flow-topology labels.

\ifdefined\HiLiftAnonymous\else
\FloatBarrier
\fi
\subsection{Reproducibility and Provenance}
\label{app:ml_provenance}

Every number in the main ML section and in this appendix is generated, not transcribed. The generator reads a frozen result index, re-derives each metric from stored sufficient statistics, verifies it against the packaged evaluator output, and only then writes the LaTeX table, the prose macros, and the audit files. Its checks include: the expected 23 records over 12 splits; a hash check of every packaged per-case metric file; exact coverage (\(1.0\) coverage and weight-coverage fraction, and scored count equal to support count) for every case and support; agreement of the checkpoint epochs with the recorded model identity; and reconstruction of the equal-case macro reduction together with the support-weighted pooled form as an internal cross-check. Any mismatch aborts asset generation.

Checkpoints are epoch 1000 for every volume model and for all but three surface models: Scarce GeoTransolver (869), Geometry Transolver (987), and OOD-deflection Transolver (943). Each record additionally carries an independent repeat evaluation; the two passes agree exactly, with a maximum absolute difference of zero across all extracted aggregate values, and their case orderings hash identically.

\ifdefined\HiLiftAnonymous\else
\paragraph{Companion artifacts and scoring version.}
The companion release~\citep{HiLiftAeroMLArtifacts2026} provides the retained outputs and scoring source for a preliminary version of the FluidsBench evaluation metrics, frozen for this study and potentially different from the final FluidsBench release. The scoring source is pinned to revision \path{e19809a194ee1cbc4b16d4212574f8f4d921cc40}. The released artifacts bind all 46 evaluated model exports (24 GeoTransolver and 22 Transolver) to their recorded SHA-256 digests and include 46 resolved configurations, 23 surface normalization files, and split selectors. Volume preprocessing uses the documented nondimensionalization without a saved normalizer.

\paragraph{Verified reproduction scope.}
CPU replay was tested in an isolated Python 3.12 environment without third-party packages across all 23 records and \(8{,}142\) case evaluations. It reduces retained per-case field statistics, predicted/truth coefficient pairs, and profile metrics into the checked aggregate values. Reconstructed metrics present in the frozen result index agree exactly, and the compared paper CSV values agree to their stored precision. This test verifies reduction of retained outputs; it does not rerun training, model inference, native force integration, or profile scoring against evaluator-owned truth. The release does not yet contain all final native inference wrappers, the executable evaluator core, the compact-profile scoring-truth/support assets, or a complete GPU environment lock. The repeat-inference evidence above belongs to the original campaign and is distinct from this package-validation test.
\fi

\subsection{Training Configuration}

\ifdefined\HiLiftAnonymous\else
\paragraph{PhysicsNeMo execution environment.}
The retained campaign records identify PhysicsNeMo framework revision \path{651c62656dd4f0ce5de9cd5afd0dbc389ae105bc}; the model exports record PhysicsNeMo 2.2.0a0. Retained runtime evidence names the \path{pytorch-26.04.sqsh} container and reports PyTorch \path{2.12.0a0+0291f960b6.nv26.04.48445190}. The framework was supplied through \texttt{PYTHONPATH}, with auxiliary dependencies from a host virtual environment. Container filenames and log banners do not establish a complete environment lock or one uniform CUDA runtime for every run. The companion release records the per-model settings, constructor arguments, retained recipe and curator overlay, and distinguishes this execution evidence from the separately cited public recipe revision.
\fi

\begin{table}[t]
\centering
\footnotesize
\setlength{\tabcolsep}{3pt}
\renewcommand{\arraystretch}{1.15}
\caption{Shared training settings and evaluation contract for the surface and volume rows.}
\label{tab:app_training_hparams_200k}
\begin{tabularx}{\textwidth}{@{}>{\raggedright\arraybackslash}p{0.18\textwidth} *{2}{>{\raggedright\arraybackslash}X}@{}}
\toprule
Quantity & Surface rows & Volume rows \\
\midrule
Training sampling & 200k surface cells per case per pass & 200k volume points per case per pass \\
Evaluation support & every retained native surface point, scored once & every valid native volume point, scored once \\
Evaluation weights & published nodal dual area & equal weight per valid point \\
Model outputs & pressure, temperature, density, velocity, wall shear & pressure, temperature, density, velocity \\
Headline quantities & \(C_p\), wall shear, \(C_L/C_D/C_M\) & \(C_p\), velocity \\
Training loss & weighted Huber & unit-weighted Huber \\
Loss weights & pressure 2.0; temperature 0.25; density 0.5; velocity 0.5; wall shear 1.0 & all 1.0 \\
Optimizer & Muon for eligible 2-D parameters; AdamW for remaining parameters & Muon for eligible 2-D parameters; AdamW for remaining parameters \\
Learning rate & $10^{-3}$ & $10^{-3}$ \\
Scheduler & StepLR(150, 0.5) & StepLR(150, 0.5) \\
Geometry features & centered centroids and normals & centered points, STL SDF/normals \\
Normalization & fields nondimensionalized, then z-score normalized for training; z-score reversed for evaluation & fields nondimensionalized and evaluated in nondimensional units; no z-score normalization \\
MAE reporting & gauge pressure and componentwise wall shear in Pa; coefficients in coefficient points & gauge pressure in Pa; componentwise velocity in \(\mathrm{m\,s^{-1}}\) \\
Load references & \(S_{\text{ref}}=297360.0\,\mathrm{in}^2\), \(C_{\text{ref}}=275.80\,\mathrm{in}\), \(\mathbf{r}_\mathrm{MRP}=(1325.9,0,177.95)\,\mathrm{in}\) & not reported \\
\bottomrule
\end{tabularx}
\end{table}

\begin{table}[t]
\centering
\footnotesize
\setlength{\tabcolsep}{3pt}
\renewcommand{\arraystretch}{1.15}
\caption{Model-specific configurations. Parameter counts report trainable parameters on the executed forward path.}
\label{tab:app_model_hparams_200k}
\begin{tabularx}{\textwidth}{@{}>{\raggedright\arraybackslash}p{0.16\textwidth} *{4}{>{\raggedright\arraybackslash}X}@{}}
\toprule
Quantity & GeoTransolver surface & Transolver surface & GeoTransolver volume & Transolver volume \\
\midrule
Local geometric input & cell centroids and normals & cell centroids and normals & volume points, SDF, and SDF normals & volume points, SDF, and SDF normals \\
Conditioning & centroid geometry and freestream context; normals in local input & broadcast freestream input & point-coordinate geometry and freestream context; SDF/SDF normals in local input & broadcast freestream input \\
Active parameters & 17.06M & 18.54M & 17.06M & 18.54M \\
Layers & 12 GALE blocks & 16 Transolver blocks & 12 GALE blocks & 16 Transolver blocks \\
Hidden size & 256 base / 352 effective & 320 & 256 base / 352 effective & 320 \\
Feed-forward size / ratio & 1408 / 4 & 1280 / 4 & 1408 / 4 & 1280 / 4 \\
Heads / slices & 8 heads, 256 slices & 8 heads, 256 slices & 8 heads, 256 slices & 8 heads, 256 slices \\
Attention / operator & geometry-aware latent attention & physics attention & geometry-aware latent attention & physics attention \\
Local features & three-scale ball query & none & SDF/SDF normals in 7-D base input, plus three-scale ball query & SDF and SDF normals in 7-D point embedding \\
\bottomrule
\end{tabularx}
\end{table}

The parameter counts in Table~\ref{tab:app_model_hparams_200k} describe the trainable parameters used by the corresponding forward path. GeoTransolver at 17.06M and Transolver at 18.54M differ by 1.48M parameters (8.7\% relative to GeoTransolver), which is the sense in which the two headline architectures are capacity matched. Neither was separately tuned for this dataset, so the comparison should be read as two similar-capacity operators evaluated under an identical protocol rather than as each architecture at its own optimum.

Measured dataset-generation, training, and inference costs are reported in Appendix~\ref{app:hpc}.

\subsection{Additional interpretation of the results}

Section~\ref{sec:ml_evaluation} discusses the aggregate interpolation, geometry, data-budget, and OOD comparisons. Two distinctions help interpret the detailed diagnostics here. First, the single-AoA $22^\circ$ models retain positive lift $R^2$ despite strongly negative velocity-profile scores: an integrated load can conceal local errors in the separated wake. Second, geometric-parameter and flow-regime shifts have different outcomes, so a single combined ``OOD'' score would obscure the purpose of the separate splits.

\paragraph{Sectional effects of training budget.}
Figures~\ref{fig:app_native_cp_ladder} and~\ref{fig:app_native_velocity_ladder} show the effects of reducing the training set on sectional pressures and velocity profiles. They compare GeoTransolver trained on \texttt{full}, \texttt{scarce}, and \texttt{super\_scarce}, holding the architecture, case, and evaluation protocol fixed. Those three splits share one test population, so the curves in each panel are evaluated on identical points as the training set falls from 1260 cases to 210 and then to 35. Training exposure and checkpoint horizon also differ, so this is not an isolated data-volume ablation. The held-out geometry is LHC015, the only geometry present in that shared test set at all three of \(4^\circ\), \(12^\circ\), and \(22^\circ\).

The degradation varies across the flow. The \texttt{full} model tracks the reference almost everywhere at \(4^\circ\) and \(12^\circ\). The \texttt{scarce} model holds the general shape but loses amplitude at the slat and flap suction peaks, which is where the sectional error concentrates before it becomes a whole-field error. The \texttt{super\_scarce} model departs over whole elements rather than at peaks, and on the velocity stations it misses the near-wall structure entirely. At \(22^\circ\), none of the three tested budgets reproduces the separated profile for this case. This result does not isolate data volume as the cause of the deficit or rule out improvements from more data: training exposure and checkpoint horizon differ across these single-seed runs, and alternative model configurations remain untested.

\paragraph{Limitations.}
Three limitations are worth restating. These are single-run results, so small differences between the two architectures are not uncertainty resolved. The evaluation is exhaustive over the native mesh but is not a single whole-mesh attention pass, so it does not answer how these models would behave with global context. And the test artifacts were inspected while the protocol was developed, so these are reproducible exploratory baselines rather than an untouched leaderboard.

\subsection{Outstanding Investigations and Future Work}

The most useful next steps suggested by these results are seed repetition on the data-efficiency ladders, so that the observed architecture inversions can be tested; completion of the \texttt{scarce} Transolver surface record and the two \texttt{medium} split families, so that both ladders are complete; and targeted work on the separated regime. Figure~\ref{fig:app_native_velocity_ladder} shows that the tested training budgets do not resolve the separated-profile discrepancy for the illustrated case. Additional data, training compute, and model configurations should be varied in controlled comparisons to determine which changes improve this regime.

\FloatBarrier
\section{Computational Cost}
\label{app:hpc}

All figures in this appendix are measured, not estimated. Dataset-generation
costs come from the scheduler accounting records of the solver jobs themselves;
training costs are integrated from the per-epoch timestamps written by each run;
inference costs are per-chunk wall-clock times recorded by the evaluators. Every
number refers to NVIDIA GB200 hardware, with four GPUs per node, and GPU-hours
are reported as allocated GPUs multiplied by elapsed wall-clock time. Training
and inference integrate only intervals of active compute, so scheduler queue
waits between the chained allocations of a long run are excluded.

\subsection{Dataset Generation}
\label{app:hpc_dataset}

Each released case is a wall-modelled large-eddy simulation run with Charles on
eight GB200 nodes, that is 32 GPUs and 288 MPI ranks. A case is produced in two
phases: an initial solution on the unadapted baseline grid of approximately
\(100\times10^6\) control volumes, followed by the adapted-grid solution that is
actually released. The adapted phase dominates, accounting for
\(742{,}738\) of the \(798{,}817\) solver GPU-hours, because it carries both the
finer grid and the long post-transient averaging window.

Table~\ref{tab:hpc_dataset} shows that generation cost is strongly
angle-dependent. The nominal adapted-grid run duration is 15 convective time
units (CTU) at \(\mathrm{AoA}\le12^\circ\) and 30--200 CTU at higher angles,
depending on how slowly the wake statistics converge. Individual cases depart
from these nominal durations. Usable averaging begins after the 10-CTU
transient reset, giving nominal averaging windows of about 5 CTU and 20--190
CTU, respectively (Appendix~\ref{app:postPro-flowStats}). The median case at
\(22^\circ\) costs about \(3.7\times\) the median case at \(4^\circ\). Across the
accounted campaign, the most expensive single case consumed \(3{,}161\)
GPU-hours against a minimum of \(24\).

\begin{table}[t]
\centering
\caption{Dataset-generation cost per case, from the scheduler records of all
\(1{,}851\) accounted solver cases on 8 GB200 nodes (32 GPUs). This is the whole
generation campaign, which is slightly larger than the release: it also covers
the additional validation and development geometries that were produced by the
same pipeline but are not part of the \(1{,}800\) released cases. Both the
baseline and adapted phases are included. The angle dependence reflects the
longer averaging window required once the flow separates.}
\label{tab:hpc_dataset}
\small
\begin{tabular}{lrrr}
\toprule
Angle of attack & Median & Min & Max \\
 & (GPU-h) & (GPU-h) & (GPU-h) \\
\midrule
\(4^\circ\)  & 150 & 29 & 811 \\
\(6^\circ\)  & 163 & 33 & 1{,}537 \\
\(8^\circ\)  & 163 & 25 & 1{,}918 \\
\(10^\circ\) & 171 & 24 & 962 \\
\(12^\circ\) & 182 & 114 & 2{,}675 \\
\midrule
\(14^\circ\) & 330 & 201 & 2{,}031 \\
\(16^\circ\) & 425 & 244 & 2{,}490 \\
\(18^\circ\) & 472 & 258 & 2{,}269 \\
\(20^\circ\) & 414 & 285 & 2{,}232 \\
\(22^\circ\) & 559 & 198 & 3{,}161 \\
\midrule
All cases & 351 & 24 & 3{,}161 \\
\bottomrule
\end{tabular}
\end{table}

The complete solver cost of the campaign is \(798{,}817\) GPU-hours, a mean of
\(432\) GPU-hours per case. Grid generation and adaptation added a further
\(14{,}296\) GPU-hours and the surface and volume export and post-processing
another \(23{,}040\), giving \(836{,}153\) GPU-hours in total. The \(1{,}800\)
released cases are \(97\%\) of that count, so the released dataset represents
close to, but slightly less than, the whole figure. This is the dominant cost of the work reported here. The 44 completed
1000-epoch runs of Section~\ref{app:hpc_training} total \(52{,}170\) GPU-hours,
and every GPU job of the machine-learning campaign together --- training,
inference, and all exploratory and superseded configurations --- accounts for
\(75{,}223\). Generating the data therefore cost about \(16\times\) the reported
training and about \(11\times\) the entire machine-learning effort, so the
dataset, not the models, is where the computational investment of this work
lies.

\subsection{Model Training}
\label{app:hpc_training}

Table~\ref{tab:hpc_training} groups the completed 1000-epoch runs of the two
headline architectures by split family, since the training population rather
than the architecture sets the cost. The GPU count per run follows the split
size: the fixed-angle splits train on 126 cases using 4 GPUs, the reduced-data
splits use 8, and the large splits use 32.

Two comparisons are worth stating explicitly. First, the two architectures are
close to each other at matched split and modality, consistent with their similar
parameter counts; Transolver is slightly cheaper on the volume and slightly more
expensive on the surface. Second, the surface runs are markedly more expensive
than the volume runs on the large splits, despite an identical 200k sampling
ceiling, because the surface model carries the additional wall-shear output and
its integrated-load bookkeeping.

\begin{table}[t]
\centering
\caption{Training cost of the completed 1000-epoch runs, integrated from
per-epoch timestamps and excluding scheduler queue waits between chained
allocations. Ranges span the split families listed; the reported minimum and
maximum are over the individual runs in each group.}
\label{tab:hpc_training}
\small
\resizebox{\textwidth}{!}{%
\begin{tabular}{llrrrr}
\toprule
Split family & GPUs & \multicolumn{2}{c}{GeoTransolver (GPU-h)} & \multicolumn{2}{c}{Transolver (GPU-h)} \\
\cmidrule(lr){3-4}\cmidrule(lr){5-6}
 & & Surface & Volume & Surface & Volume \\
\midrule
Fixed angle (126 train cases)      & 4  & 74--152      & 58--111      & 70--95       & 89--104 \\
Reduced data (35--210 train cases) & 8  & 267--296     & 125--163     & 272--320     & 103--151 \\
Large splits (788--1260 cases)     & 32 & 1{,}490--4{,}570 & 856--1{,}638 & 3{,}343--4{,}684 & 807--1{,}048 \\
\midrule
Median over all completed runs     & -- & 296          & 163          & 320          & 151 \\
\bottomrule
\end{tabular}%
}
\end{table}

Because the surface and volume networks are trained separately, a complete
surrogate for one split is the sum of two runs. On the full split that is
\(4{,}308+1{,}638=5{,}946\) GPU-hours for GeoTransolver and
\(4{,}473+1{,}048=5{,}521\) for Transolver: the cost of roughly fourteen
simulated cases, or \(0.7\%\) of the campaign that produced the training data.

\subsection{Inference}
\label{app:hpc_inference}

The distinction between a single sampled context and a complete-native sweep is
the largest single cost factor in the evaluation protocol, and
Table~\ref{tab:hpc_inference} quantifies it. One 200k context is a single
forward pass and takes well under a second on one GB200 GPU. Scoring every
retained native entity of the same case requires between 687 and 1{,}205 such
passes, because the native supports hold 137--145 million surface points and
221--239 million valid volume points, so the exhaustive sweep costs three orders
of magnitude more than the sampled pass it replaces.

The per-context cost is essentially constant across chunks, angles, and
modalities: the measured mean for GeoTransolver was 0.454--0.455\,s per chunk in
every one of the six streams reported here, which is why the complete-native
totals scale almost exactly with native support size. Transolver, which carries
no local-feature branch, is roughly 1.4--1.9 times faster per context.

\begin{table}[t]
\centering
\caption{Inference cost per held-out case on one GB200 GPU, measured for the
held-out geometry LHC006. One context is a single 200k forward pass, the unit
used during training. The complete-native sweep visits every retained native
entity exactly once in bounded contexts of at most 200k entities, and its cost is
reported both as the number of contexts required and as elapsed wall-clock time.}
\label{tab:hpc_inference}
\small
\resizebox{\textwidth}{!}{%
\begin{tabular}{llrrrr}
\toprule
Model & Support & Entities & One context & Contexts & Complete native \\
 & & (millions) & (s) & required & (min) \\
\midrule
GeoTransolver & Surface & 137--145 & 0.45 & 687--734    & 5.2--5.6 \\
GeoTransolver & Volume  & 221--239 & 0.45 & 1{,}107--1{,}205 & 8.4--9.1 \\
Transolver    & Surface & 137--145 & 0.32 & 687--734    & 3.7--3.9 \\
Transolver    & Volume  & 221--239 & 0.24 & 1{,}107--1{,}205 & 4.5--4.8 \\
\bottomrule
\end{tabular}%
}
\end{table}

Two consequences follow. A complete surface and volume evaluation of one case
with both architectures costs roughly 22--24 GPU-minutes, so the 23 reported
records containing \(8{,}142\) case--model--split evaluations represent a substantial evaluation budget
in their own right, though still a small fraction of either training or dataset
generation. And because the sweep is an exhaustive traversal in bounded contexts
rather than one whole-mesh attention pass, its cost grows linearly in native
support size, so the protocol remains tractable on meshes far larger than those
released here.

\FloatBarrier
\section{Dataset description}
\label{app:dataset}

\subsection{Access to dataset}
\label{app:dataset-access}

The dataset is publicly hosted on Hugging Face at
\url{https://huggingface.co/datasets/nvidia/HiLiftAeroML}, with no access charge.
The repository occupies approximately 66.9~TB as stored on the Hub and is substantially larger after archive extraction. Selective downloads allow users to inspect metadata and individual cases before obtaining a larger subset. The examples below use an immutable repository revision to make the downloaded files reproducible; this access example does not identify the revision used in every training run.

\paragraph{Example 1: Download the small metadata files.}
Install the Hugging Face command-line client, then download the dataset card, geometry and load summaries, and frozen split manifest. Authentication is optional for this public dataset; \texttt{hf auth login} can be used to obtain authenticated rate limits for larger downloads.

\lstset{basicstyle=\ttfamily\scriptsize,breaklines=true,breakatwhitespace=false,postbreak=\mbox{\textcolor{gray}{$\hookrightarrow$}\space},columns=fullflexible}%
\begin{lstlisting}
python -m pip install -U huggingface_hub hf_xet
HILIFT_REV=bbec30bcfc6103309c1375c5228b3ad0a586bfaf

hf download nvidia/HiLiftAeroML \
  README.md geo_values_all.csv force_mom_all.csv \
  splits/manifest.json splits/README.md \
  --repo-type dataset --revision "$HILIFT_REV" \
  --local-dir ./hiliftaeroml_data
\end{lstlisting}

\paragraph{Example 2: Preview a single-case download.}
The following command lists the files and download size for one case without downloading its fields. Remove \texttt{--dry-run} to download that case, including its surface and volume archives, geometry, quadrature sidecar, reference quantities, and diagnostic files. Change \texttt{HILIFT\_CASE} to select another geometry--AoA pair. The root \texttt{data\_quality/} directory provides the volume-row exclusion metadata described in the dataset card and should also be obtained when processing volume fields.

\begin{lstlisting}
HILIFT_CASE=geo_LHC006_AoA_4
hf download nvidia/HiLiftAeroML \
  --repo-type dataset --revision "$HILIFT_REV" \
  --include "${HILIFT_CASE}/*" \
  --local-dir ./hiliftaeroml_data --dry-run
\end{lstlisting}

\subsection{Hosting and maintenance}
The dataset is hosted in the NVIDIA organization on Hugging Face and maintained by the authors. Corrections and updates are recorded in the dataset README and repository revision history. Users can report issues to the dataset maintainers at \href{mailto:contact@caemldatasets.org}{contact@caemldatasets.org}. The project page, \url{https://caemldatasets.org/hiliftaeroml/}, provides dataset information and download guidance. Users should record the immutable repository revision associated with their downloaded files when reporting results.

\subsection{Licensing terms}
\label{app:dataset-license}

The dataset is provided with the Creative Commons CC-BY v4.0 license\footnote{\url{https://creativecommons.org/licenses/by/4.0/deed.en}}.
The license grants the user the right to \textbf{share} the work, e.g.
by copying and redistributing the material in any medium or format for any purpose, which includes redistribution for commercial purposes.
Likewise, the material can be \textbf{adapted} by remixing or transforming it, or building upon the material for any purpose.
In case of redistribution, you must give appropriate credit to the original authors, which includes providing the names of the creators and attribution parties, a copyright notice, a license notice, a disclaimer notice, and a link to the material.
You must also indicate if you modified the material and retain an indication of previous modifications.
You may do so in any reasonable manner, but not in any way that suggests the licensor endorses you or your use (``Attribution'' clause).
No warranties are given. The license may not give you all of the permissions necessary for your intended use.
For example, other rights such as publicity, privacy, or moral rights may limit how you use the material.
The dataset card declares CC-BY-4.0 at \url{https://huggingface.co/datasets/nvidia/HiLiftAeroML}. The full license terms are available at
\url{https://creativecommons.org/licenses/by/4.0/legalcode.en}.

\subsection{Intended use \& potential impact}
The dataset was created with the following intended uses:

\begin{itemize}
\item Development and testing of data-driven and physics-inspired AI surrogate models for the prediction of external aerodynamics quantities (lift, drag, pressure, velocity) on high-lift aircraft configurations
\item For academia, an additional dataset to test functionality of ML architectures in another use-case i.e beyond automotive bluff bodies (e.g AhmedML \cite{ashton2024ahmedCorrected} and WindsorML \cite{ashton2024windsorCorrected} ).
For an aerospace company, it can be a useful dataset that is similar in size and complexity to an internal non-public dataset, i.e an aerospace company's own data.
\item As a `challenge' test-case at future conferences/workshops to benchmark the performance of different ML approaches on an openly licensed high-lift aircraft dataset.
This is aimed for the 6th High-Lift CFD Prediction Workshop.
\item The dataset was partially created for the 6th High-Lift Prediction Workshop \footnote{https://aiaa-hlpw.org/HLPW/} workshop, as a test case for the AI/ML technical working group, to allow for the assessment of different ML approaches on an identical open-source dataset.
\item Large-scale dataset for the study of flow physics over full aircraft, i.e potential non-ML use-case.
\end{itemize}

The potential impact could be:

\begin{itemize}
\item Establishing an industry-standard benchmark for the testing of AI methods for the aerospace external aerodynamics community.
\item Allowing for fairer testing of large-scale CFD versus ML approaches, i.e training and inference time on non-canonical problems.
\item Addressing the lack of high-fidelity, openly licensed training data for ML models of high-lift aircraft aerodynamics.
\end{itemize}

\subsection{Versioning and citation}

The accompanying paper is available at \url{https://arxiv.org/abs/2605.19565}. Dataset files are versioned through the Hugging Face repository history. Cite the paper and record the dataset revision used; the access examples above pin revision \texttt{bbec30bcfc6103309c1375c5228b3ad0a586bfaf}.

\subsection{Documented corrections and data validity}
\label{app:dataset-quality}

The pinned release documents six replacements of solver-monitor mean coefficients by integration of the released time-averaged surface pressure and wall-shear fields: \texttt{geo\_LHC012\_AoA\_16}, \texttt{geo\_LHC018\_AoA\_16}, \texttt{geo\_LHC028\_AoA\_18}, \texttt{geo\_LHC028\_AoA\_22}, \texttt{geo\_LHC129\_AoA\_4}, and \texttt{geo\_LHC172\_AoA\_22}. The substitutions affect \texttt{cd}, \texttt{cl}, \texttt{cm}, \texttt{clp}, \texttt{clv}, \texttt{cdp}, and \texttt{cdv} in the per-case files and aggregate force summary. The \href{https://huggingface.co/datasets/nvidia/HiLiftAeroML/blob/bbec30bcfc6103309c1375c5228b3ad0a586bfaf/force_mom_surface_overrides_v1.json}{force/moment correction record} gives the original and replacement values, integration conventions, and source hashes. Other cases retain their solver-monitor means.

For these six cases, the retained \texttt{*\_stdev}, \texttt{*\_stderr}, and \texttt{*\_ci95} columns and convergence plots describe the original solver monitors. They were not recomputed for the substituted means and must not be used as uncertainty estimates for those means.

The \href{https://huggingface.co/datasets/nvidia/HiLiftAeroML/blob/bbec30bcfc6103309c1375c5228b3ad0a586bfaf/data_quality/README.md}{volume data-quality record} identifies 1,232,817 invalid all-zero rows among 419,416,158,837 raw volume rows (approximately $0.000294\%$), across 1,768 of the 1,800 cases. These rows must be excluded before analysis or scoring. Their case-local raw point IDs, coordinates, and provenance are published under \texttt{data\_quality/}. The raw Float32 rule \texttt{avg(P) != 0.0}, applied before normalization, selects the retained rows; it must not be replaced by a test on derived $C_p$. All exported fields on the excluded rows are documented as zero, but the release does not assign a cause to the zero fill.

\subsection{Details of provided data}
\label{app:dataset-dataDesc}

In the dataset, each folder (e.g \verb|geo_LHC025_AoA_6|, \verb|geo_LHC025_AoA_8|, ..., \verb|geo_LHCi_AoA_j|, etc.) corresponds to a different geometry and angle of attack, where "i" is the geometry ID (ranging from 001 to 180) and j is the angle of attack that ranges from 4 to 22 in intervals of 2 degrees. All run folders feature the same structure:
{\scriptsize\begin{verbatim}
geo_LHC025_AoA_6/
|
|- boundary_geo_LHC025_AoA_6.vtu.tgz
|- boundary_dual_area_geo_LHC025_AoA_6.npy
|- force_mom_geo_LHC025_AoA_6.csv
|- geo_LHC025_AoA_6.stl
|- geo_LHC025_AoA_6.stp
|- geo_values_geo_LHC025_AoA_6.csv
|- img_wss_LHC025_AoA_6.png
|- plot_CD_geo_LHC025_AoA_6.png
|- plot_CL_geo_LHC025_AoA_6.png
|- plot_CM_geo_LHC025_AoA_6.png
|- ref_values_geo_LHC025_AoA_6.csv
|- volume_geo_LHC025_AoA_6.vtu.tgz
\end{verbatim}}

A brief description of the contents in each file, including the file format and the file size, is given in the Tab.~\ref{tab:dataset-output-files} and Tab.~\ref{tab:dataset-output-files-more}. Tab.~\ref{tab:dataset-output-quantities} provides a list of output flow variables, obtained by time-averaging the unsteady flow after discarding the initial transient.

\begin{itemize}
    \item \textbf{Volume field:} The complete, three-dimensional and time-averaged flow field is provided. The most commonly analysed quantities in aerospace aerodynamics were stored, including first and second order flow statistics (see Tab.~\ref{tab:dataset-output-quantities}).
    \item \textbf{Surface field:} The complete, time-averaged flow field on the aircraft surface is provided. All flow quantities necessary to compute the integral force coefficients along with time-averaged surface pressure fluctuations are included.
    \item \textbf{Surface quadrature:} A one-dimensional \texttt{float32} NPY sidecar supplies one mass-lumped dual area per native boundary point, in exact raw-VTU \texttt{PointData} order. These weights are evaluation metadata rather than an additional CFD target and permit area-weighted native-point metrics without reconstructing the surface quadrature.
    \item \textbf{Force coefficients:} Time-averaged force and moment coefficients are also provided.
    \item \textbf{Flow visualisations:} Image files of flow separation and skin-friction (e.g., \verb|img_wss_LHCi_AoA_j.png|) on the aircraft surface are provided for every case. They are intended to give an impression of the flow field quickly and conveniently, without having to process the raw data first. Examples of the plots are given in Fig.~\ref{fig:isoqtop4appendix} through Fig.~\ref{fig:isoqtop22appendix}.
    \item \textbf{Convergence plots:} Image files tracking the transient and instantaneous convergence of the Drag ($C_D$), Lift ($C_L$), and Pitching Moment ($C_M$) coefficients over time (CTU) are provided, with running means and nominal CTU-based bands as defined in Appendix~\ref{app:postPro-flowStats}.
\end{itemize}

The release uses standard text and image formats, open-source VTK files (i.e. *.vtu), and NumPy NPY arrays for the surface quadrature sidecars. The VTK output files can be loaded in the most common 3D data visualisation tools, e.g. using the open source software ParaView\footnote{\url{https://www.paraview.org/}}. The data can also be further post-processed with Python scripts using the corresponding VTK and NumPy modules.

\clearpage
\begin{table}[htb]
  \caption{Description of the main components of the dataset}
  \label{tab:dataset-output-files}
  \centering
\setlength{\tabcolsep}{3.5pt}%
\begin{tabular}{p{0.35\linewidth} p{0.12\linewidth} p{0.12\linewidth} p{0.3\linewidth}}
    \toprule
    \cmidrule(r){1-4}
    Output     & Size     & Format & Description  \\
    \midrule
    geo\_LHCi\_AoA\_j.stl & 197 MB  & stl & surface mesh (tris) of the HiLiftAeroML geometry ($\approx$ 600k cells)  \\
    \midrule
    geo\_LHCi\_AoA\_j.stp & 48 MB  & stp & surface geometry of the HiLiftAeroML geometry  \\
    \midrule  
    \path{boundary_geo_LHCi_AoA_j.vtu.tgz} & 13 GB (21GB unzipped) & vtu & time-averaged flow quantities on the HiLiftAeroML ($\approx$ 32M cells) \\
    \midrule
    \path{boundary_dual_area_geo_LHCi_AoA_j.npy} & $\approx$ 0.5 GB & npy & native-point mass-lumped surface areas [\(\mathrm{in}^2\)] in raw boundary-VTU point order \\
    \midrule
    volume\_geo\_LHCi\_AoA\_j.vtu.tgz     & 23 GB (86GB unzipped) & vtu & time-averaged flow quantities within the domain volume ($\approx$ 200M cells)  \\
    \midrule
    ref\_values\_geo\_LHCi\_AoA\_j.csv     & 4 KB       & csv & reference values such as area,Q,AoA of each geometry and AoA \\
    \midrule
    geo\_values\_geo\_LHCi\_AoA\_j.csv     & 4 KB       & csv & reference geometry values used to define the particular geometry via the DoE method \\
    \midrule
    force\_mom\_geo\_LHCi\_AoA\_j.csv  & 4 KB       & csv & time-averaged drag, lift, moment and pressure/viscous drag lift coefficients \\
    \midrule
    img\_wss\_LHCi\_AoA\_j.png  & $\approx$ 1 MB & png & visual rendering of the wall shear stress / skin-friction on the aircraft surface \\
    \midrule
    plot\_CD\_geo\_LHCi\_AoA\_j.png  & $\approx$ 250 KB & png & convergence plot of the Drag Coefficient ($C_D$) over time \\
    \midrule
    plot\_CL\_geo\_LHCi\_AoA\_j.png  & $\approx$ 250 KB & png & convergence plot of the Lift Coefficient ($C_L$) over time \\
    \midrule
    plot\_CM\_geo\_LHCi\_AoA\_j.png  & $\approx$ 250 KB & png & convergence plot of the Pitching Moment Coefficient ($C_M$) over time \\
    \bottomrule
  \end{tabular}
\end{table}

\begin{table}[htb]
  \caption{Description of the additional components of the dataset outside of those per run}
  \label{tab:dataset-output-files-more}
  \centering
\setlength{\tabcolsep}{3.5pt}%
\begin{tabular}{p{0.35\linewidth} p{0.12\linewidth} p{0.12\linewidth} p{0.3\linewidth}}
    \toprule
    \cmidrule(r){1-4}
    Output     & Size     & Format & Description  \\
        \midrule
    geo\_values\_all.csv     & 18.7 kB       & csv & reference geometry values used to define the particular geometry via the DoE method for all runs \\
        \midrule
    force\_mom\_all.csv  & 282.2 kB       & csv & time-averaged drag, lift, moment and pressure/viscous drag lift coefficients for all runs \\
        \midrule
    splits/Makefile & 510 Bytes & make & Build script for generating/managing splits \\
        \midrule
    splits/README.md & 5.48 kB & md & Markdown documentation for the dataset splits \\
        \midrule
    splits/README.pdf & 165 kB & pdf & PDF documentation for the dataset splits \\
        \midrule
    splits/generate\_splits.py & 17 kB & py & Python script to generate dataset splits \\
        \midrule
    splits/manifest.json & 411.5 kB & json & Manifest file containing the split definitions \\
        \midrule
    splits/preamble.tex & 235 Bytes & tex & LaTeX preamble for the documentation \\
        \midrule
    splits/stall\_onset.png & 140 kB & png & Image visualization of stall onset \\
        \midrule
    splits/visualize\_stalled\_cases.py & 3.13 kB & py & Python script to visualize stalled cases \\
        \midrule
    \path{surface_dual_areas/manifest.json} & 9.64 MB & json & complete 1,800-case inventory, payload/source hashes, topology, area-closure validation, and provenance for surface quadrature sidecars \\
    \bottomrule
  \end{tabular}
\end{table}

\begin{table}[htb]
    \caption{Reference quantities, solver definitions, and baseline values used in the dataset.}
    \label{tab:dataset-refQuantCoeffs}
    \centering
    \setlength{\tabcolsep}{3.5pt}%
\begin{tabular}{p{0.2\linewidth} p{0.2\linewidth} p{0.25\linewidth} p{0.25\linewidth}}
        \toprule
        Variable & Symbol & Value / Range & Units \\
        \midrule
        Reference Chord & $C_{ref}$ (MAC) & $275.80$ & $[\si{in}]$ \\
        Reference Area & $S_{ref}$ & $297,360.0$ & $[\si{in^2}]$ \\
        Reference Span & $b_{ref}$ & $1156.75$ & $[\si{in}]$ \\
        Moment Center & $\vec{x}_{ref}$ & $(1325.9, 0.0, 177.95)$ & $[\si{in}]$ \\
        \midrule
        Mach Number & $M$ & $0.2$ & $[-]$ \\
        Reference Velocity & $U_{ref}$ & $2679.505$ & $[\si{in/s}]$ \\
        Reference Q & $Q_{ref}$ & $4.937856$ & $[\mathrm{slug}/(\mathrm{in}\,\mathrm{s}^2)]$ \\
        Reynolds Number & $Re$ & $1.6 \times 10^{6}$ & $[-]$ \\
        Angle of Attack & $\alpha$ & $4 - 22$ & $[^{\circ}]$ \\
        \midrule
        Temperature & $T_{ref}$ & $518.67$ & $[\si{R}]$ \\
        Atmospheric Pressure & $p_{atm}$ & $14.696$ & $[\si{psi}]$ \\
        Specific Gas Const. & $R_{gas}$ & $1716.594$ & $[\mathrm{ft}^2/(\mathrm{s}^2\,{}^\circ\mathrm{R})]$ \\
        Specific Gas Const. & $R_{gas}$ & $247189.536$ & $[\mathrm{in}^2/(\mathrm{s}^2\,{}^\circ\mathrm{R})]$ \\
        Specific Heat Ratio & $\gamma$ & $1.4$ & $[-]$ \\
        \bottomrule
    \end{tabular}
\end{table}

\clearpage

\begin{table}[htb]
    \caption{List of output quantities available in the dataset files. Quantities are either time-averaged (\texttt{avg}) or root-mean-square (\texttt{rms}) values.}
    \label{tab:dataset-output-quantities}
    \centering
    \setlength{\tabcolsep}{3.5pt}%
\begin{tabular}{p{0.12\linewidth} p{0.15\linewidth} p{0.32\linewidth} p{0.32\linewidth}}
        \toprule
        \multicolumn{4}{c}{\textbf{Volume Data (volume\_i.vtu) -- Point Data}}                   \\
        Symbol       & Units                 & Field Name                                      & Description  \\
        \toprule
        $ID_{node}$  & $[-]$                 & \texttt{NodeID}                                 & Unique vertex identifier \\
        \midrule
        $\bar{p}, p'_{rms}$ & $[\mathrm{slug}/(\mathrm{in}\,\mathrm{s}^2)]$ & \texttt{avg(P), rms(P)}                         & Static pressure \\
        \midrule
        $\bar{T}, T'_{rms}$ & $[\si{R}]$                 & \texttt{avg(T), rms(T)}                         & Static temperature \\
        \midrule
        $\bar{\rho}, \rho'_{rms}$ & $[\si{slug/in^3}]$   & \texttt{avg(rho), rms(rho)}                     & Density \\
        \midrule
        $\bar{\mathbf{u}}, \mathbf{u}'_{rms}$ & $[\si{in/s}]$          & \texttt{avg(u), rms(u)}                         & Velocity vector \\
        \midrule
        $\bar{\mu}_{sgs}, \mu'_{rms}$ & $[\mathrm{slug}/(\mathrm{in}\,\mathrm{s})]$ & \texttt{avg(mu\_sgs), rms(mu\_sgs)}             & Subgrid-scale viscosity \\
        \midrule
        $\overline{u'_i u'_j}$ & $[\si{in^2/s^2}]$      & \texttt{rey(u)}                                 & Resolved Reynolds stress tensor \\
        \toprule
        \multicolumn{4}{c}{\textbf{Volume Data (volume\_i.vtu) -- Cell Data}}                   \\
        \toprule
        $ID_{cell}$  & $[-]$                 & \texttt{CellID}                                 & Unique octree cell identifier \\
        \toprule
        \multicolumn{4}{c}{\textbf{Surface Data (boundary\_i.vtu) -- Point Data}}                   \\
        Symbol       & Units                 & Field Name                                      & Description  \\
        \toprule
        $\bar{\tau}_{w}, \tau'_{w,rms}$ & $[\mathrm{slug}/(\mathrm{in}\,\mathrm{s}^2)]$ & \texttt{AVG(TAU\_WALL), RMS(TAU\_WALL)}          & Wall shear stress magnitude \& vector components (0,1,2) \\
        \midrule
        $y^+, y^{+\prime}_{rms}$ & $[-]$                 & \texttt{AVG(Y\_PLUS), RMS(Y\_PLUS)}             & Non-dimensional wall distance \\
        \midrule
        $\mathbf{n}$ & $[-]$                 & \texttt{N\_BF}                                  & Boundary face normal vector \\
        \midrule
        $\bar{p}_{proj}$ & $[\mathrm{slug}/(\mathrm{in}\,\mathrm{s}^2)]$ & \texttt{PROJ(AVG(P)), PROJ(RMS(P))}             & Projected static pressure (Avg \& RMS) \\
        \midrule
        $\bar{T}_{proj}$ & $[\si{R}]$                 & \texttt{PROJ(AVG(T)), PROJ(RMS(T))}             & Projected static temperature (Avg \& RMS) \\
        \midrule
        $\bar{\rho}_{proj}$ & $[\si{slug/in^3}]$     & \texttt{PROJ(AVG(RHO)), PROJ(RMS(RHO))}         & Projected density (Avg \& RMS) \\
        \midrule
        $\bar{\mathbf{u}}_{proj}$ & $[\si{in/s}]$         & \texttt{PROJ(AVG(U)), PROJ(RMS(U))}             & Projected velocity vector (Avg \& RMS) \\
        \midrule
        $\bar{\mu}_{sgs, proj}$ & $[\mathrm{slug}/(\mathrm{in}\,\mathrm{s})]$ & \texttt{PROJ(AVG(MU\_SGS)), PROJ(RMS(...))}     & Projected SGS viscosity (Avg \& RMS) \\
        \midrule
        $\overline{u'_i u'_j}_{proj}$ & $[\si{in^2/s^2}]$      & \texttt{PROJ(REY(U))}                           & Projected Reynolds stress tensor \\
        \bottomrule
    \end{tabular}
\end{table}

\clearpage
\subsection{Device-positioning schedules}
\label{app:geometry_schedules}

Building a relevant database of flow solutions utilizing the CRM-HL as the reference geometry requires that a model be parameterized to encompass some set of geometric perturbations. In this work, the leading edge slats and trailing edge flaps are parameterized independently in a manner that encompasses the set of already defined reference positions. \\

\begin{figure}[!htb]
\centering
\includegraphics[width=0.86\linewidth]{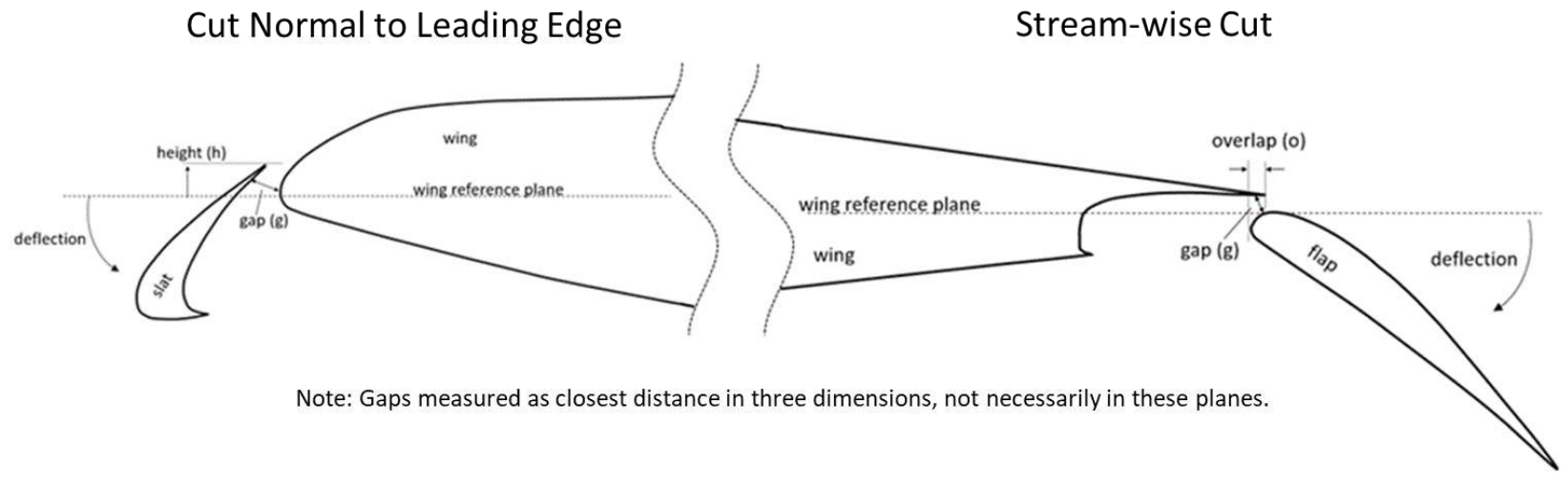}
\caption{Sectional views of Leading and Trailing Edge Device Positioning Parameters\label{fig:CRMHL_pos_main}}
\end{figure}

\noindent
The leading-edge high-lift system features a slat, whose deployment relative to the main wing is defined by its deflection angle, gap, and height as shown in Fig. \ref{fig:CRMHL_pos_main}. The parametric study variation includes two variables on the leading-edge slat. Deflection is typically the dominant variable, and is varied between 10$^\circ$ and 35$^\circ$ relative to the wing reference plane. Gap between the slat trailing edge and wing-under-slat-surface typically varies as a function of position, where it is fully sealed at takeoff positioning (22$^\circ$), and opened to a reference gap at the landing position (30$^\circ$). This gap schedule is followed for the present study, but also multiplied by a parametric gap multiplier which varies between 0.5 and 1.5.  The third variable, height, is typically a function of deployment angle, with it being highest in its stowed (0$^\circ$) position, and lowest at the fully deployed (30$^\circ$) position. This schedule is followed without variation, making for a total of 4 independent slat parameters -- Inboard slat deflection and gap, and outboard slat deflection and gap. \\

\noindent
At the trailing edge, single slotted flaps are employed, and their geometric settings are characterized by deflection angle, gap, and overlap relative to the main wing element. Similar to the slat, the flap deflection is the most dominant variable, and is allowed a range of 10$^\circ$ through 45$^\circ$. This range fully captures the most shallow takeoff deflection (10$^\circ$) and deepest landing deflections (43$^\circ$). A baseline gap schedule versus deflection is implicitly defined by evaluating the reference takeoff and landing positions. This schedule is also multiplied by a parametric multiplier with a range from 0.5 to 1.5. Similarly, an overlap schedule is also defined by the reference positioning set, but left to strictly follow the reference schedule rather than parameterized. Similar to slats, the trailing edge flaps are perturbed by 4 total parameters -- Inboard flap deflection and gap, and outboard flap deflection and gap. Parametric ranges are summarized in Table~\ref{table:dataset_parameters} (left).

\subsection{Aerodynamic variation across the parameter space}

The eight geometry parameters and the ten-angle operating sweep are summarized in Section~\ref{sec:geometry_parameterization}; Appendix~\ref{app:geometry_schedules} gives the deployment schedules. Figure~\ref{fig:CRMHL_pos_main} illustrates the device positioning, and Table~\ref{table:dataset_parameters} gives the parameter ranges. The following results describe the aerodynamic variation across this design space.

\begin{figure}[!htb]
\begin{center}
\includegraphics[width=0.8\textwidth]{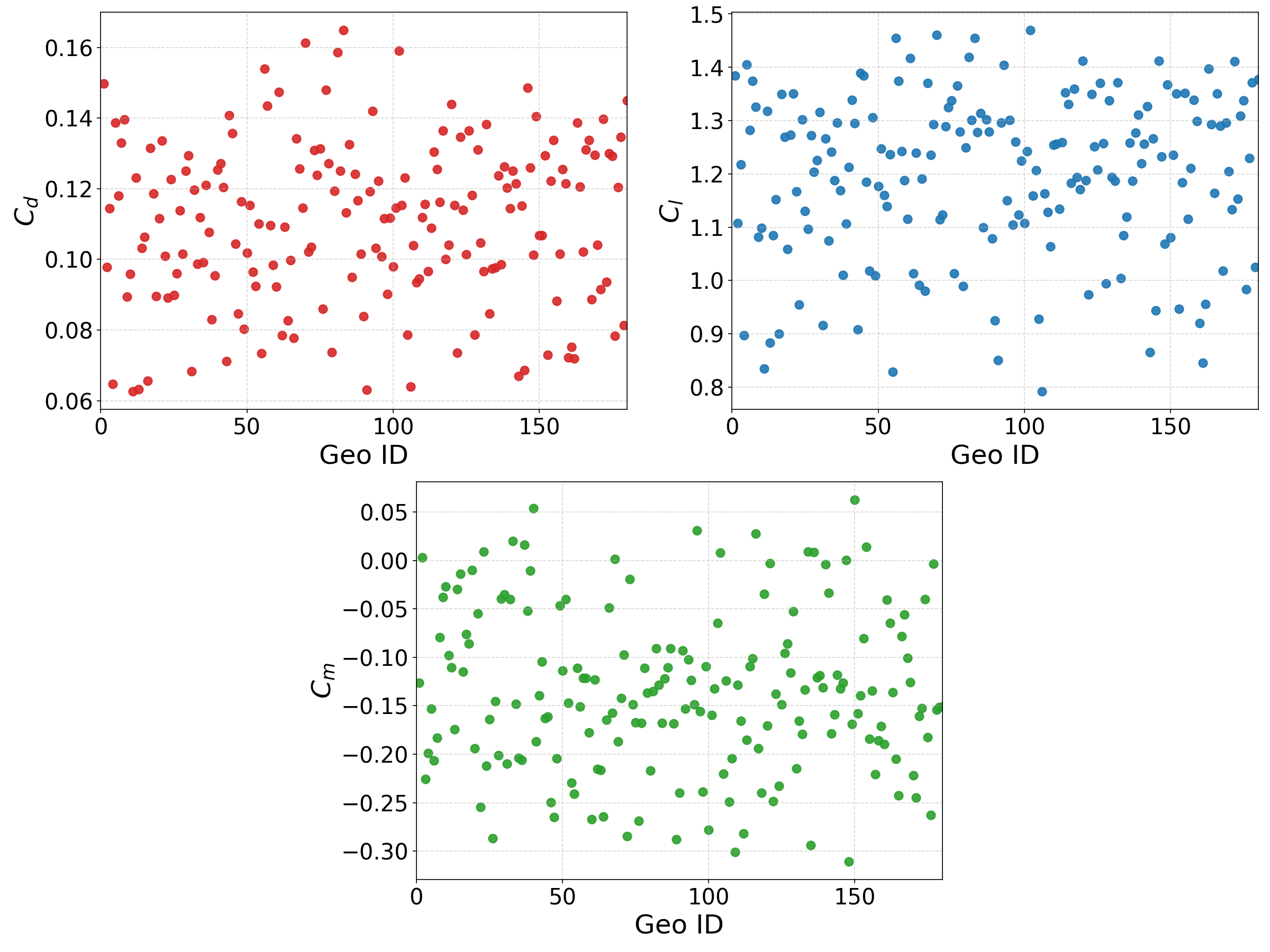}
\caption{Drag, Lift and Moment Coefficient vs Geo ID number for 4 degrees AoA \label{fig:forces4}}
\end{center}
\end{figure}

\begin{figure}[!htb]
\begin{center}
\includegraphics[width=0.8\textwidth]{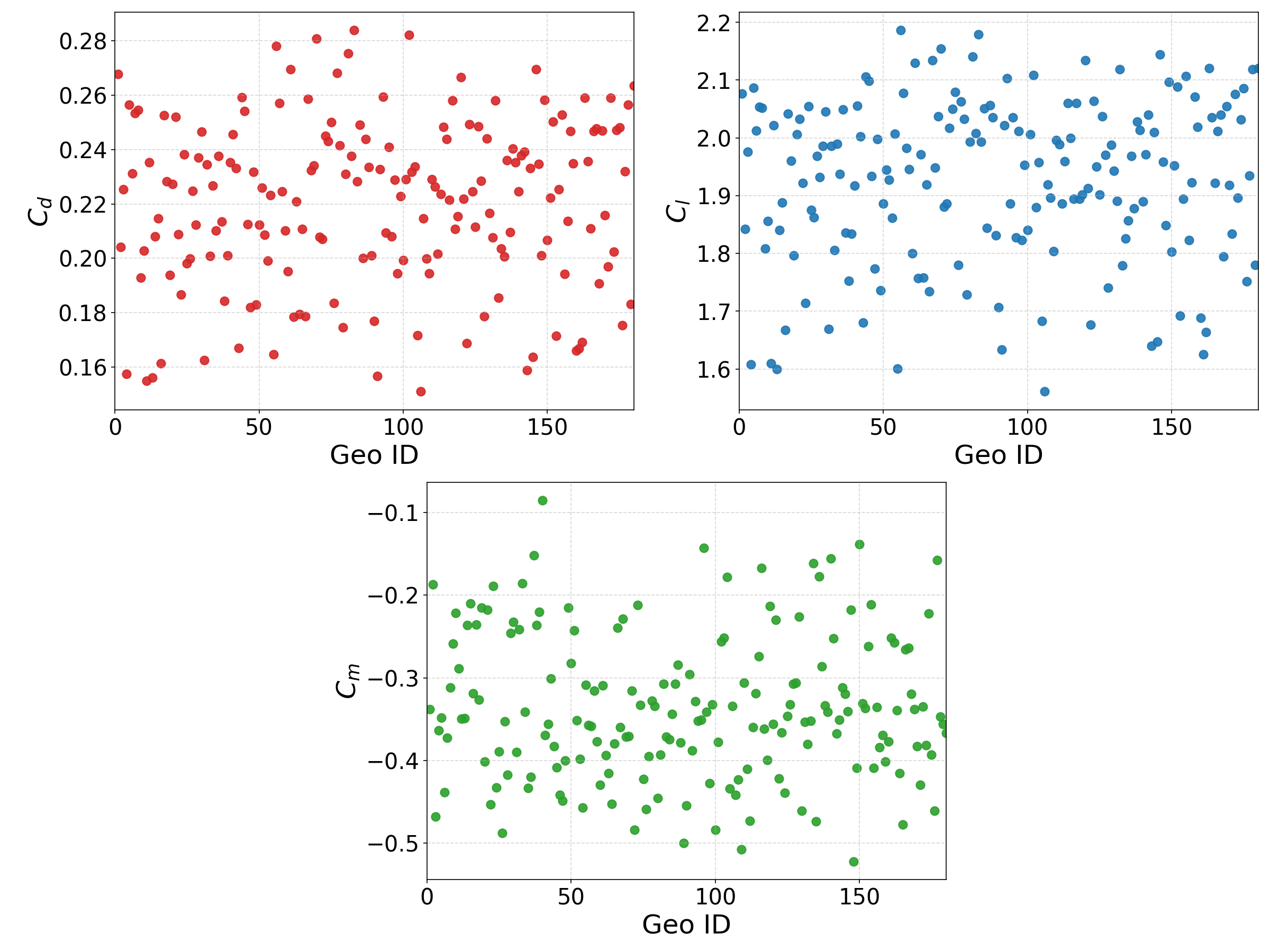}
\caption{Drag, Lift and Moment Coefficient vs Geo ID number for 12 degrees AoA \label{fig:forces12}}
\end{center}
\end{figure}

\begin{figure}[!htb]
\begin{center}
\includegraphics[width=0.8\textwidth]{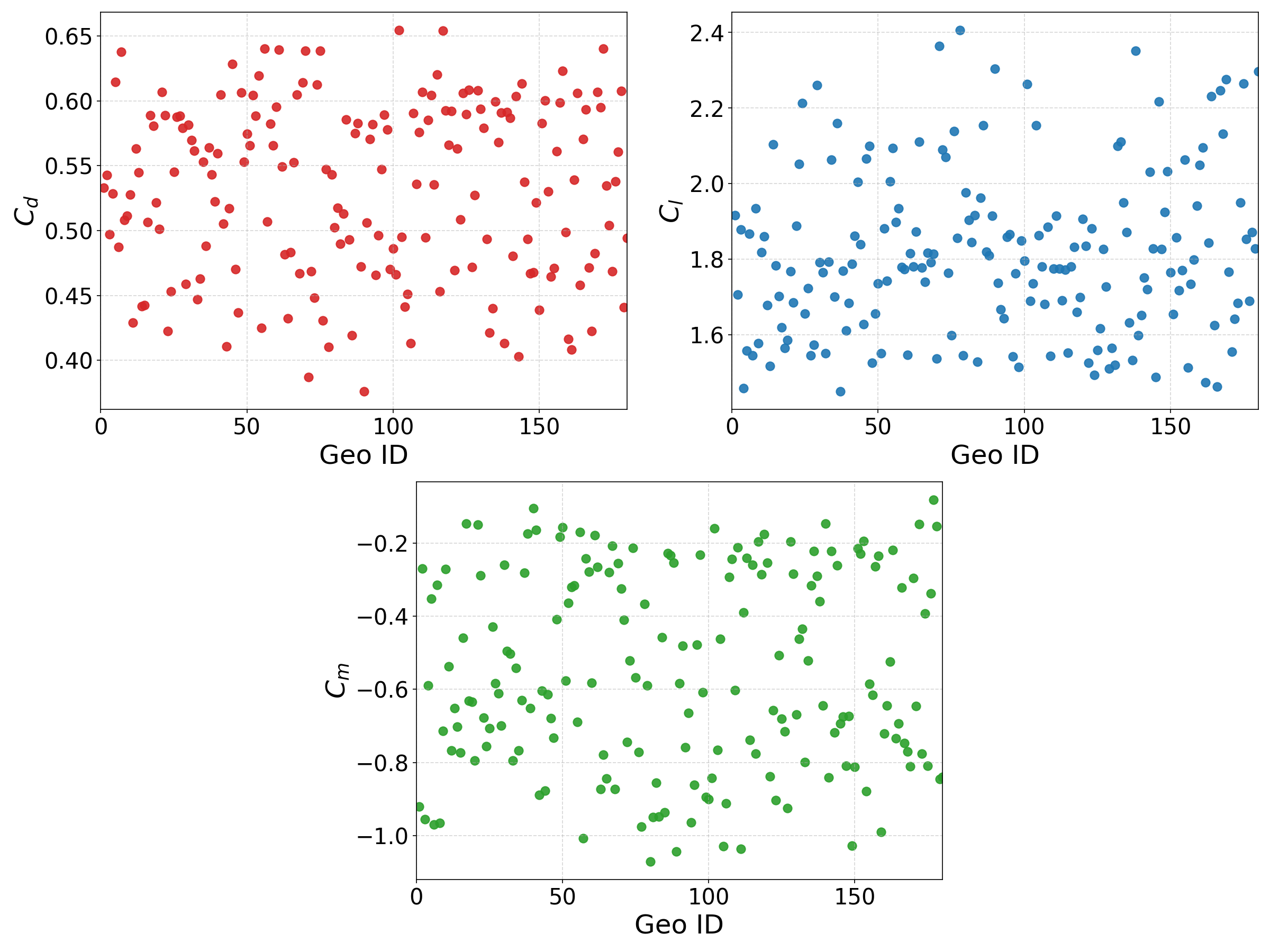}
\caption{Drag, Lift and Moment Coefficient vs Geo ID number for 22 degrees AoA \label{fig:forces22}}
\end{center}
\end{figure}

Fig.~\ref{fig:forces4}, Fig.~\ref{fig:forces12} and Fig.~\ref{fig:forces22} show the variation of drag, lift, and pitching moment coefficients for each geometry ID at three distinct angles of attack ($4^\circ$, $12^\circ$, and $22^\circ$). 
The data reveals a broad range of aerodynamic performance across the 180 geometries. 
For instance, at $AoA=4^\circ$ (Fig.~\ref{fig:forces4}), the lift coefficient ($C_L$) varies from approximately 0.79 to 1.47, while at $AoA=22^\circ$ (Fig.~\ref{fig:forces22}), it spans approximately 1.45 to 2.41 in the pinned release's force summary. This wider range reflects differing high-angle responses across the sampled configurations.
This diversity confirms that the parametric variations in slat and flap settings successfully generate a rich design space covering both attached and separated flow regimes.

\begin{figure}[p]
    \centering
    
    \begin{subfigure}[t]{0.48\textwidth}
        \flushleft \includegraphics[width=0.35\textwidth]{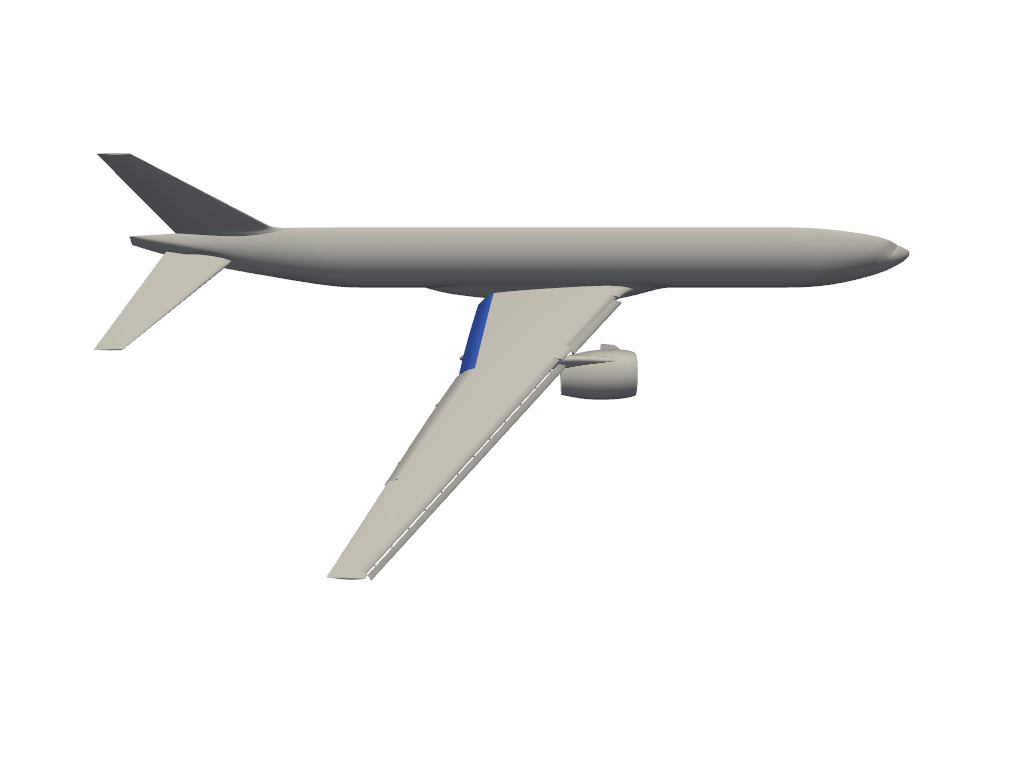} \\
        \vspace{-0.5ex} %
        \centering \includegraphics[width=1\textwidth]{images/datasetimages/Combined_All_AoA_IB_Flap_Deflection_vs_Cl.png}
        \caption{Inboard Flap (All AoA)}
    \end{subfigure}
    \hfill
    \begin{subfigure}[t]{0.48\textwidth}
        \flushleft \includegraphics[width=0.35\textwidth]{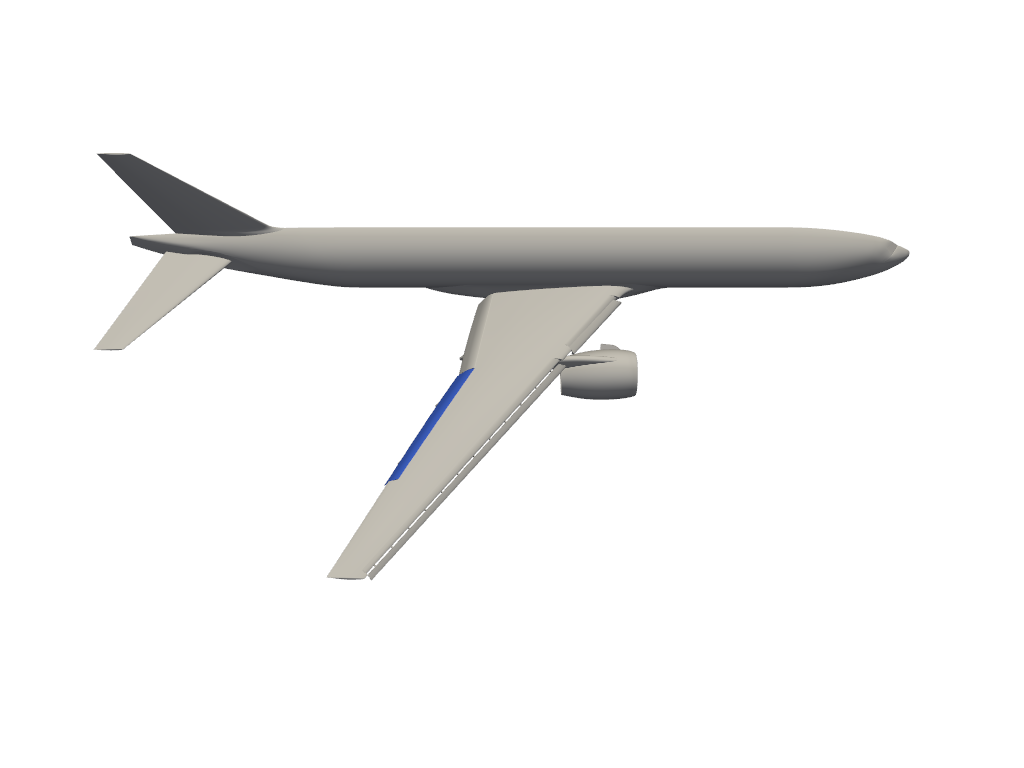} \\
        \vspace{-0.5ex}
        \centering \includegraphics[width=1\textwidth]{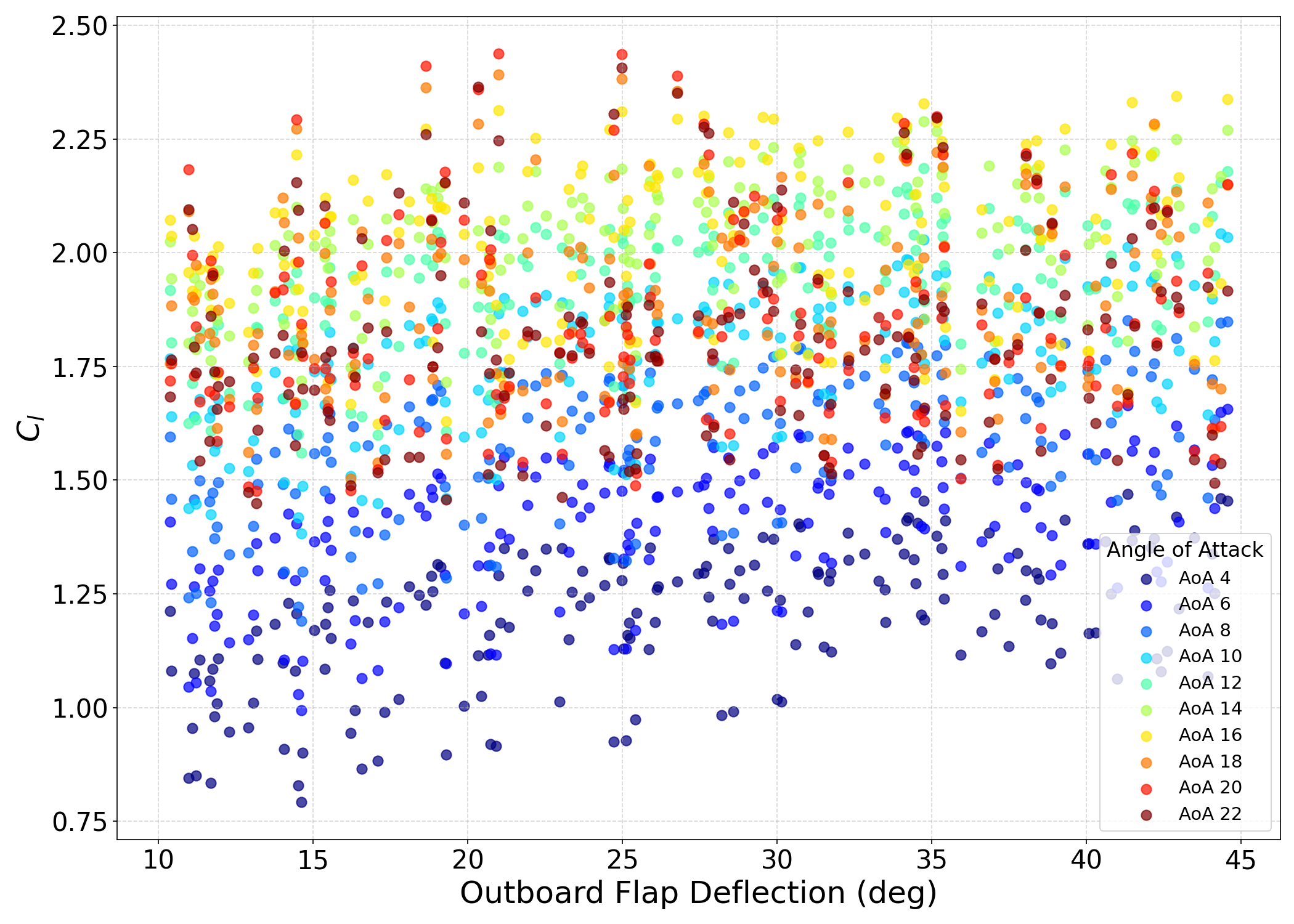}
        \caption{Outboard Flap (All AoA)}
    \end{subfigure}

    \vspace{4ex} %

    \begin{subfigure}[t]{0.48\textwidth}
        \flushleft \includegraphics[width=0.35\textwidth]{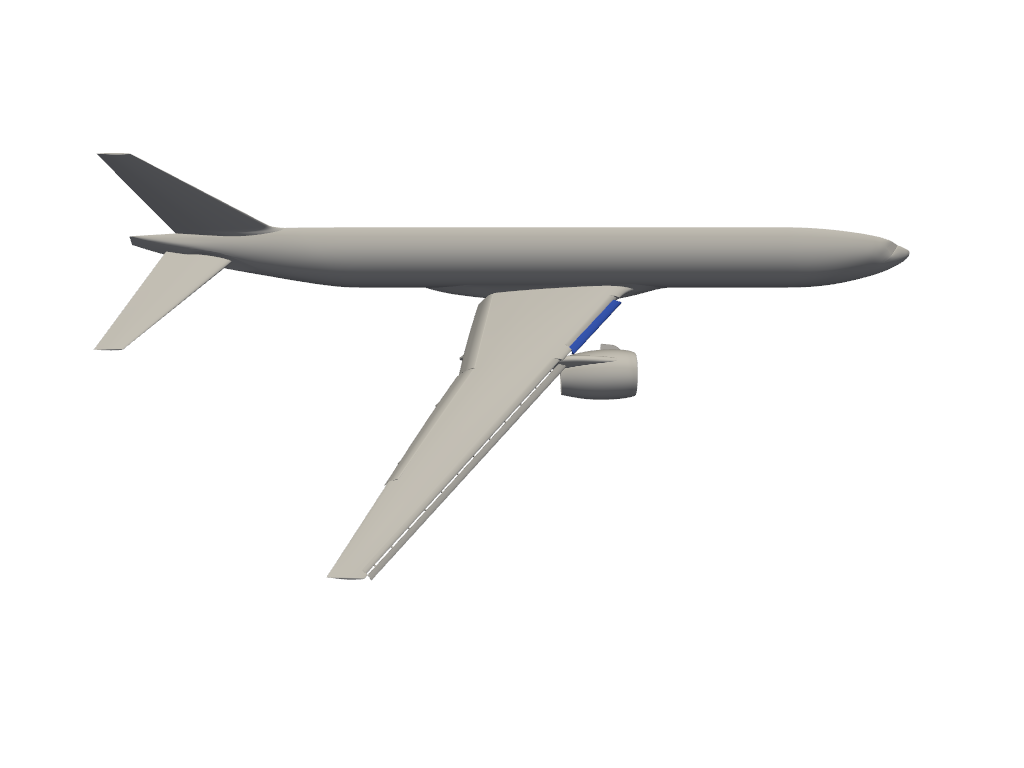} \\
        \vspace{-0.5ex}
        \centering \includegraphics[width=1\textwidth]{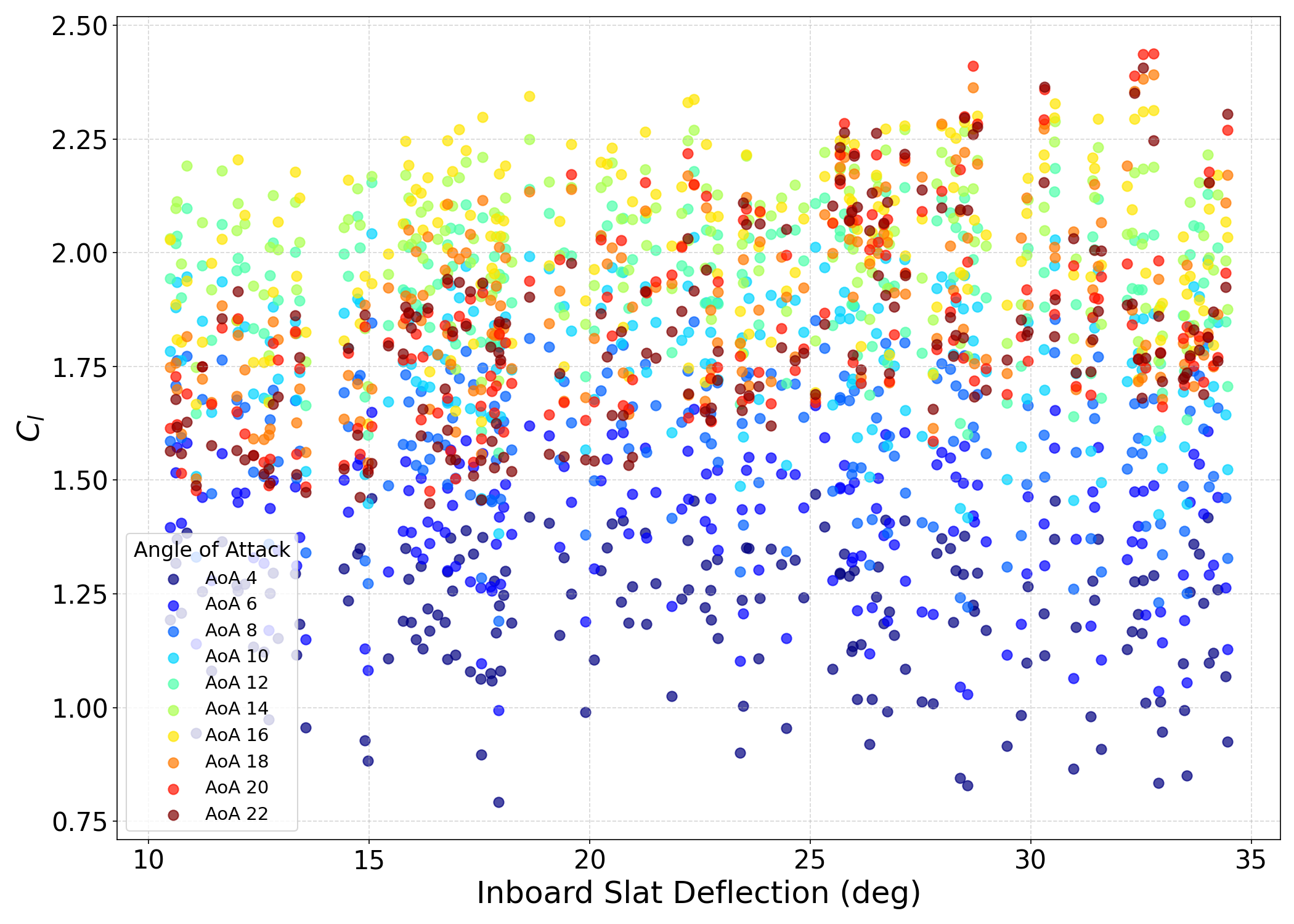}
        \caption{Inboard Slat (All AoA)}
    \end{subfigure}
    \hfill
    \begin{subfigure}[t]{0.48\textwidth}
        \flushleft \includegraphics[width=0.35\textwidth]{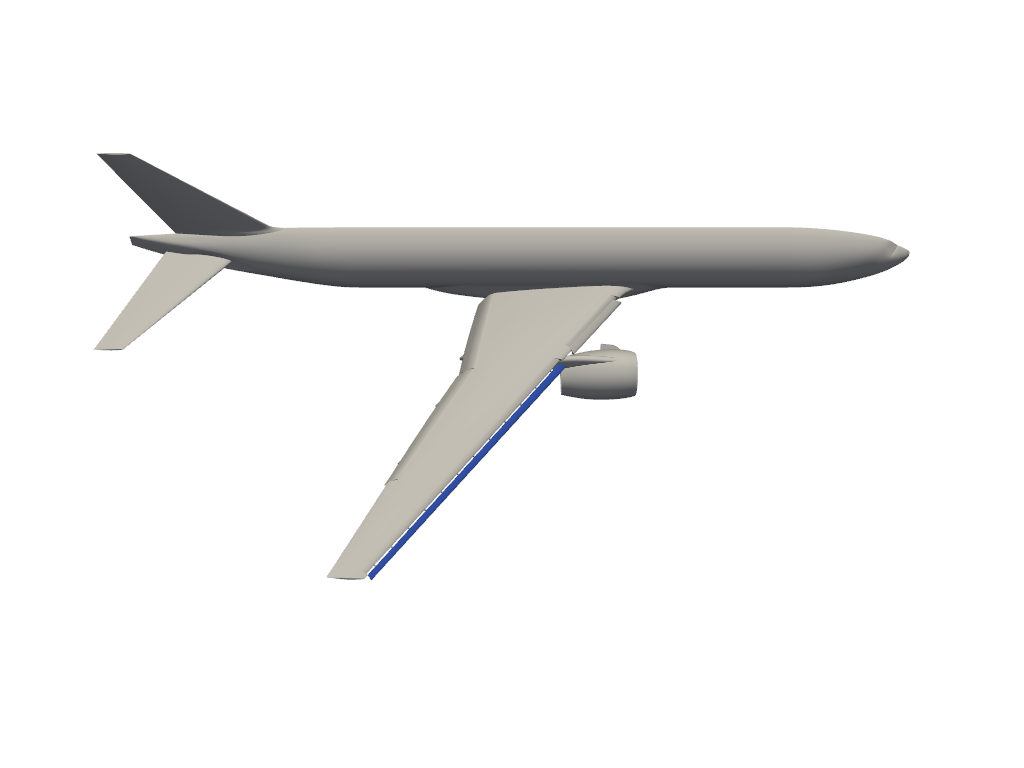} \\
        \vspace{-0.5ex}
        \centering \includegraphics[width=1\textwidth]{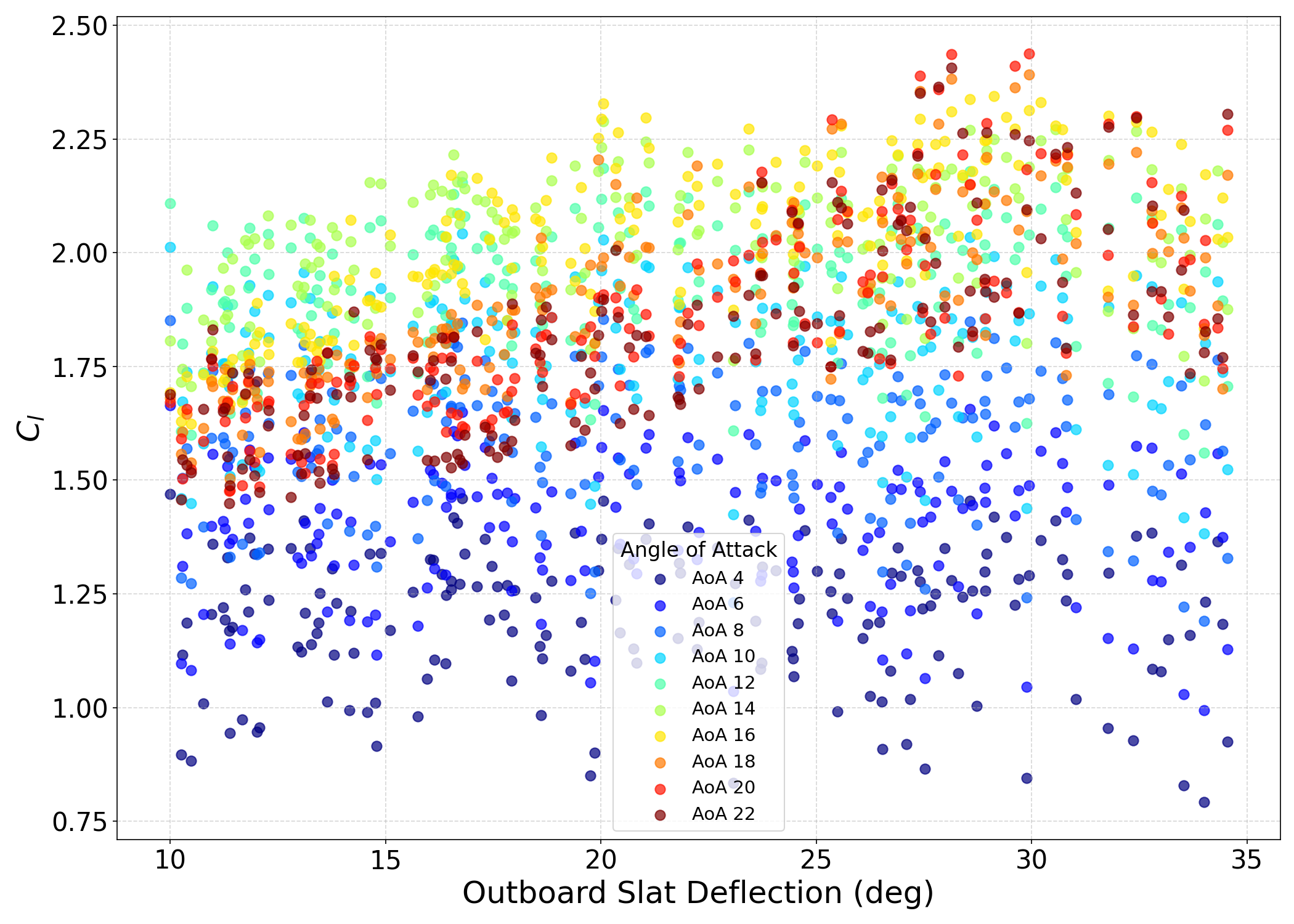}
        \caption{Outboard Slat (All AoA)}
    \end{subfigure}

    \vspace{2ex}
    \caption{Variation of force coefficients with selected geometry parameters over all samples and AoA in the dataset. The top-left inset in each quadrant visualises the geometry parameter being varied.}
    \label{fig:forces-vs-geomParsall-2x2}
\end{figure}

\begin{figure}[p]
    \centering
    
    \begin{subfigure}[t]{0.48\textwidth}
        \centering
        \raggedright \includegraphics[width=0.9\textwidth]{images/highlight_IBFLAP.png} \\ %
        \vspace{-3ex} %
        \centering \includegraphics[width=1\textwidth]{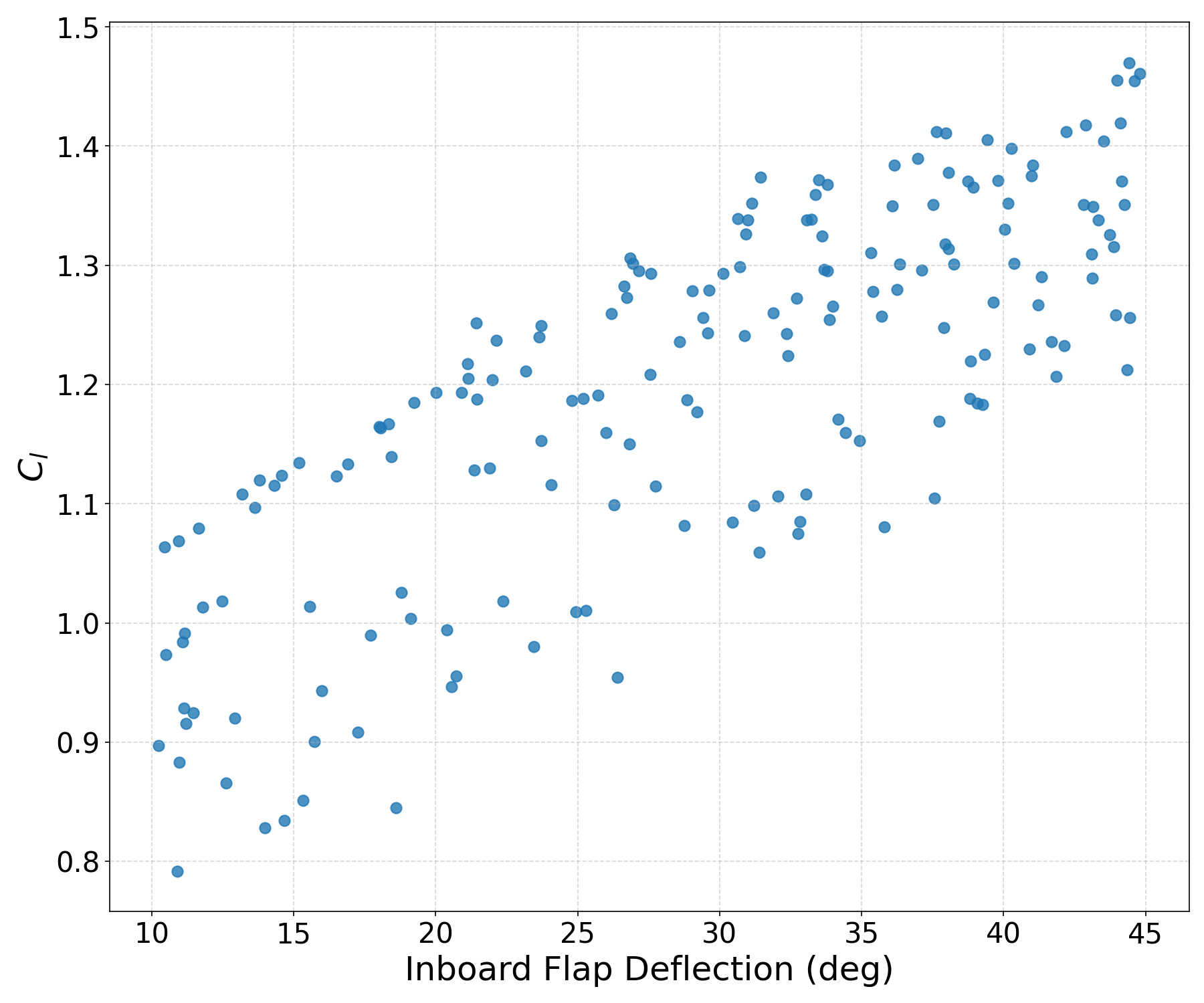}
        \caption{Inboard Flap Deflection}
    \end{subfigure}
    \hfill
    \begin{subfigure}[t]{0.48\textwidth}
        \centering
        \raggedright \includegraphics[width=0.9\textwidth]{images/highlight_OBFLAP.png} \\
        \vspace{-3ex}
        \centering \includegraphics[width=1\textwidth]{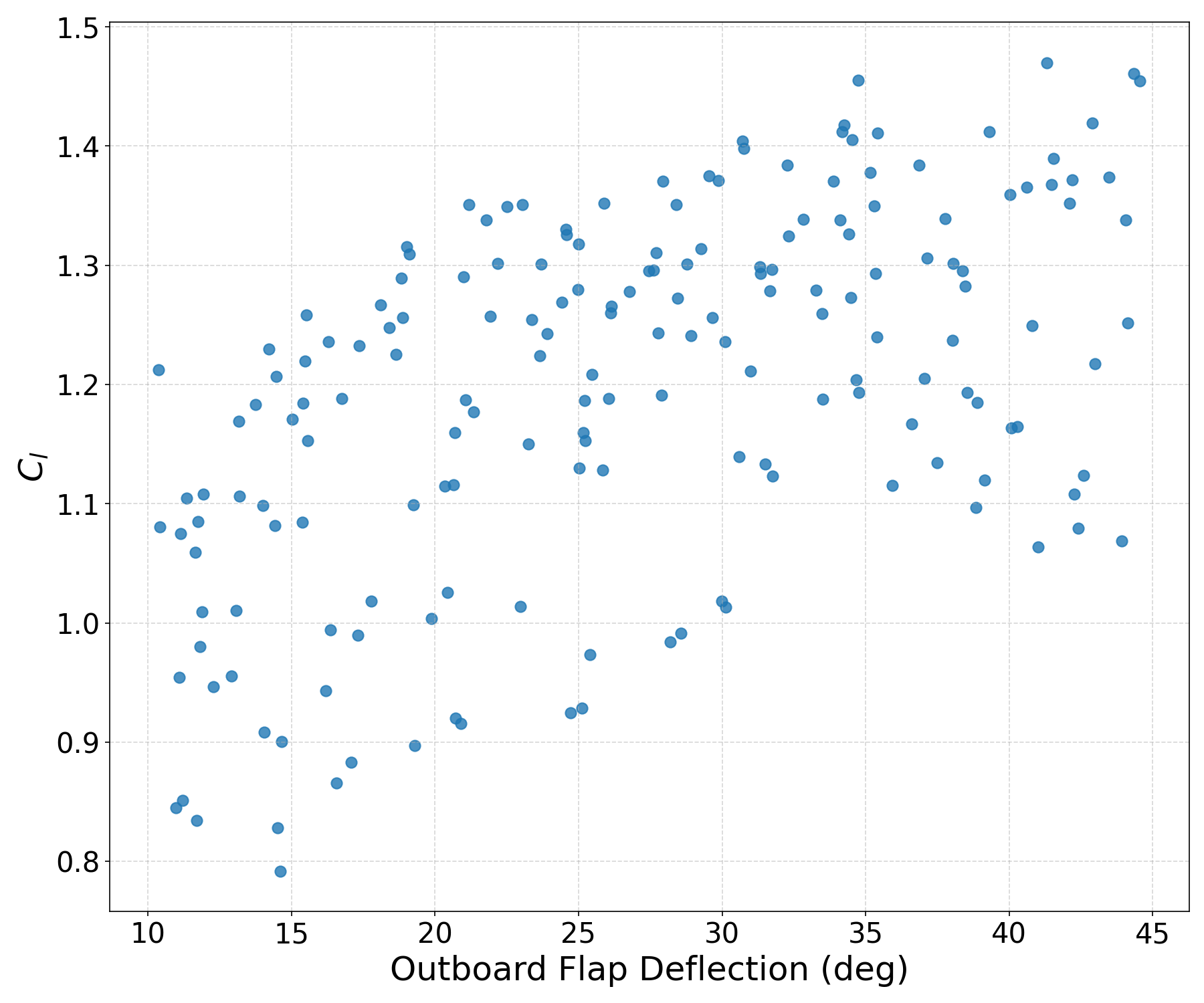}
        \caption{Outboard Flap Deflection}
    \end{subfigure}

    \vspace{4ex} %

    \begin{subfigure}[t]{0.48\textwidth}
        \centering
        \raggedright \includegraphics[width=0.9\textwidth]{images/highlight_IBSLAT.png} \\
        \vspace{-3ex}
        \centering \includegraphics[width=1\textwidth]{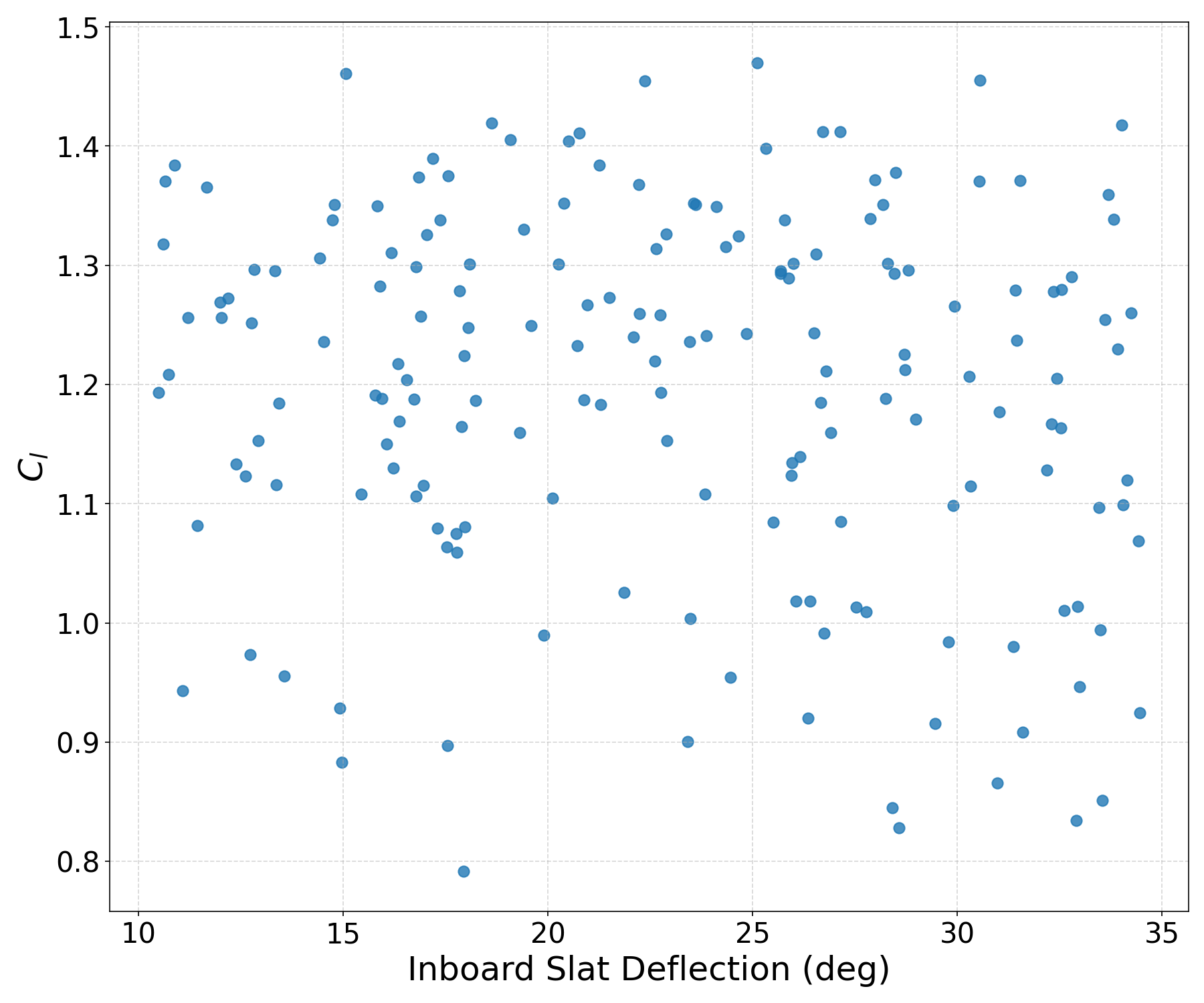}
        \caption{Inboard Slat Deflection}
    \end{subfigure}
    \hfill
    \begin{subfigure}[t]{0.48\textwidth}
        \centering
        \raggedright \includegraphics[width=0.9\textwidth]{images/highlight_OBSLAT.png} \\
        \vspace{-3ex}
        \centering \includegraphics[width=1\textwidth]{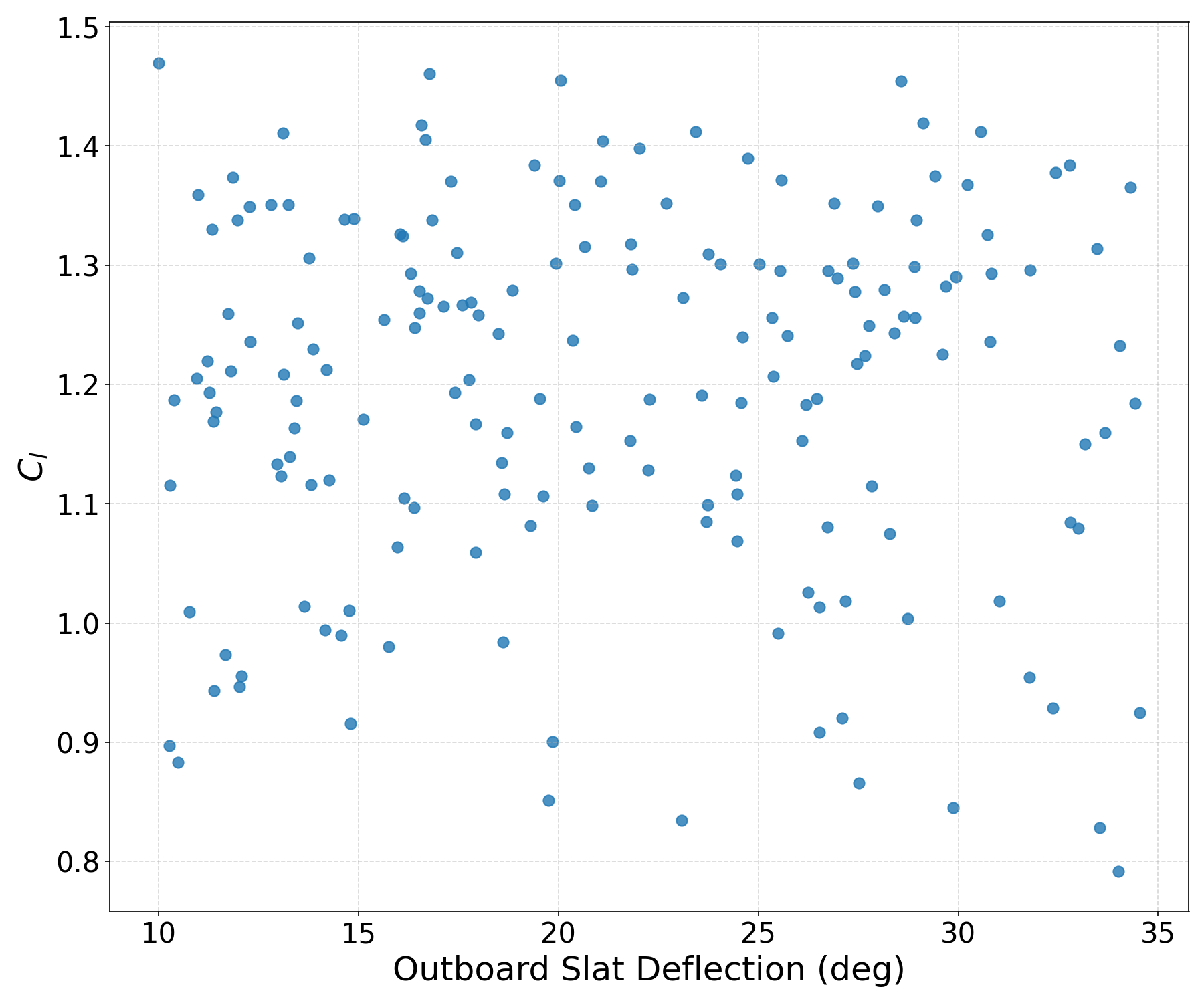}
        \caption{Outboard Slat Deflection}
    \end{subfigure}

    \vspace{2ex}
    \caption{Variation of force coefficients with selected geometry parameters (2x2 layout). The top-left inset in each quadrant indicates the highlighted geometry component.}
    \label{fig:forces-vs-geomPars-2x2}
\end{figure}

\begin{figure}[p]
    \centering
    
    \begin{subfigure}[t]{0.48\textwidth}
        \flushleft \includegraphics[width=0.35\textwidth]{images/highlight_IBFLAP.png} \\
        \vspace{-0.5ex} 
        \centering \includegraphics[width=1\textwidth]{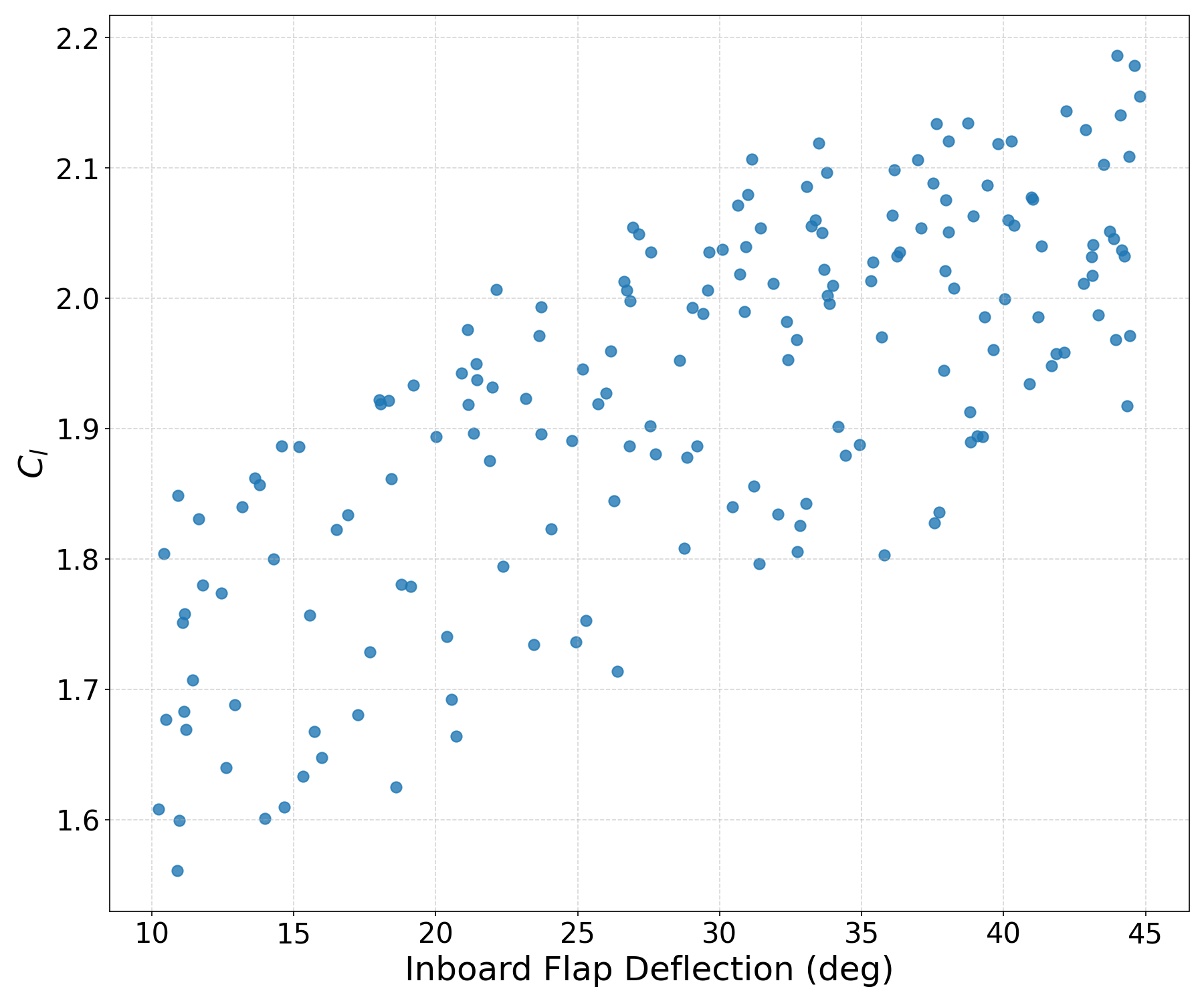}
        \caption{Inboard Flap (12$^\circ$ AoA)}
    \end{subfigure}
    \hfill
    \begin{subfigure}[t]{0.48\textwidth}
        \flushleft \includegraphics[width=0.35\textwidth]{images/highlight_OBFLAP.png} \\
        \vspace{-0.5ex}
        \centering \includegraphics[width=1\textwidth]{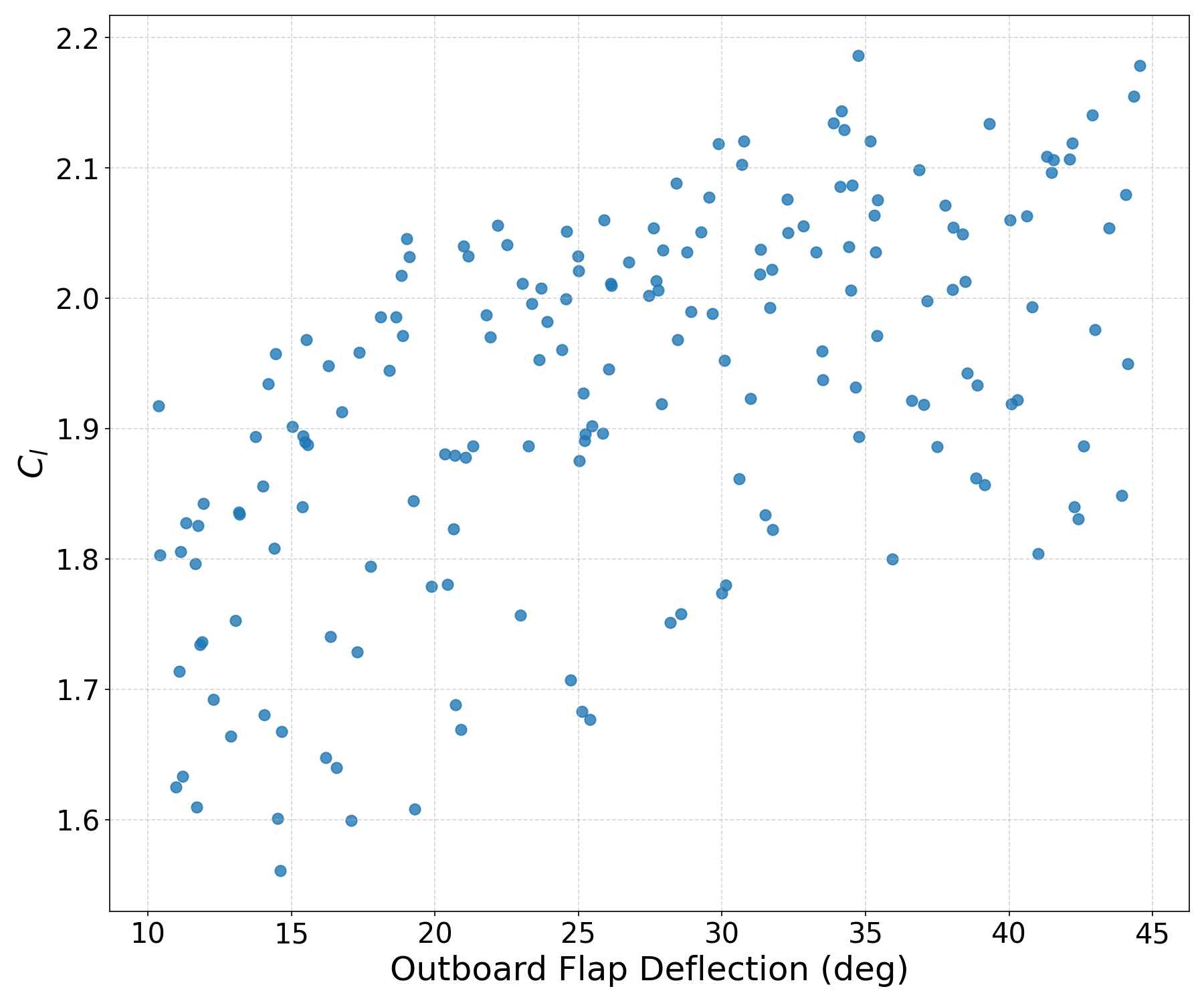}
        \caption{Outboard Flap (12$^\circ$ AoA)}
    \end{subfigure}

    \vspace{4ex} %

    \begin{subfigure}[t]{0.48\textwidth}
        \flushleft \includegraphics[width=0.35\textwidth]{images/highlight_IBSLAT.png} \\
        \vspace{-0.5ex}
        \centering \includegraphics[width=1\textwidth]{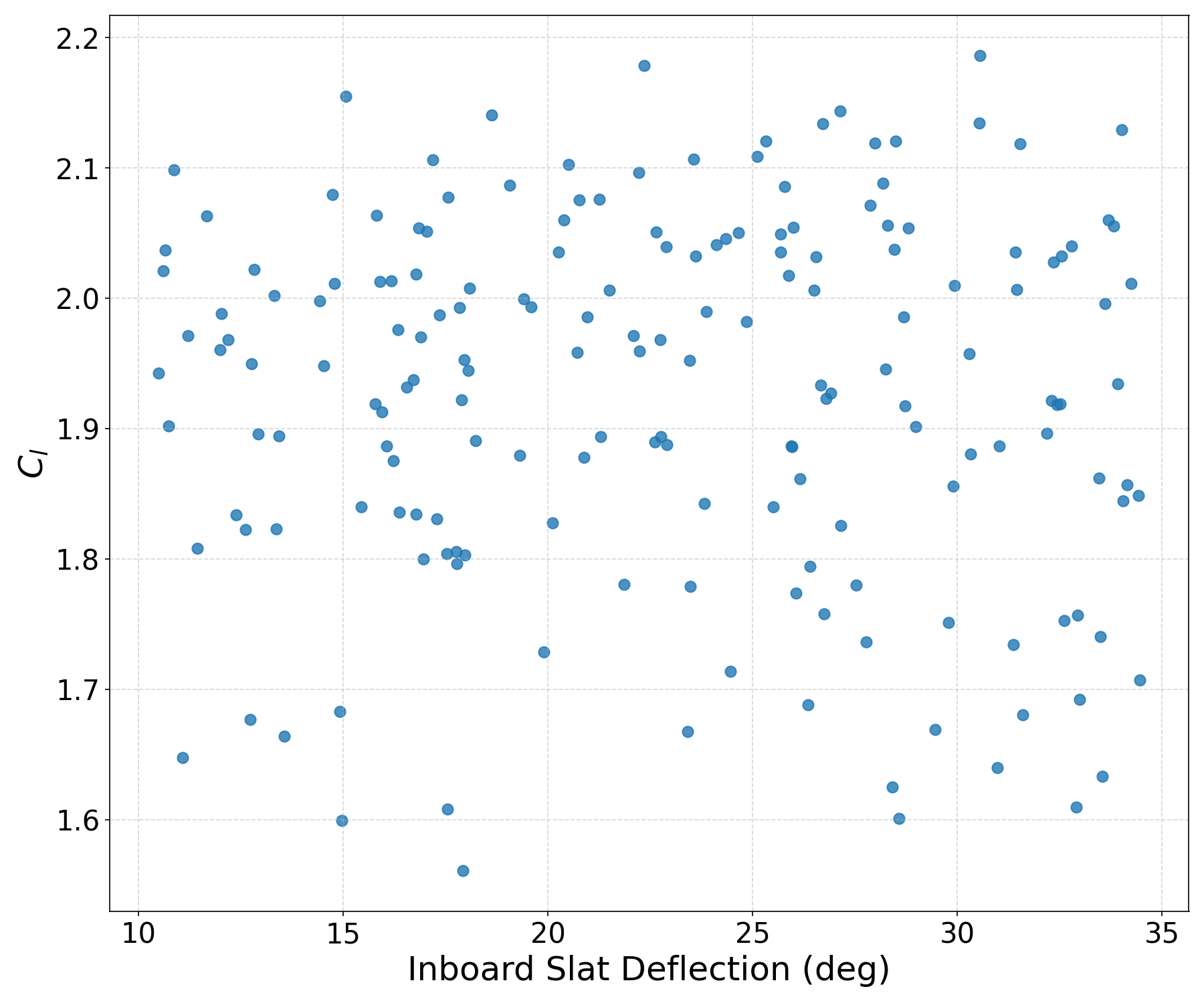}
        \caption{Inboard Slat (12$^\circ$ AoA)}
    \end{subfigure}
    \hfill
    \begin{subfigure}[t]{0.48\textwidth}
        \flushleft \includegraphics[width=0.35\textwidth]{images/highlight_OBSLAT.png} \\
        \vspace{-0.5ex}
        \centering \includegraphics[width=1\textwidth]{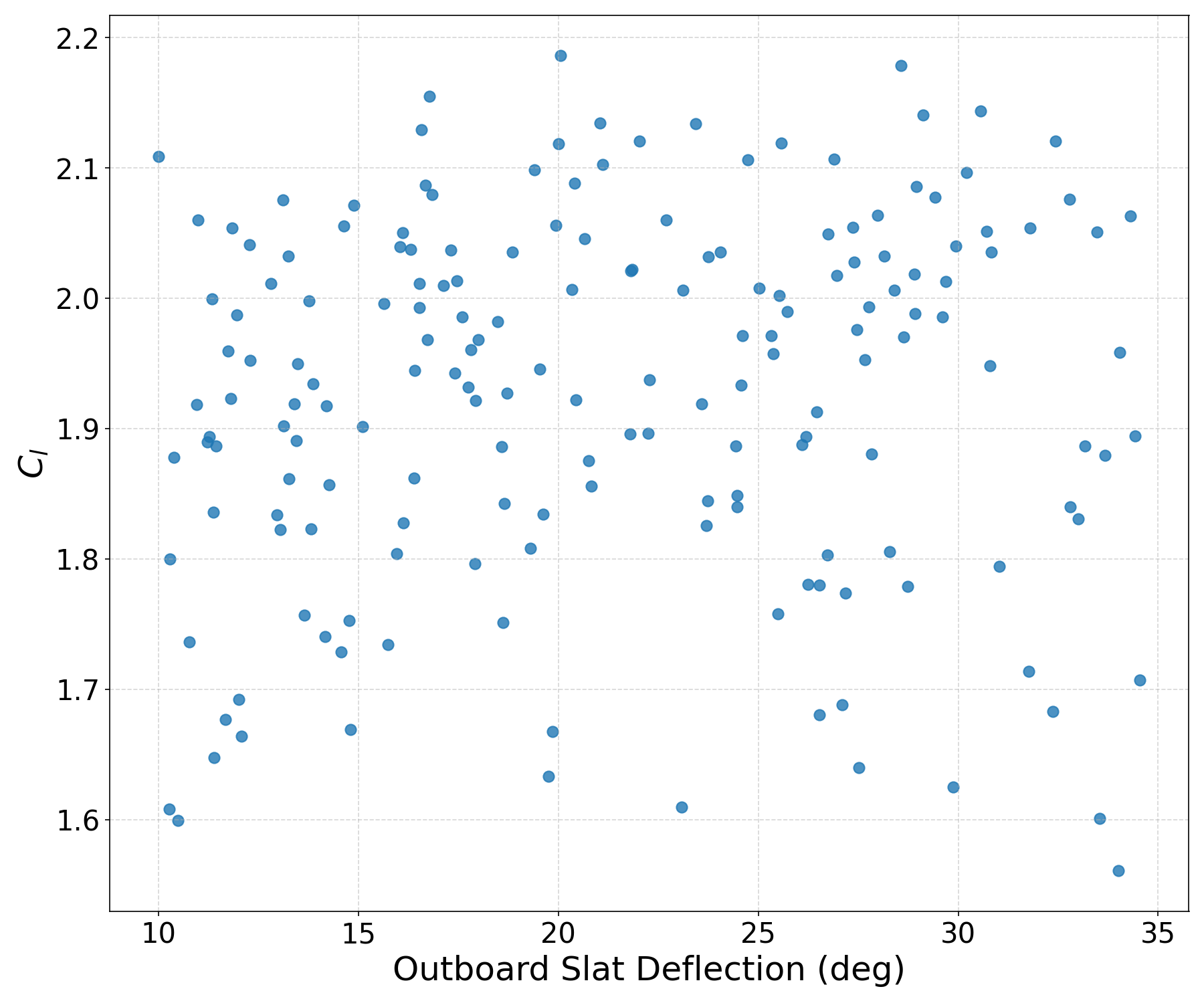}
        \caption{Outboard Slat (12$^\circ$ AoA)}
    \end{subfigure}

    \vspace{2ex}
    \caption{Variation of force coefficients with selected geometry parameters over all samples and 12 degrees AoA in the dataset. The top-left inset visualises the geometry configuration reference.}
    \label{fig:forces-vs-geomPars12}
\end{figure}

\begin{figure}[p]
    \centering
    
    \begin{subfigure}[t]{0.48\textwidth}
        \flushleft \includegraphics[width=0.35\textwidth]{images/highlight_IBFLAP.png} \\
        \vspace{-0.5ex} 
        \centering \includegraphics[width=1\textwidth]{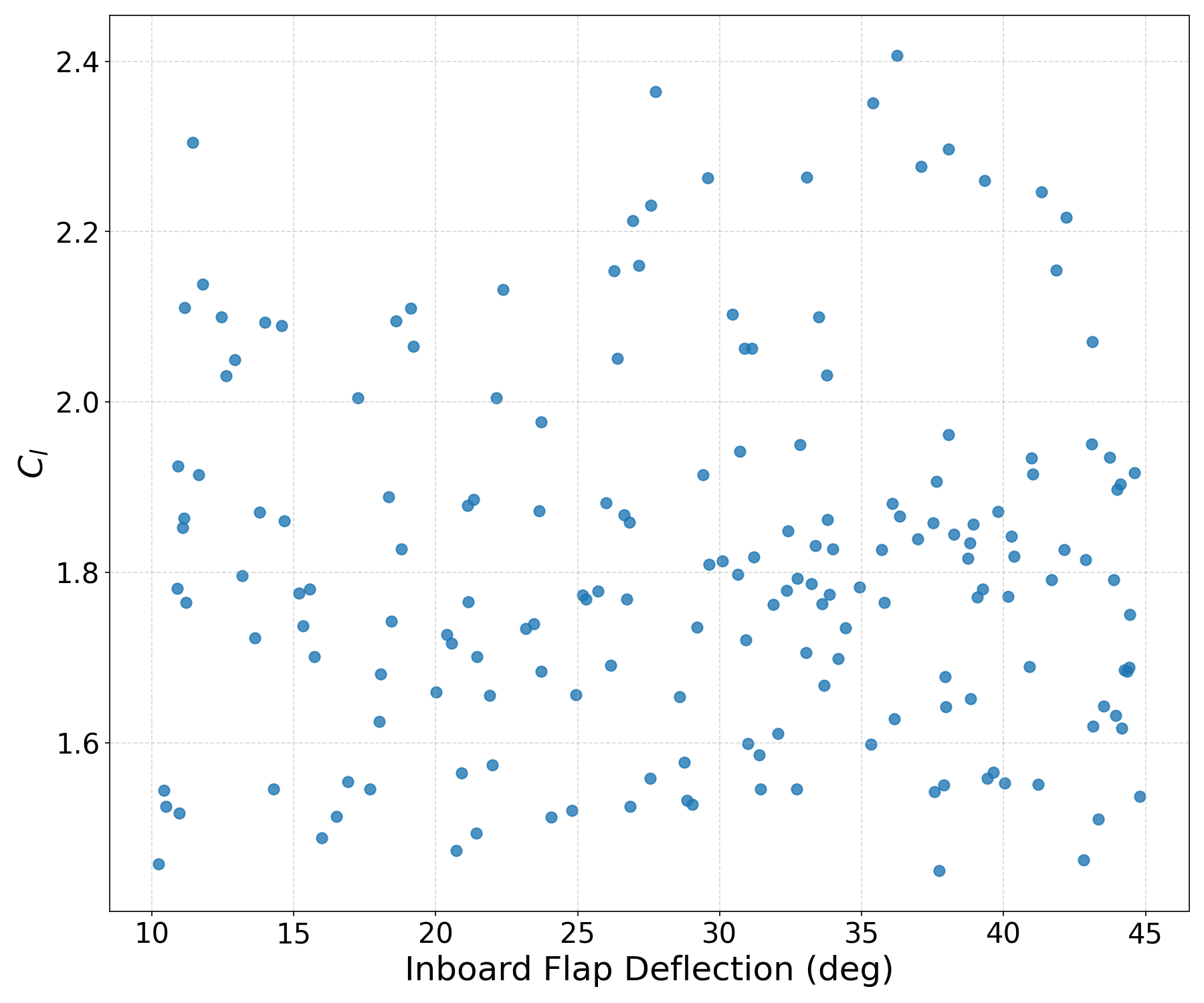}
        \caption{Inboard Flap (22$^\circ$ AoA)}
    \end{subfigure}
    \hfill
    \begin{subfigure}[t]{0.48\textwidth}
        \flushleft \includegraphics[width=0.35\textwidth]{images/highlight_OBFLAP.png} \\
        \vspace{-0.5ex}
        \centering \includegraphics[width=1\textwidth]{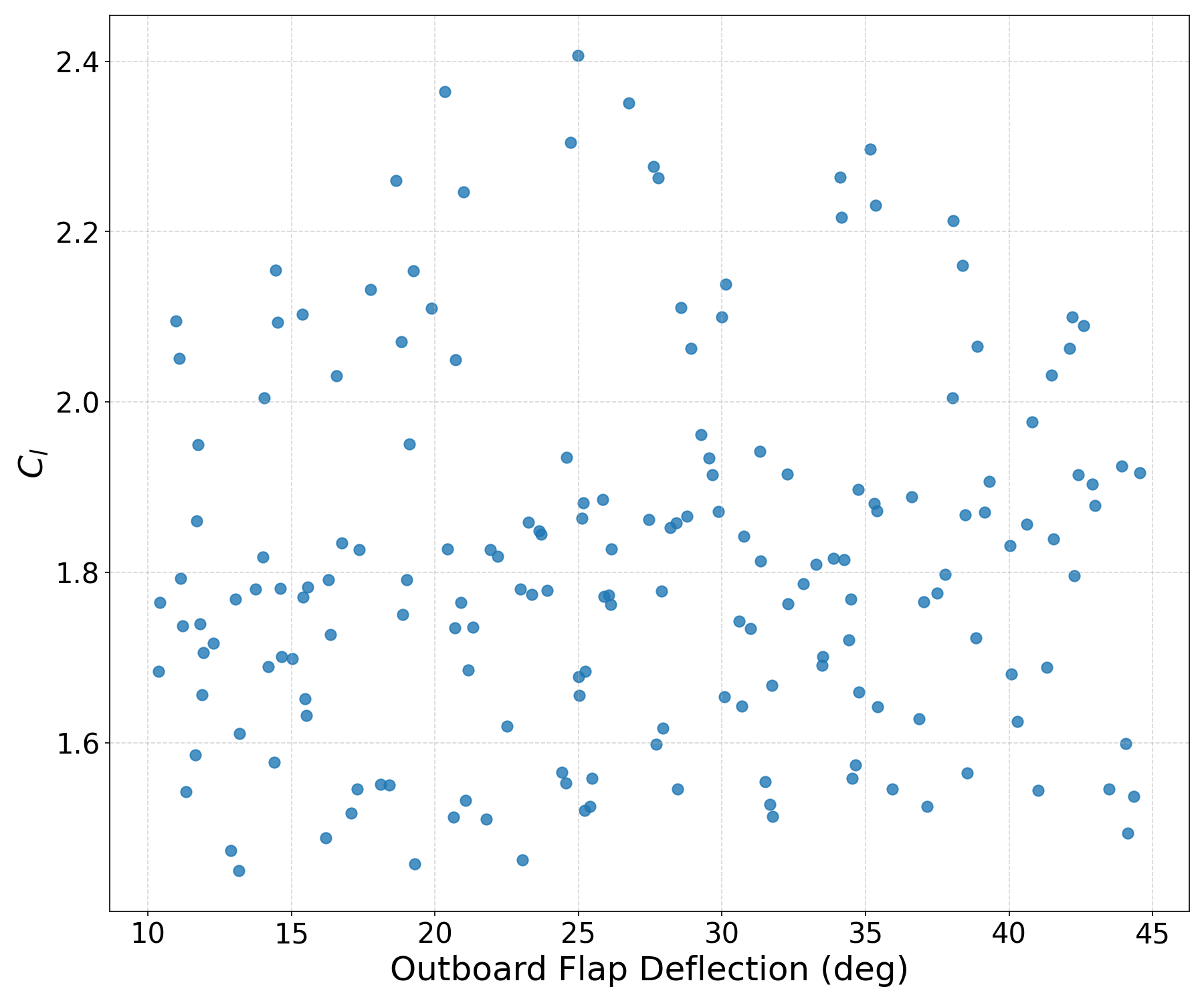}
        \caption{Outboard Flap (22$^\circ$ AoA)}
    \end{subfigure}

    \vspace{4ex} %

    \begin{subfigure}[t]{0.48\textwidth}
        \flushleft \includegraphics[width=0.35\textwidth]{images/highlight_IBSLAT.png} \\
        \vspace{-0.5ex}
        \centering \includegraphics[width=1\textwidth]{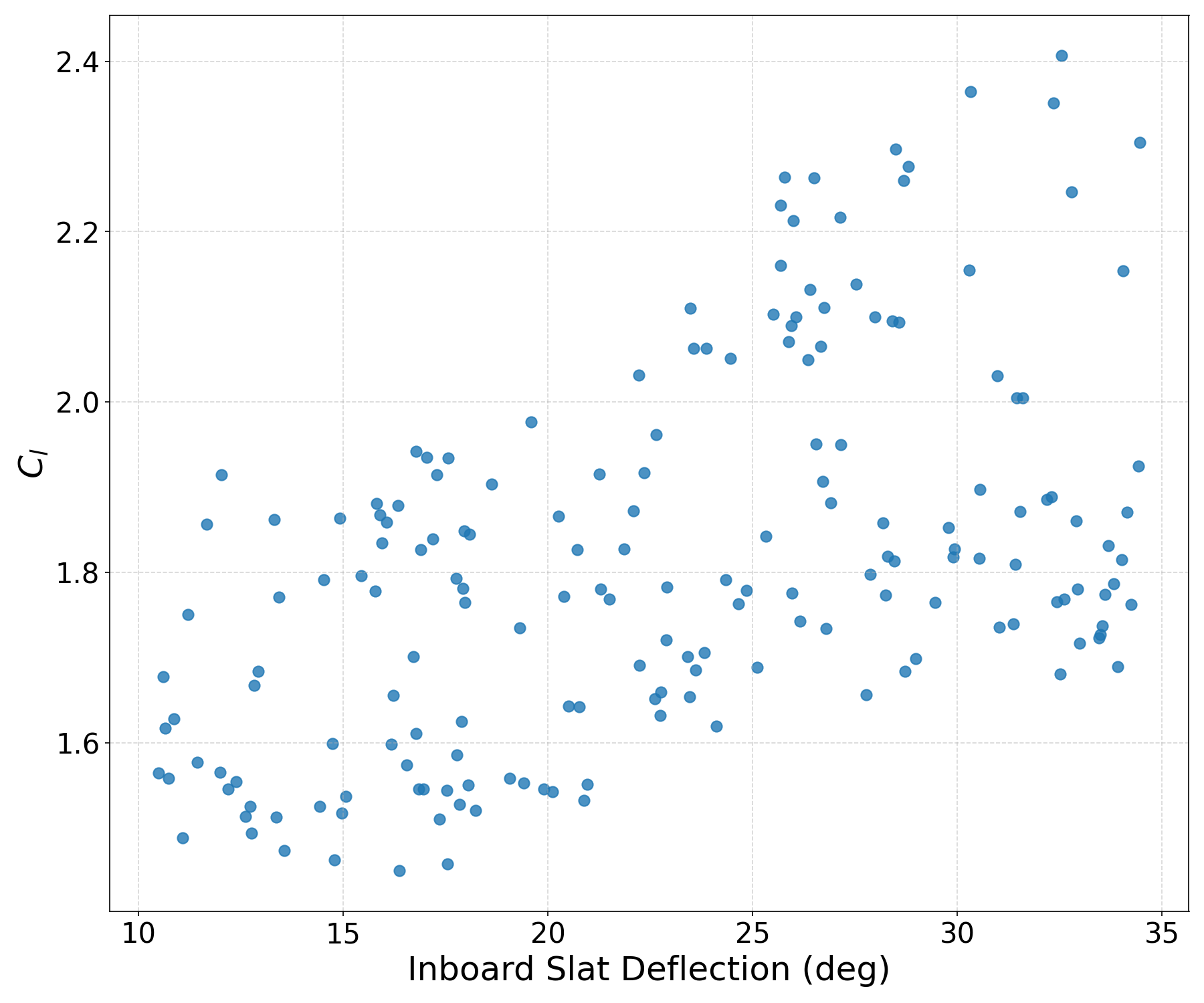}
        \caption{Inboard Slat (22$^\circ$ AoA)}
    \end{subfigure}
    \hfill
    \begin{subfigure}[t]{0.48\textwidth}
        \flushleft \includegraphics[width=0.35\textwidth]{images/highlight_OBSLAT.png} \\
        \vspace{-0.5ex}
        \centering \includegraphics[width=1\textwidth]{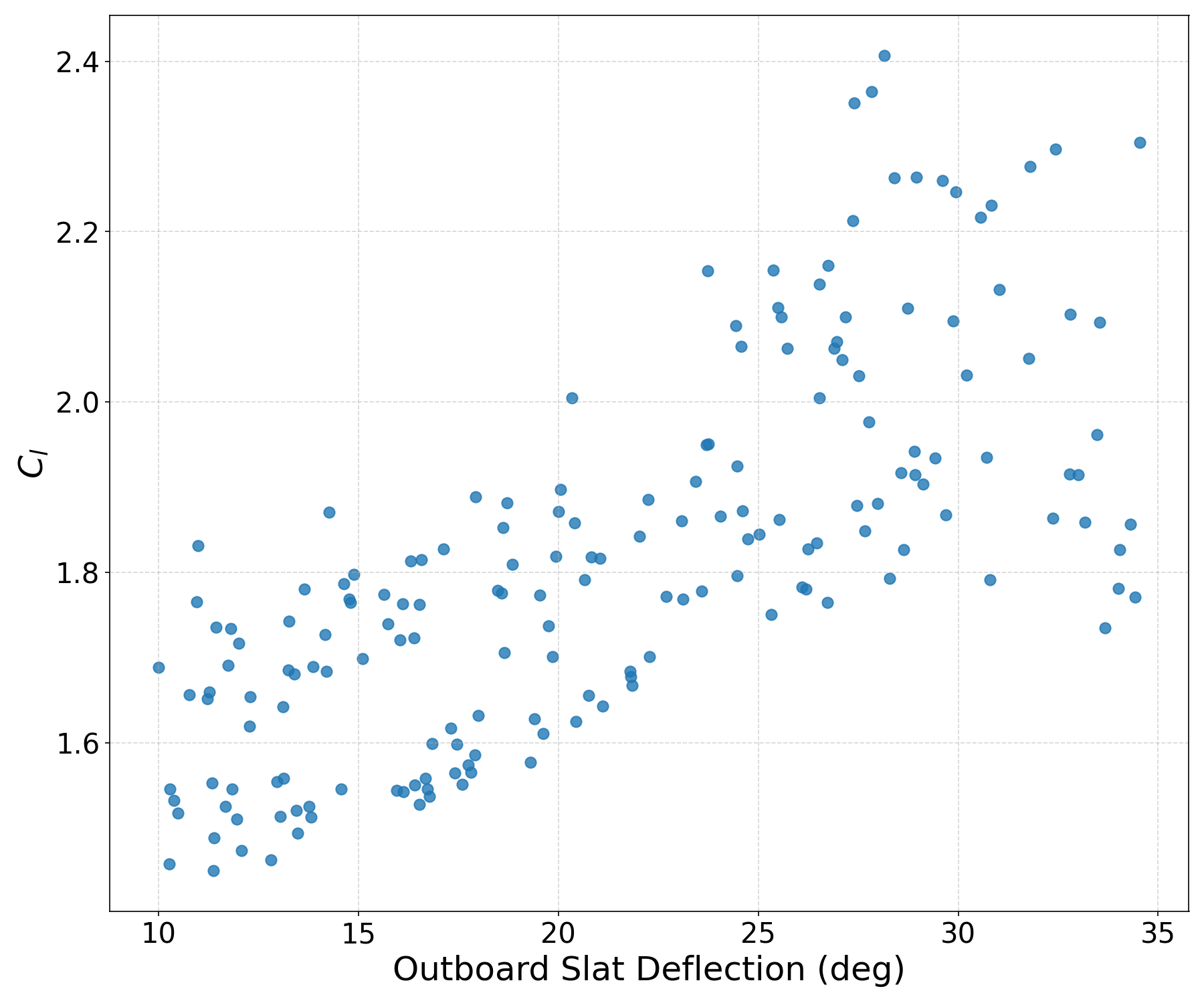}
        \caption{Outboard Slat (22$^\circ$ AoA)}
    \end{subfigure}

    \vspace{2ex}
    \caption{Variation of force coefficients with selected geometry parameters over all samples and 22 degrees AoA in the dataset. The top-left inset visualises the geometry configuration reference.}
    \label{fig:forces-vs-geomPars22}
\end{figure}

To further illustrate the dependencies between geometric parameters and aerodynamic performance, Fig.~\ref{fig:forces-vs-geomParsall-2x2} and Fig.~\ref{fig:forces-vs-geomPars-2x2} plot the Lift Coefficient ($C_L$) against key deflection parameters.
Several clear trends emerge.
First, there is a primary positive correlation between the inboard flap deflection and $C_L$, which is consistent with the increase in effective camber provided by flap deployment.
Second, a distinct stratification by Angle of Attack is visible (Fig.~\ref{fig:forces-vs-geomParsall-2x2}), where the spacing between AoA bands compresses at the highest angles ($20^\circ-22^\circ$), reflecting the non-linear aerodynamic behavior associated with the onset of stall.
These plots demonstrate that the dataset captures the expected physics of high-lift devices, providing a robust ground truth for ML model training.

\noindent
A closer examination of the parametric dependencies in Fig.~\ref{fig:forces-vs-geomParsall-2x2} through Fig.~\ref{fig:forces-vs-geomPars22} reveals a fundamental shift in aerodynamic sensitivity as the aircraft progresses through the flight envelope.
At lower angles of attack ($\alpha=4^{\circ}$ and $12^{\circ}$), the lift coefficient ($C_L$) is primarily driven by Inboard and Outboard Flap deflections. 
Physically, this behavior aligns with the linear aerodynamic regime where flow is largely attached; here, lift generation is dominated by the increase in circulation resulting from the added effective camber of the trailing-edge flaps. 
In this regime, the slats play a secondary role, as the leading-edge suction peaks are not yet severe enough to induce separation. 
However, as the angle of attack increases to $\alpha=22^{\circ}$ (the near-stall/post-stall regime), the dominant sensitivity shifts markedly toward the Inboard and Outboard Slat deflections. 
At these high angles, the flow physics are governed by the stability of the boundary layer near the leading edge. 
While aggressive flap settings provide the \textit{potential} for higher lift via increased camber, that potential is only realizable if the slat successfully mitigates the strong leading-edge pressure gradients to prevent massive flow separation. 
Consequently, at $\alpha=22^{\circ}$, the slat configuration acts as the limiting factor for aerodynamic performance, resulting in the higher correlation between slat deflection and $C_L$ observed in the dataset.

\subsubsection{Flow fields}

\begin{figure}[t]
\vskip 0.2in
\begin{center}
\centerline{\includegraphics[width=\textwidth]{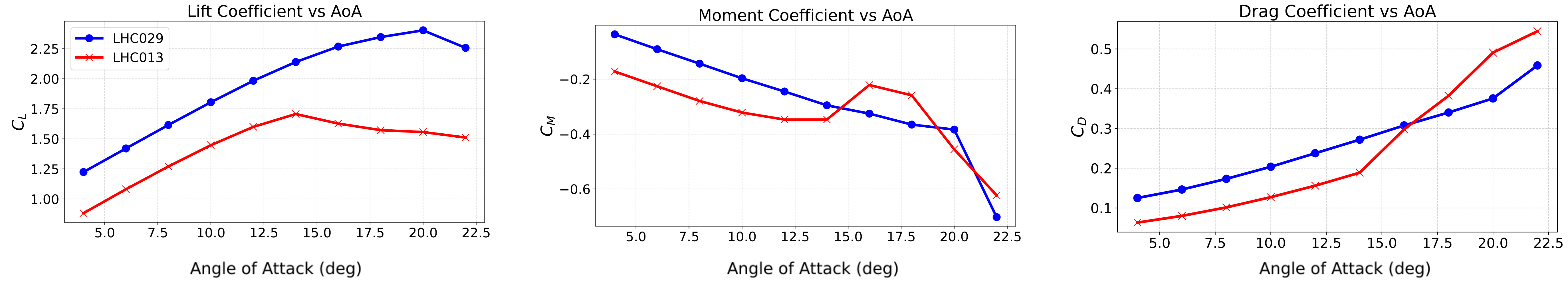}}
\caption{Comparison of integrated forces (lift, pitching moment, and drag) across the angle of attack sweeps for the two configurations (LHC029 and LHC013).}
\label{fig:highlow}
\end{center}
\vskip -0.2in
\end{figure}

To demonstrate the dataset's capability to capture diverse flow physics across the design space, we compare two distinct geometric configurations: LHC013 and LHC029 (see Table~\ref{tab:geom_data_vertical} for parameter details).
These cases illustrate the sensitivity of the flow field to different high-lift settings.
LHC013 represents a configuration with lower high-lift deflections (inboard flap $\approx 11^\circ$), whereas LHC029 features significantly more aggressive settings (inboard flap $\approx 39^\circ$).
As shown in Fig.~\ref{fig:highlow}, the higher deflection settings of LHC029 yield significantly higher lift coefficients across the linear range.
However, this performance benefit incurs a substantial penalty in drag ($C_D$) and pitching moment ($C_M$).
At higher angles of attack, LHC013 exhibits stall behavior that actually results in higher drag than the attached flow of LHC029, illustrating the complex trade-offs captured in the dataset.

\begin{figure}[!htb]
\centering
\includegraphics[width=0.73\linewidth]{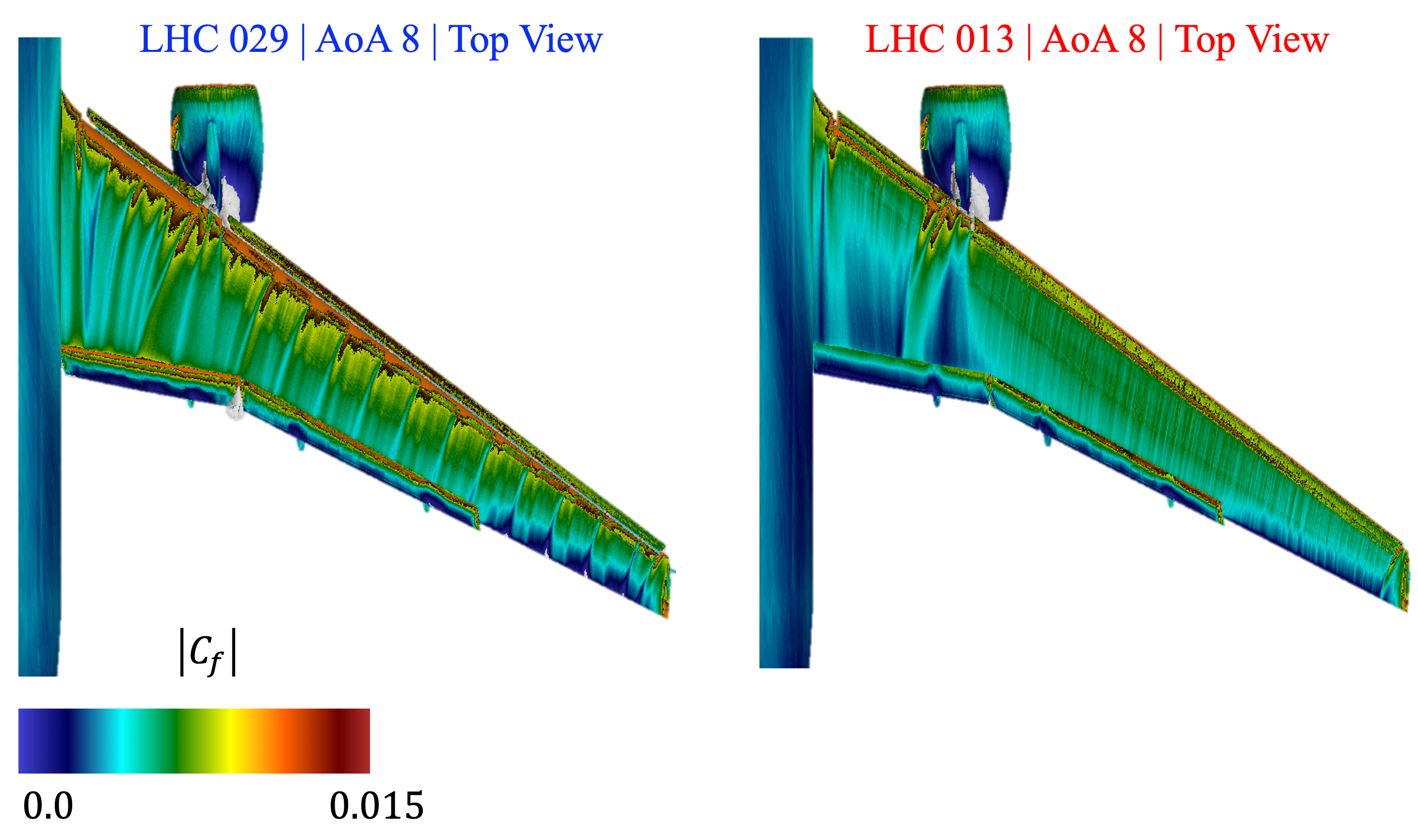}\par
\smallskip
\includegraphics[width=0.73\linewidth]{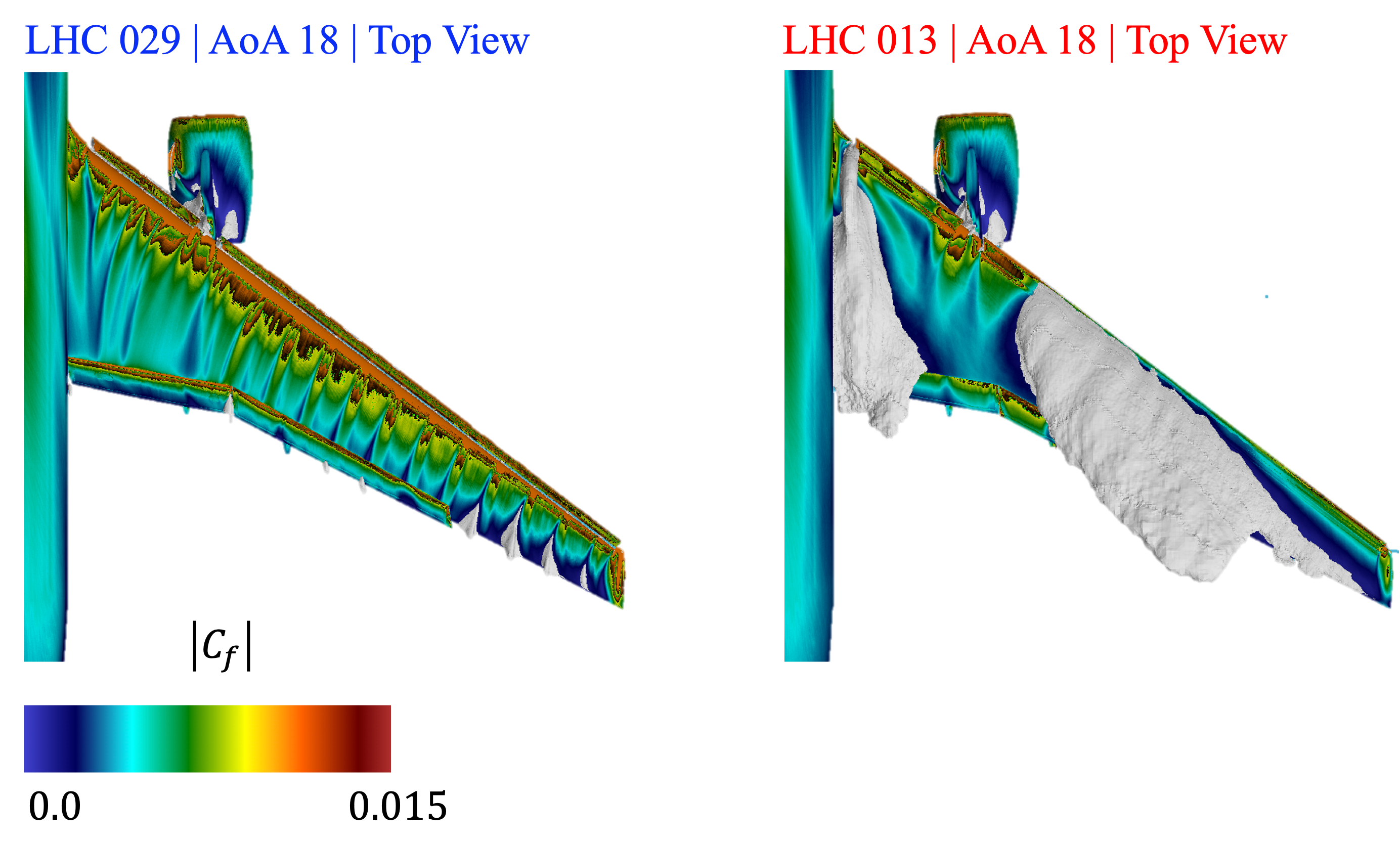}
\caption{Comparison of absolute value of the skin-friction on the wing-surfaces, $|C_f|$, at two specific AoA, $\alpha=8^\circ$ and $18^\circ$, for the two configurations (LHC029 and LHC013), illustrating two different flow characteristics. The white isosurface denotes the $u_x/U_\infty     \approx 0$ region where $x$ is the body-axis coordinate and $U_\infty$ denotes the freestream velocity magnitude.}
\label{fig:highlow2_main}
\end{figure}

\begin{figure}[htb]
    \centering
    \includegraphics[width=0.62\textwidth]{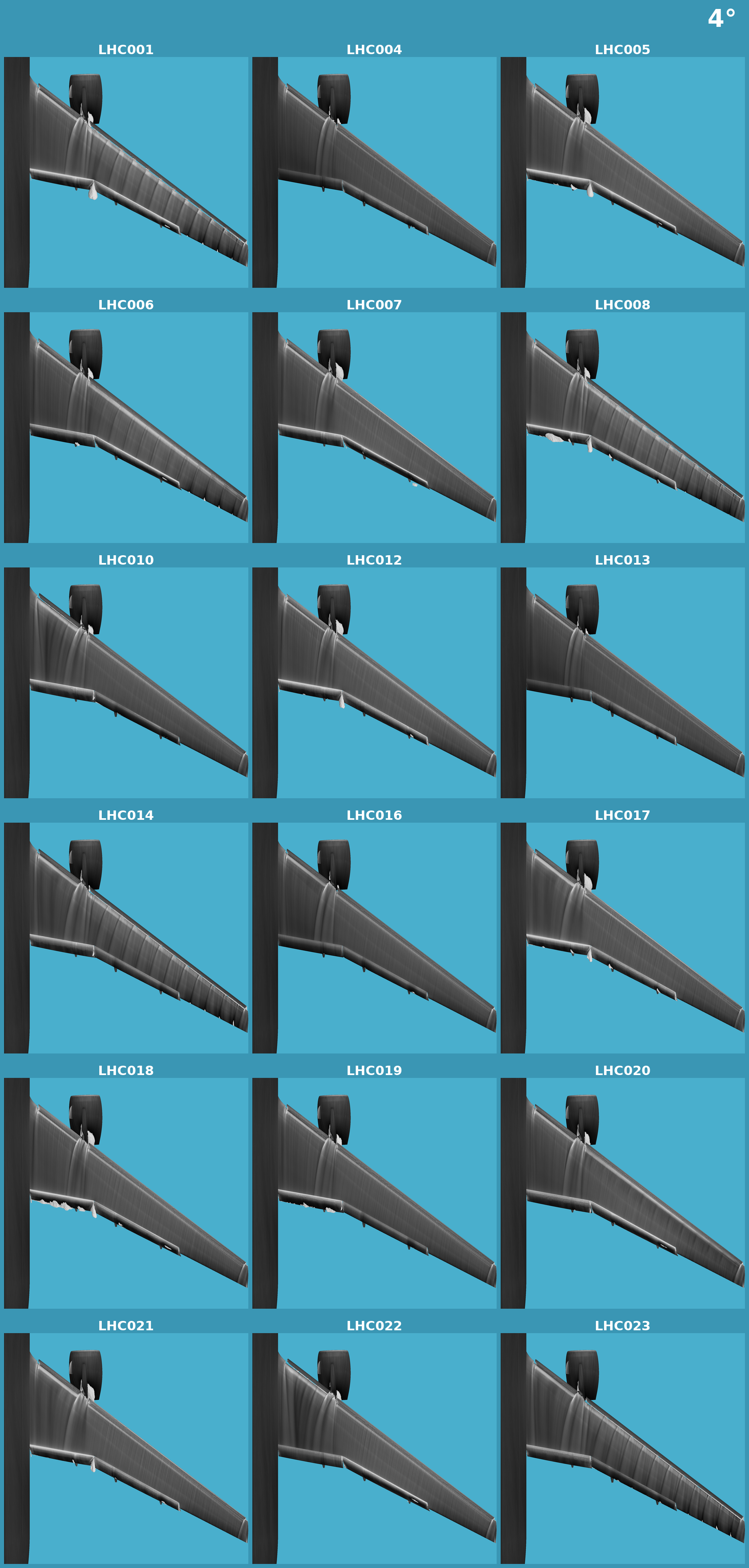}
    \caption{Iso-surfaces of negative velocity (i.e flow separation) for runs 1 to 20 at 4 degrees AoA - top view}
\label{fig:isoqtop4appendix}
\end{figure}

\begin{figure}[htb]
    \centering
    \includegraphics[width=0.9\textwidth]{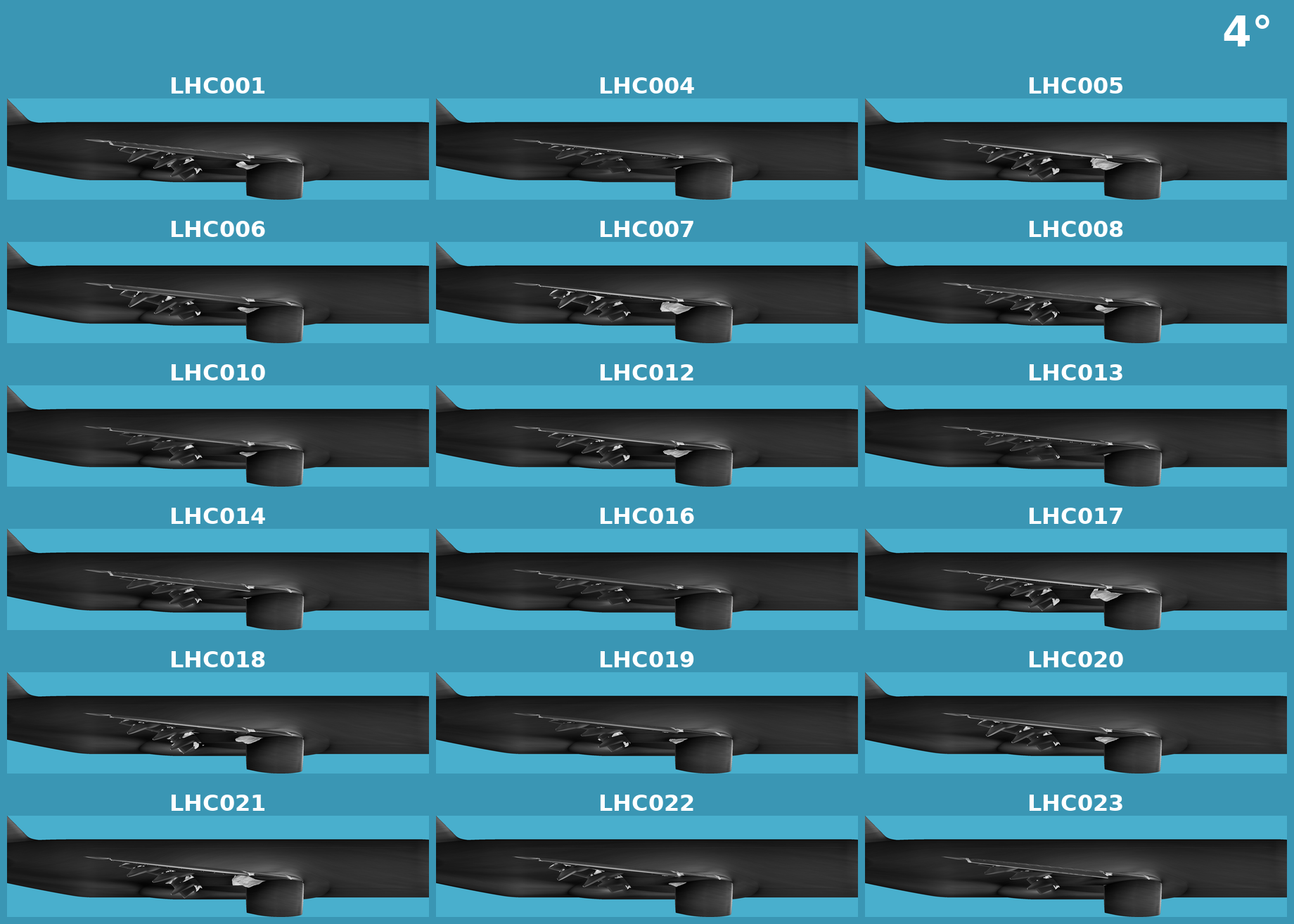}
    \caption{Iso-surfaces of negative velocity (i.e flow separation) for runs 1 to 20 at 4 degrees AoA - side view}
\label{fig:isoqside4appendix}
\end{figure}

\begin{figure}[htb]
    \centering
    \includegraphics[width=0.9\textwidth]{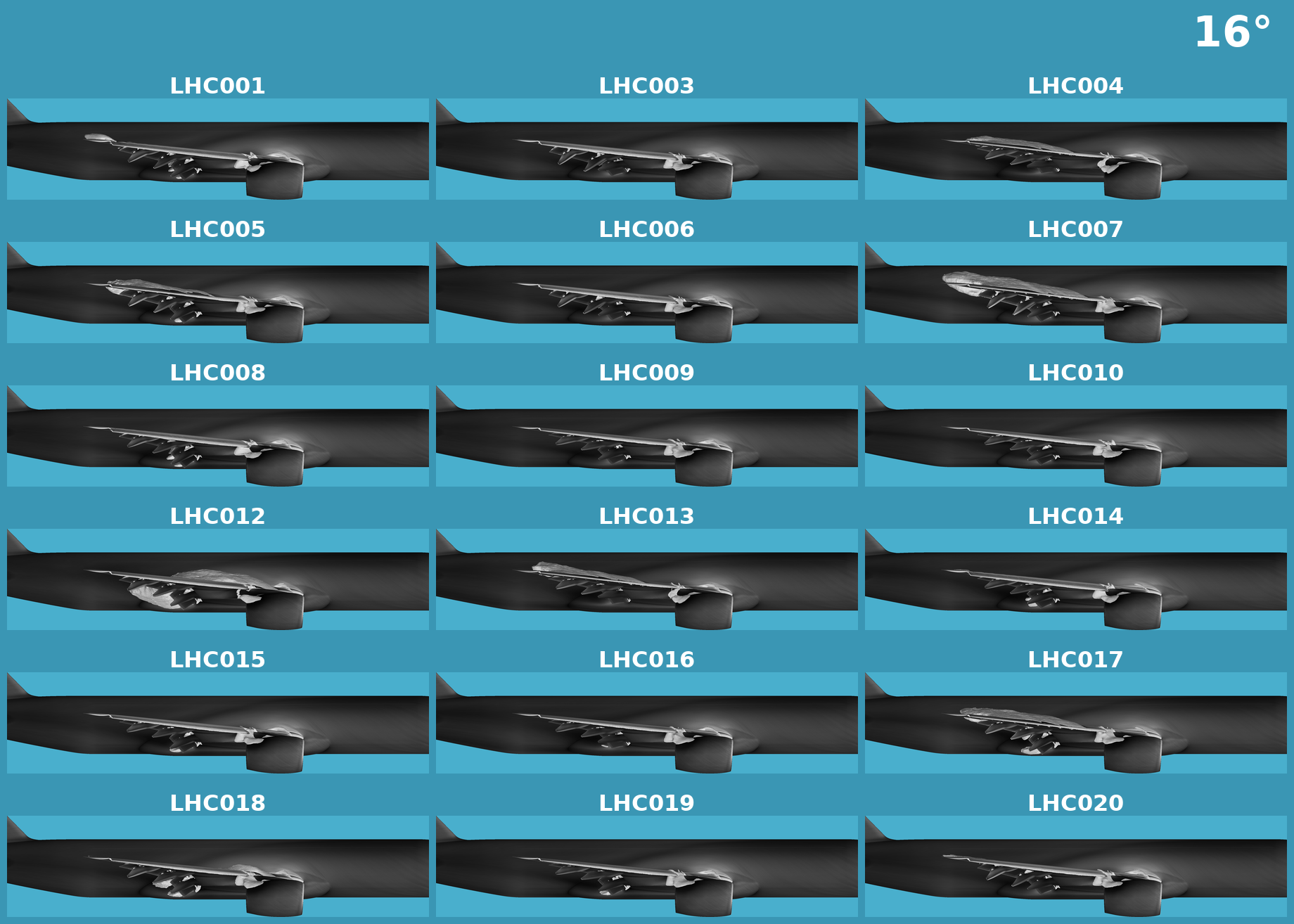}
    \caption{Iso-surfaces of negative velocity (i.e flow separation) for runs 1 to 20 at 16 degrees AoA - side view}
\label{fig:isoqside16appendix}
\end{figure}

\begin{figure}[htb]
    \centering
    \includegraphics[width=0.62\textwidth]{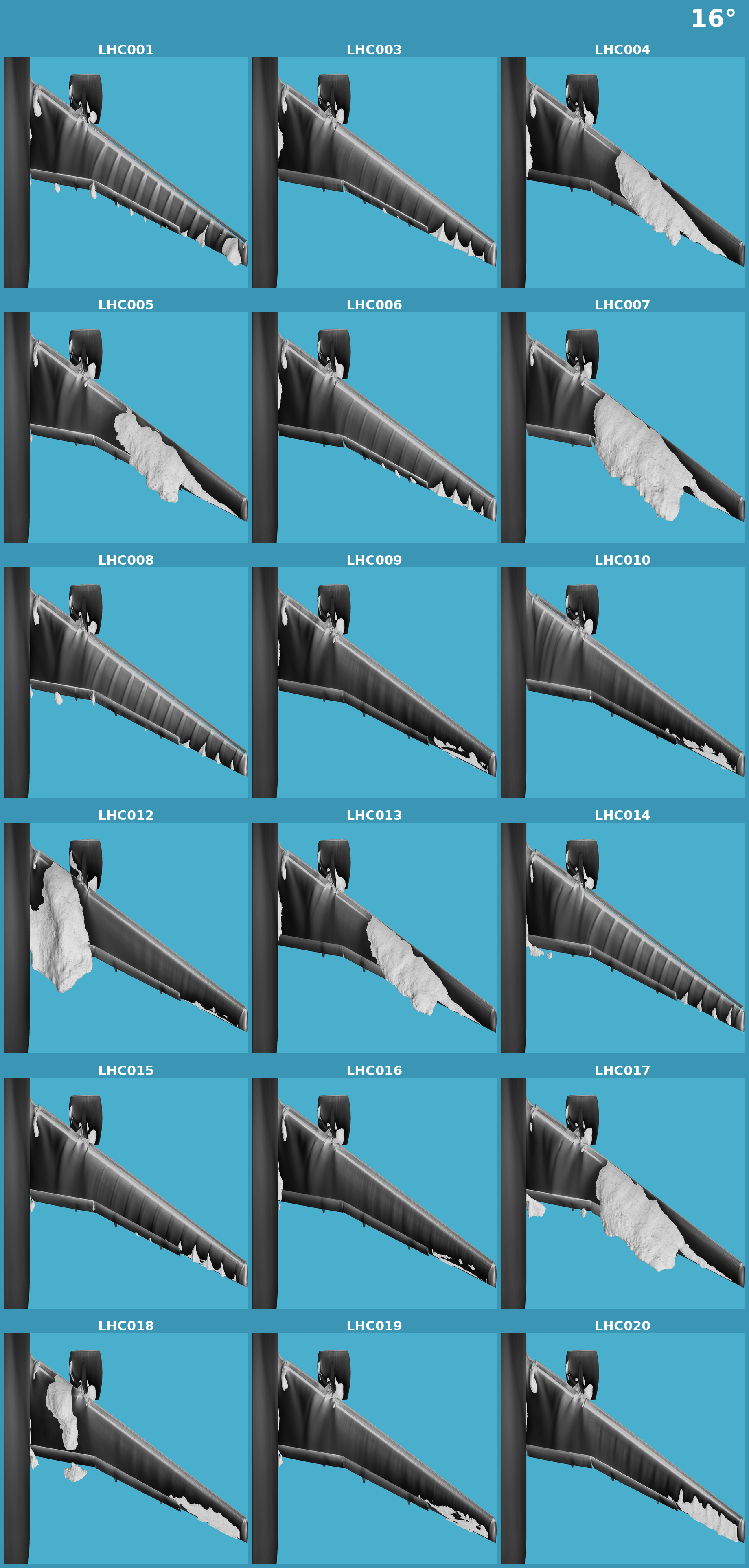}
    \caption{Iso-surfaces of negative velocity (i.e flow separation) for runs 1 to 20 at 16 degrees AoA - top view}
\label{fig:isoqtop16appendix}
\end{figure}

\begin{figure}[htb]
    \centering
    \includegraphics[width=0.9\textwidth]{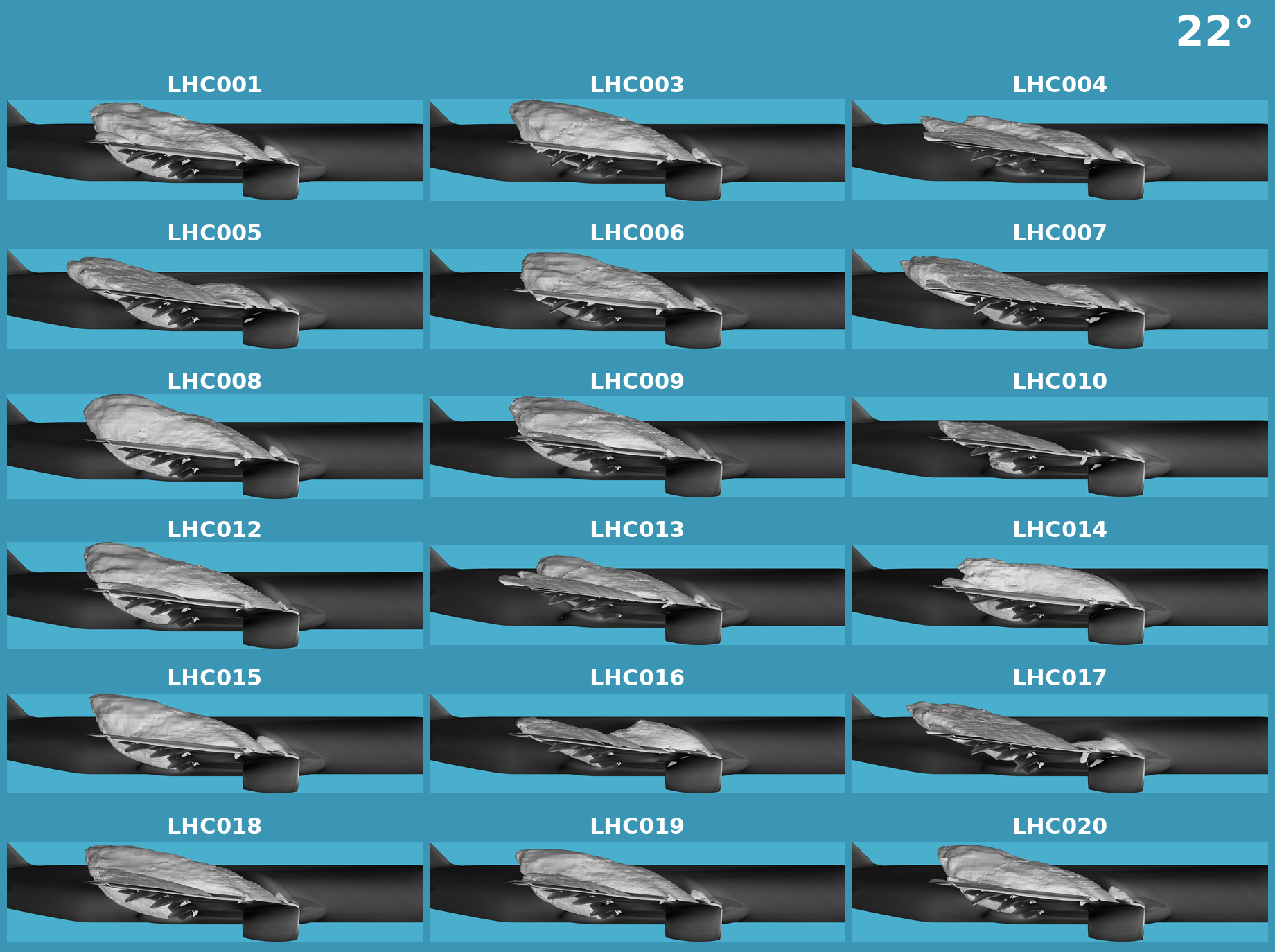}
    \caption{Iso-surfaces of negative velocity (i.e flow separation) for runs 1 to 20 at 22 degrees AoA - side view}
\label{fig:isoqside22appendix}
\end{figure}

\begin{figure}[htb]
    \centering
    \includegraphics[width=0.62\textwidth]{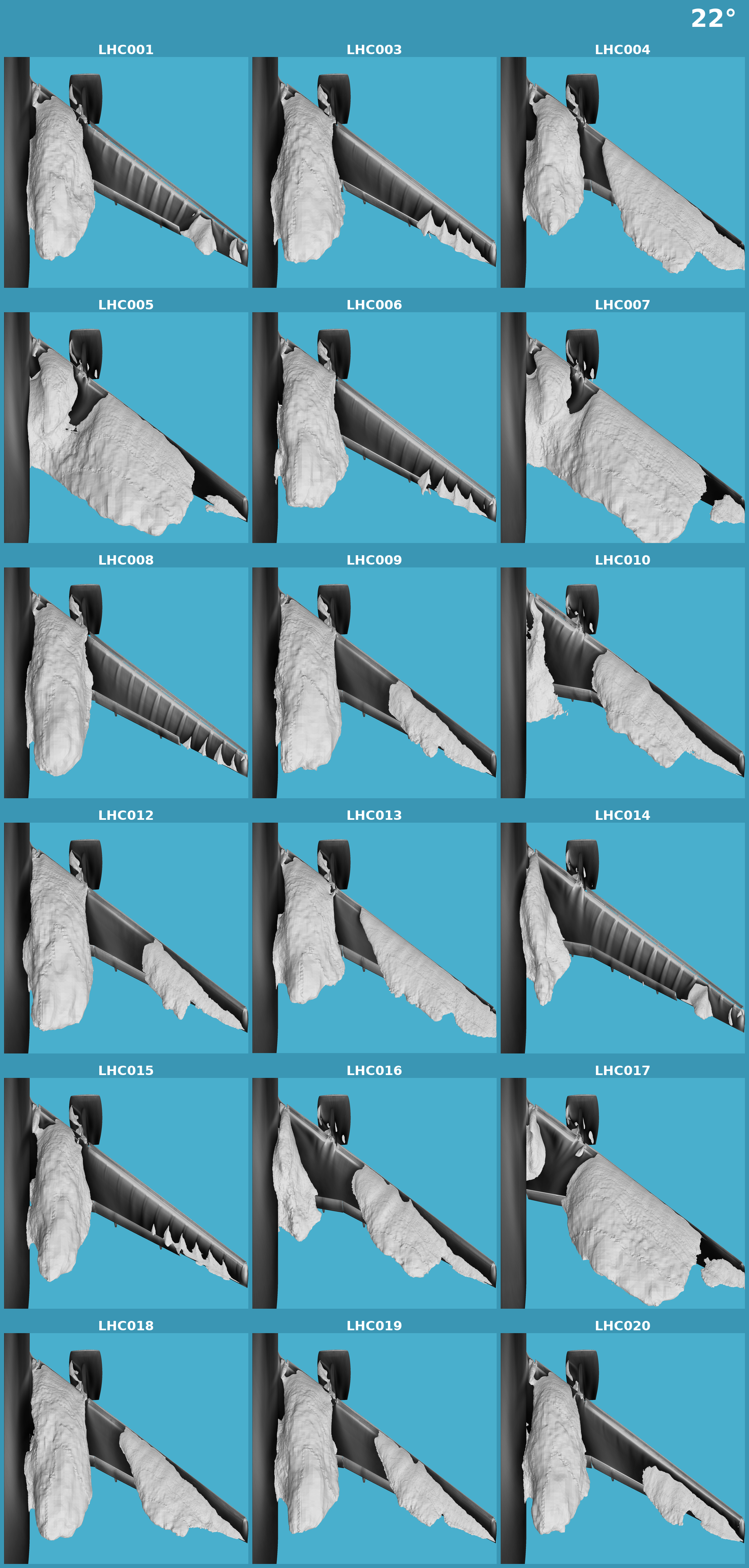}
    \caption{Iso-surfaces of negative velocity (i.e flow separation) for runs 1 to 20 at 22 degrees AoA - top view}
\label{fig:isoqtop22appendix}
\end{figure}

The differences in the flow physics are further visualized in Fig.~\ref{fig:highlow2_main}, which compares the skin-friction coefficient ($|C_f|$) on the wing surfaces.
At $\alpha=18^\circ$, LHC029 maintains attached flow over a wider region of the wing element compared to LHC013, which shows large-scale flow separation (indicated by the white isosurfaces of zero streamwise velocity).
The progression of this separation is systematically captured in the dataset. 
Figures \ref{fig:isoqtop4appendix} through \ref{fig:isoqtop22appendix} display iso-surfaces of negative streamwise velocity for the first 20 runs at $AoA=4^\circ$, $16^\circ$, and $22^\circ$.
At $4^\circ$ (Fig.~\ref{fig:isoqtop4appendix}), the flow is largely attached across all geometries. 
By $16^\circ$ (Fig.~\ref{fig:isoqtop16appendix}), significant separation pockets appear, particularly on the outboard wing sections. 
Finally, at $22^\circ$ (Fig.~\ref{fig:isoqtop22appendix}), many configurations exhibit massive separation and deep stall, presenting a challenging prediction task for surrogate models.

\begin{table}[h]
    \centering
    \begin{tabular}{lrr}
        \toprule
        \textbf{Metric} & \textbf{LHC013} & \textbf{LHC029} \\
        \midrule
        IB\_Flap\_Deflection       & 10.97 & 39.34 \\
        OB\_Flap\_Deflection       & 17.09 & 18.65 \\
        IB\_Flap\_Gap\_Multiplier  & 1.33  & 0.74  \\
        OB\_Flap\_Gap\_Multiplier  & 1.46  & 1.06  \\
        IB\_Slat\_Deflection       & 14.97 & 28.69 \\
        OB\_Slat\_Deflection       & 10.48 & 29.61 \\
        IB\_Slat\_Gap\_Multiplier  & 1.11  & 1.50  \\
        OB\_Slat\_Gap\_Multiplier  & 1.03  & 1.42  \\
        \bottomrule
    \end{tabular}
    \caption{Geometric Parameters for LHC013 and LHC029}
    \label{tab:geom_data_vertical}
\end{table}

\newpage

\newpage
\clearpage
\section{Datasheet}
\subsection{Motivation}

\begin{itemize}

\item \textbf{For what purpose was the dataset created?} The dataset was created to facilitate the development and testing of machine learning methods for high-lift aircraft aerodynamics. It addresses the lack of high-fidelity, scale-resolving CFD training data for complex 3D airframes in the high-lift regime.

\item \textbf{Who created the dataset (e.g., which team, research group) and on behalf of which entity (e.g., company, institution, organization)?} The dataset was created by a collaboration between NVIDIA, Cadence Design Systems, and The Boeing Company.

\item \textbf{Who funded the creation of the dataset?} Computing resources were provided by Cadence, NVIDIA, the Texas Advanced Computing Center (TACC), and the Swiss National Supercomputing Centre (CSCS). Additional support was provided by the Oak Ridge Leadership Computing Facility (ORNL).

\end{itemize}

\subsection{Distribution}

\begin{itemize}

\item \textbf{Will the dataset be distributed to third parties outside of the entity (e.g., company, institution, organization) on behalf of which the dataset was created?} Yes, the dataset is open to the public.

\item \textbf{How will the dataset will be distributed (e.g., tarball on website, API, GitHub)?} The dataset is free to download from HuggingFace.
\ifdefined\HiLiftAnonymous\else
The companion score-reproduction package and paper checkpoints are distributed separately through the Figshare artifact release~\citep{HiLiftAeroMLArtifacts2026}.
\fi

\item \textbf{When will the dataset be distributed?} The dataset is already available to download via HuggingFace.

\item \textbf{Will the dataset be distributed under a copyright or other intellectual property (IP) license, and/or under applicable terms of use (ToU)?} The dataset is licensed under CC-BY-4.0, as declared in the dataset card. Appendix~\ref{app:dataset-license} links to the license terms.
\ifdefined\HiLiftAnonymous\else
The companion checkpoint weights are also licensed under CC-BY-4.0; software retains the existing licenses identified in the artifact release.
\fi

\item \textbf{Have any third parties imposed IP-based or other restrictions on the data associated with the instances?} No.

\item \textbf{Do any export controls or other regulatory restrictions apply to the dataset or to individual instances?} No.

\end{itemize}

\subsection{Maintenance}

\begin{itemize}

\item \textbf{Who will be supporting/hosting/maintaining the dataset?} The dataset is hosted on HuggingFace and maintained by the authors (primarily NVIDIA).

\item \textbf{How can the owner/curator/manager of the dataset be contacted (e.g., email address)?} The dataset maintainers can be contacted at contact@caemldatasets.org.

\item \textbf{Is there an erratum?} Yes. The release documents six force/moment-mean corrections and invalid all-zero volume rows. Appendix~\ref{app:dataset-quality} identifies the affected cases and links to the correction and data-quality records.

\item \textbf{Will the dataset be updated (e.g., to correct labeling errors, add new instances, delete instances)?} Yes, the dataset may be updated to address errors or provide extra functionality.

\item \textbf{If the dataset relates to people, are there applicable limits on the retention of the data associated with the instances?} N/A

\item \textbf{Will older versions of the dataset continue to be supported/hosted/maintained?} The repository revision history records changes. Users should record the immutable revision used for their experiments; Appendix~\ref{app:dataset-access} gives a revision-pinned download example.

\item \textbf{If others want to extend/augment/build on/contribute to the dataset, is there a mechanism for them to do so?} Users can modify and redistribute the data under the terms of the CC-BY-4.0 license.

\end{itemize}

\subsection{Composition}

\begin{itemize}

\item \textbf{What do the instances that comprise the dataset represent?} Each instance represents the high-fidelity aerodynamic field of the NASA Common Research Model High-Lift (CRM-HL) aircraft configuration.

\item \textbf{How many instances are there in total?} There are 1,800 samples in total, comprising 180 geometry variants each simulated at 10 angles of attack.

\item \textbf{Does the dataset contain all possible instances or is it a sample?} The dataset is a sample of 180 geometry variants generated using Latin Hypercube Sampling from a parametric design space. It covers 10 angles of attack ranging from $4^{\circ}$ to $22^{\circ}$ to capture pre-stall and post-stall regimes.

\item \textbf{What data does each instance consist of?} Each instance consists of the CAD geometry (.stp/.stl), time-averaged volume and surface fields (e.g., pressure, velocity, shear stress), and integrated aerodynamic forces and moments.

\item \textbf{Is there a label or target associated with each instance?} Each instance is indexed by its geometry ID and Angle of Attack (AoA).

\item \textbf{Is any information missing from individual instances?} Yes. Invalid all-zero volume rows occur in 1,768 of the 1,800 cases and must be masked. Their raw point IDs and coordinates are provided in the data-quality records; Appendix~\ref{app:dataset-quality} gives the counts and validity rule.

\item \textbf{Are relationships between individual instances made explicit?} Yes. Each geometry ID occurs at ten angles of attack, from $4^\circ$ to $22^\circ$ in $2^\circ$ increments. Case names and metadata identify the shared geometry and the operating condition.

\item \textbf{Are there recommended data splits?} Yes. The released \texttt{splits/manifest.json} defines 14 deterministic split families, described in Appendix~\ref{app:detailedml}. Use \texttt{geometry} for unseen-geometry evaluation, \texttt{full} for case-level interpolation, their nested variants for data efficiency, the single-AoA splits for fixed-condition comparisons, and \texttt{aoa}, \texttt{deflection}, and \texttt{stall} for the specified distribution shifts. The \texttt{full} split can share geometry IDs across partitions at different angles, whereas the geometry-based splits keep all conditions of a geometry together.

\item \textbf{Are there any errors, sources of noise, or redundancies in the dataset?} Finite-duration LES averages contain statistical noise, particularly in separated flow. The release also includes the force/moment corrections and invalid volume rows described in Appendix~\ref{app:dataset-quality}. For the six corrected cases, retained monitor standard deviations, CTU-based standard errors and bands, and convergence plots are not uncertainty estimates for the substituted means. The averaging procedure is described in Appendix~\ref{app:postPro-flowStats}.

\item \textbf{Is the dataset self-contained?} Yes, the dataset is self-contained.

\item \textbf{Does the dataset contain data that might be considered confidential?} No.

\item \textbf{Does the dataset contain data that might be considered offensive?} No.

\end{itemize}

\subsection{Collection Process}

\begin{itemize}

\item \textbf{How was the data associated with each instance acquired?} The data was generated using high-fidelity Wall-Modeled Large Eddy Simulation (WMLES) on solution-adapted grids.

\item \textbf{What mechanisms or procedures were used to collect the data?} Simulations were performed using the ``Fidelity Charles'' explicit finite-volume flow solver and ``Fidelity Stitch'' Voronoi mesher. The methodology was validated against wind tunnel experimental data for the CRM-HL.

\item \textbf{If the dataset is a sample from a larger set, what was the sampling strategy?} The geometric parameters were sampled using Latin Hypercube Sampling to ensure coverage of the design space.

\item \textbf{Who was involved in the data collection process?} The data generation was carried out by the authors from NVIDIA, Cadence, and Boeing.

\item \textbf{Over what timeframe was the data collected?} The data was generated between October 2025 and January 2026.

\item \textbf{Were any ethical review processes conducted?} An ethical review was not considered necessary as the data is synthetic engineering data.

\end{itemize}

\subsection{Preprocessing/cleaning/labeling}

\begin{itemize}

\item \textbf{Was any preprocessing/cleaning/labeling of the data done?} The raw time-dependent LES solution was time-averaged, and volume data was exported in an Octree format. The release documents six force/moment-mean substitutions and identifies invalid all-zero volume rows to be masked before use; see Appendix~\ref{app:dataset-quality}.

\item \textbf{Was the ``raw'' data saved in addition to the preprocessed/cleaned/labeled data?} The full time-history of the 3D fields was not saved due to excessive storage requirements.

\item \textbf{Is the software that was used to preprocess/clean/label the data available?} The CFD generation workflow uses commercial software (Fidelity Charles). Public training and inference components, including HiLift configurations, are available in the \href{https://github.com/NVIDIA/physicsnemo/tree/d25b6361e16457370bd2c0558add625cb367c245/examples/cfd/external_aerodynamics/unified_external_aero_recipe}{NVIDIA PhysicsNeMo external-aerodynamics recipe}. This link identifies a public recipe revision, rather than a complete record of the preprocessing, checkpoints, and evaluator used for every result reported here.
\ifdefined\HiLiftAnonymous\else
The companion artifact release~\citep{HiLiftAeroMLArtifacts2026} additionally provides the evaluated model exports, resolved configurations, normalization and split records, retained recipe sources, and executable CPU score replay. Its preliminary FluidsBench scoring version, verified reproduction scope, and remaining native-evaluator and runtime gaps are documented in Appendix~\ref{app:ml_provenance} and the release documentation.
\fi

\end{itemize}

\subsection{Uses}

\begin{itemize}

\item \textbf{Has the dataset been used for any tasks already?} Yes, the dataset has been used to benchmark GeoTransolver and Transolver ML architectures.

\item \textbf{Is there a repository that links to any or all papers or systems that use the dataset?} The dataset page on HuggingFace and the associated website (caemldatasets.org) will serve as hubs for this information.

\item \textbf{What (other) tasks could the dataset be used for?} The dataset allows for the exploration of complex high-lift flow physics, such as slat wakes and confluent boundary layers, beyond just ML surrogate training.

\item \textbf{Is there anything about the composition of the dataset that might impact future uses?} The simulations are at a fixed Mach number (0.2) and Reynolds number ($1.6 \times 10^6$), which are specific to wind-tunnel conditions rather than full flight envelope.

\item \textbf{Are there tasks for which the dataset should not be used?} The dataset assumes fully turbulent flow (using wall modeling) and does not explicitly model laminar-turbulent transition, which may limit its use for specific transition-dominated regimes.

\end{itemize}

\end{document}